\documentclass{article}

\usepackage[english]{babel}

\usepackage[a4paper,top=2cm,bottom=2cm,left=3cm,right=3cm,marginparwidth=1.75cm]{geometry}

\usepackage{amssymb}
\usepackage{siunitx}
\PassOptionsToPackage{hyphens}{url}\usepackage{hyperref}
\usepackage{cleveref}
\usepackage[utf8]{inputenc}
\usepackage[right]{lineno}
\usepackage{csquotes}
\usepackage{booktabs}
\usepackage{longtable}
\usepackage{adjustbox}
\usepackage{array}
\usepackage{url}
\usepackage{bm}
\usepackage{titlesec}
\usepackage{authblk}
\usepackage{xcolor} % Load the xcolor package for color options

\usepackage{subfigure}
\titleformat{\subsection}
  {\mdseries\itshape\large} % Medium series, italic shape, and large font size
  {\thesubsection}{1em}{} % Numbering, spacing, and the section title itself

\usepackage[english]{babel}
\usepackage{natbib}
\crefformat{figure}{#2Figure~#1#3}
\Crefformat{figure}{#2Figure~#1#3}
\crefformat{table}{#2Table~#1#3}
\Crefformat{table}{#2Table~#1#3}
\crefformat{section}{#2Section~#1#3}
\Crefformat{section}{#2Section~#1#3}

\author[1]{Chady Ghnatios}
\author[2]{Thibault Xavier}
\author[3]{Anatoly S. Prokushkin}
\author[4]{Laurent Orgogozo}
\author[5]{Francisco Chinesta}

\affil[1]{Department of Mechanical Engineering, University of North Florida, Jacksonville, Florida, USA
chady.ghnatios@unf.edu (corr. author)}
\affil[2]{Laboratory TBI - Toulouse Biotechnology Institute, Bio \& Chemical Engineering, INRAE, UMR 792, Toulouse, France}
\affil[3]{V.N. Sukachev Institute of Forest SB RAS, Krasnoyarsk, Russia}
\affil[4]{Laboratory GET - Geoscience Environnement Toulouse, CNRS, UMR 5563, Toulouse, France}
\affil[5]{Laboratory PIMM, Arts et Metiers Institute of Technology, CNRS, CNAM, Paris, France}

\title{Permafrost augmented simulation and forecast using hybrid modeling through spectral neural networks}

\begin{document}
\maketitle

\begin{abstract}
\textbf{Purpose} - Global warming is impacting several crucial Earth systems, including the permafrost. This study investigates the possibility of performing fast and reliable simulation and forecasting of permafrost behavior, addressing a critical gap in the existing literature related to high computation times and energy costs of the simulations.

\textbf{Methodology} - A numerical cryohydrogeological study on water and energy transfer in a forested, continuous permafrost site has been conducted using parallel high-performance computing to simulate the permafrost dynamics. The numerical results reasonably replicate the measured soil temperature profiles and the dynamics of infiltrated waters. This work uses a spectral network to achieve a fast, reliable, and stable surrogate model for the simulations to forecast over long periods of time. The surrogate is also enhanced by assimilating experimental data through the development of a hybrid modeling framework.

\textbf{Findings} - The proposed approach improves the physical simulation while maintaining the stability of the dynamic model over long periods of time. The results reveal that the proposed hybrid modeling remains stable in integration and forecasting. The final hybrid solution demonstrates a significant improvement in simulation accuracy when tested against a test set of measurements. Predictions of the permafrost behavior in the coming decade are later made based on the CMIP6 climate change scenarios. 

\textbf{Originality} - This research offers a novel perspective on permafrost modeling that extends the current understanding of its behavior under different climate change scenarios. It shows a higher-than-expected temperature increase of the permafrost. Moreover, the stable forecasting of climate change effects is demonstrated. 
\\
\\
%add 6 keywords
\end{abstract}
%\linenumbers

\section{Introduction}\label{sec:introduction}
Permafrost is the soil permanently frozen in depth, covering a quarter of Northern Hemisphere lands. Due to climate warming, the permafrost is experiencing fast and widespread thawing \citep{nitze2018}, and this induces essential impacts in the Arctic, both on the environment (e.g., water resources, \citep{walvoord2016}) and on societies (e.g., infrastructure destabilization \citep{hjort2022}). These permafrost thaw impacts are expected to generate significant additional financial costs for maintaining key human activities. In fact, additional investments ranging from 200 to 600 billion euros will be required by 2085 to maintain normal human activity \citep{JIN2024102791}. Moreover, permafrost thaw will likely trigger critical climatic feedback \citep{Miner2022}. Thus, anticipating permafrost thaw by numerical simulations is paramount for ensuring the resilience of Arctic environments, societies, and activities while controlling the associated costs.

Meanwhile, numerical simulation of permafrost dynamics, using computational fluid dynamics methods (CFD), is highly complex and challenging due to the strong nonlinearities and couplings involved in the related physics. Since these simulations require extremely fine spatial and temporal discretizations, high-performance computing is needed \citep{Painter2013,Karavaev-2021,ORGOGOZO2023108541,Xavier2024-xi}. However, an improvement of several orders of magnitude in performance is necessary for operational applications. For instance, the century-watershed-scale simulations performed in \citep{Xavier2024-xi} consumed about half a million CPU hours on the IRENE-ROME TGCC supercomputer each and are associated with high environmental impact in terms of High-performance computing (HPC) energy consumption. This high computation cost prevents performing a large number of simulations and thus applying the methodology to numerous sites in order to achieve a comprehensive sensitivity and uncertainties analysis. This limitation also hinders the use of large-scale data assimilation techniques. 

On the other hand, the democratization of machine learning technologies paved the way to novel possibilities in data-driven modeling of large-scale problems \citep{annurev2020}. Recent works aimed to combine machine learning and data models to forecast environmental temperatures in different locations around the globe \citep{Guo2023,Guo2024}, or pollutant concentrations \citep{atmos15121432,toxics14010044}. However, time forecasting and stability over large time intervals remained a tricky issue in time-dependent problems, especially for nonlinear phenomena \citep{AZIZI2025134969}, with memory length being identified as an important factor in the prediction \citep{Guo2025}. Being highly nonlinear, the permafrost simulation and forecasting are not currently being tackled using typical data-driven technologies. On the other hand, novel artificial intelligence technologies \citep{chen2018neural,Mounayer2026}, and their coupling with established model order reduction technologies \citep{Mounayer2026,Ghnatios2024} are promising fast and reliable simulations and forecasting. The combination of model order reduction technologies (MOR), used for learning a stable projection base, and the machine learning techniques for stable time integration schemes appeared to be an appealing approach in tackling forecasting for large-scale problems \citep{en16155790}. Recent developments either used a linear reduced basis with a non-linear time integration scheme using neural networks \citep{ARESDEPARGA2026118443}, or a non-linear construction of the latent followed by a decomposition in the latent space \citep{mounayer2025rankreductionautoencoders}. These technologies proved being useful for the prediction of typical use cases, and stable when used in forecasting applications.

When it comes to permafrost modeling using machine learning applications, machine learning is presented as a revolutionary tool, able to tackle issues not addressed before in permafrost modeling \citep{Memis2025}. For instance, \citep{Chance2024} tested several machine learning models to predict the temperature of the permafrost on the northern slope of Alaska. Model hyperparameter tuning was performed on the available data, and the results are shown without forecasting or extrapolation. The impact of the time history taken as input into the prediction was also assessed. \citep{https://doi.org/10.1029/2023WR036700} combined InSAR and ML models to quantify the surface deformation in discontinuous permafrost regions. They also built machine learning models with fine-tuning of hyperparameters based on synthetic data to predict permafrost deformation. From the engineering point of view, \citep{Manos2025} used novel machine learning technology to create an enhanced map of the Alaska region and coupled it with the infrastructure damage cost related to permafrost melting. These studies show that data-driven modeling is developing rapidly in the direction of improving the resilience of geotechnical infrastructure in cold regions but is not ready yet for large-scale forecasting. All the aforementioned works consist of instantaneous predictions and are not yet coupled with a stable algorithm able to forecast the behavior over centuries.

In this work we aim to develop a high-fidelity, high-efficiency simulation pipeline for permafrost thermal and hydrological modeling, using the hybrid twin approach \citep{Chinesta2020,Ghnatios2024}, for the first time in the field of geosciences. Our research hypothesis is that the new modeling paradigm of the hybrid twin \citep{partialobservations2022,Champaney2022} can be successfully applied to high-performance computing permafrost simulation \citep{https://doi.org/10.1002/ppp.1995,ORGOGOZO2023108541,Xavier2024-xi}. This implementation improves and tremendously accelerates the current permafrost modeling capabilities. In order to test this hypothesis, we propose a proof of concept built on the environmental monitoring site of the Kulingdakan in Central Siberia, which has already been extensively studied by means of numerical twinning \citep{https://doi.org/10.1002/ppp.1995,ORGOGOZO2023108541,Xavier2024-xi}, and which is thus an ideal study case for applying for the first time the hybrid twin concept to permafrost modeling and to evaluate the associated outcomes. 

The manuscript starts with a review of the available experimental data in section \ref{expdatasec}, and the numerical modeling developed in OpenFOAM in section \ref{cfdsec}. The surrogate modeling used for the acceleration of the simulation is detailed in section \ref{surrogatedetailsec}, and the hybrid modeling technology is illustrated in \ref{hybridsec}. The results of the surrogate model and hybrid models are shown in sections \ref{surrogateresultssec} and \ref{hybridresultssec}, respectively. Later, the forecasting until the year 2100 is shown in section \ref{forecastingsec}.  Finally, the manuscript ends with discussions and conclusions in sections \ref{discussionsec} and \ref{conclusionsec} respectively.

\section{Material and Methods}\label{mmsec}
This section reviews the used experimental measurements and numerical methods to develop the hybrid modeling of the permafrost thermal fields' evolution throughout the thickness.

\subsection{Study site}\label{expdatasec}
The Kulingdakan watershed, located in the continuous-permafrost zone in Central Siberia (64.31° N, 100.28° E), is covered with pristine boreal larch forest. It has been monitored for the study of boreal processes for more than twenty years. Two horizons constitute the soil in the first few meters: an organic horizon (litter and peat) and a mineral horizon (mainly rocky and/or gravely loam). A detailed biophysiogeographical description of this watershed may be found, for instance, in \citep{doi:10.1139/x05-150}. The mean annual temperature is -8 °C, and the annual total precipitation is 400 mm (1999-2014 meteorological data of the Tura station, about 5 km away from the Kulingdakan watershed). The stream flows east to west and divides the $\approx 40 km^2$ watershed into two more or less rectangular slopes of equal area, the north-facing slope and the south-facing slope. The Kulingdakan watershed has been the subject of previous high-performance modeling studies \citep{https://doi.org/10.1002/ppp.1995,Xavier2024-xi}, using the PermaFOAM solver \citep{ORGOGOZO2023108541}, using a numerical twin approach. Strong hydrological, thermal, and pedological contrasts between the north-facing slope and the south-facing slope stem from the difference in solar radiation induced by the aspect, which impacts primary production and evapotranspiration \citep{Gentsch2011,https://doi.org/10.1029/2007GB002938,Viers2013}. Thus, all the subsequent analyses will be differentiated with respect to the aspects. For instance, we separate the analysis for the north-facing slope on one hand and for the south-facing slope on the other hand.

\subsection{Computational Fluid Dynamics simulations}\label{cfdsec}
To develop a surrogate model, this study mainly draws on climate simulations conducted using the PermaFOAM solver \citep{ORGOGOZO2023108541,Xavier2024-xi}. The aim here is to summarize the key findings and parameters of these simulations. 

The solved physical equations and the numerical implementation of permaFoam are described in details in \citep{ORGOGOZO2023108541}. It consist in an OpenFOAM solver for a set of two coupled, non linear equations of heat and water transfer in variably saturated, variably frozen porous media, which uses an iterative linearisation procedure and a finite volume scheme. Regarding the physical and computational paramaters used in the centennial permaFoam simulations, they can be found in the S2 part of the supplementary material \citep{supplement1}, and also attached as supplement to this work for completion. The PermaFOAM solver is designed to simulate heat and water transfer within a porous medium, taking into account liquid water-ice phase changes. Each cell takes into account four phases (air, liquid water, solid water and solid medium) using volumetric fraction to represent each phase. Simulations are transient and can be run in 1D, 2D or 3D domains. The solver includes complex physical phenomena such as water uptake by roots or freeze/thaw, which can pose particular challenges to numerical solvers due to the strong nonlinearities involved. Built on the OpenFOAM software framework, it benefits from excellent performances when using massively parallel computational resources. However, simulating small-scale processes with a mechanistic approach requires adapted mesh sizes of the order of few $mm$, and time steps of typically around one second. Under these constraints, simulating the impact of climate change (2014-2100) on a slope of the Kulingdakan watershed represented by a 2D transect discretized with approximately a 50 million cell mesh required one month of simulation time on the IRENE-ROME supercomputer, which emphasizes the need for surrogate modeling.

\begin{figure}
    \centering
    \includegraphics[width=1\linewidth]{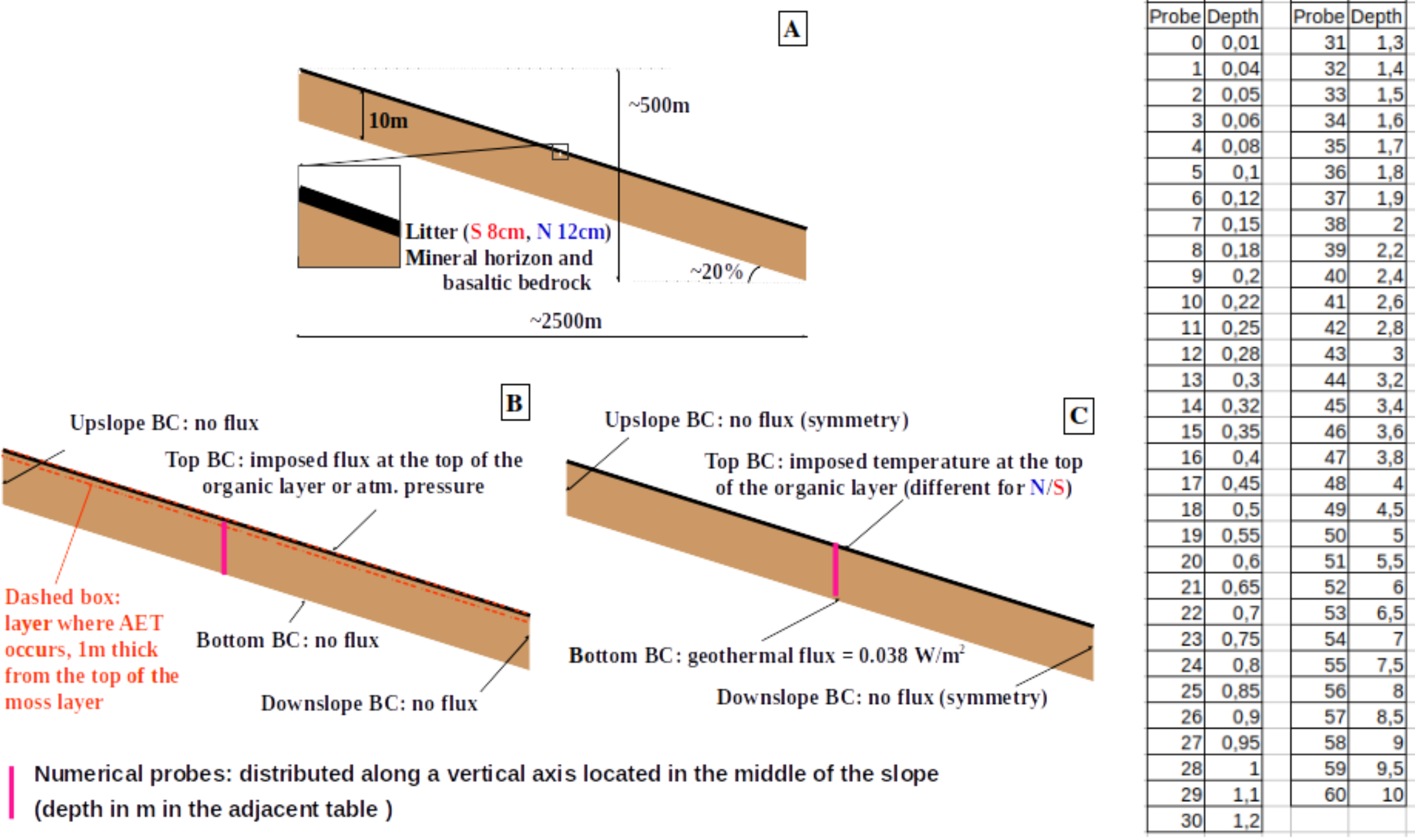}
    \caption{Domains of modeling and boundary conditions: (A) Description of the domains: a parallelepipedic representation of a watershed slope; (B) Hydrological boundary conditions (BC): No fluid flux inputs from the side and bottom boundaries, and a precipitation-based prescribed flux condition on the top; (C) Thermal boundary conditions (BC): Symmetry boundary conditions on the side edges, air temperature on the top, and geothermal heat flux on the bottom; (table): depth of the numerical probes: Probe number and depth position in meters (m). Source: Authors own work.}
    \label{fig1thibault}
\end{figure}

Among the variables involved in permafrost study, soil temperature is of particular interest. It enables the freeze-thaw line to be tracked and the thermal response of the soil to be investigated. \citep{Xavier2024-xi} demonstrate that the timescale of the soil's thermal response is strongly dependent on the intensity of the external forcing change and the soil depth. At a depth of 2 meters, it can take several years for a steady response to be reached to a forcing applied to the soil surface. This history effect, which can last several years, combined with a thermal response timescale that varies according to depth, poses a particular challenge for surrogate modeling. To study the thermal response of the soil, 61 numerical probes were placed along a vertical line located in the middle of the 10-meter-thick soil parallelepiped that constitutes the numerical domain representing a slope of the Kulindakan watershed, with a higher probe density close to the surface, where the gradients are prominent. Figure \ref{fig1thibault} presents the geometry, boundary conditions, and the location of the probes in the simulation domain. This domain is a simplified parallelepipedic representation of a slope of 2500 m and 10 m of thickness, with a vertical drop of 500 m. We consider only the first 10 m of soil since this depth of modeling domain is enough to use a fixed geothermal flux boundary condition at the bottom without damaging the physical relevance of the thermal dynamics simulated close to the soil surface \citep{supplement1}. The top soil boundary conditions used for the centennial permaFoam simulations are derived from the climatic projections of atmospheric conditions through a process presented in the S1 part of the supplementary material \citep{supplement1}. The location of the selected probes reproduces the location of the available experimental measurements. The probes allow computing the average temperature for each day of the simulation. Considering the 61 probes and the 86-year period, each of the eight datasets used for this study (four climate scenarios and two slopes) contains 1.9 million data points of daily averaged soil temperature. Analogous datasets are produced simultaneously for ice content, liquid water content, total water content, potential evapotranspiration sink term and actual evapotranspiration sink term.

The study also uses an experimental dataset based on daily observation data from 2003 to 2005, alongside with PermaFOAM numerical simulations for the same period. This enabled the  PermaFOAM-based numerical twin of the Kulingdakan watershed to be validated under current conditions, with the difference between the modeling and the experimental survey being quantified \citep{https://doi.org/10.1002/ppp.1995}. This dataset enables us to improve the surrogate modeling approach with a hybrid model to better predict the soil's thermal response under different climate scenarios.

\section{Surrogate model of PermaFOAM}\label{surrogatedetailsec}
The first step in the simulation process is to create a surrogate model to accelerate the simulation performed by the OpenFOAM-based tool PermaFOAM. The objective of the surrogate modeling is to provide a fast representation of the quantities of interest, which are mainly the thermal fields through the thickness.

\subsection{Surrogate modeling formulation}
To evaluate the thermal fields, first a singular value decomposition is performed to extract the relevant reduced basis from the available simulation. Without any loss of generality, a set of $N_{probes}=61$ probes is selected from the PermaFOAM simulations for all time steps. A thermal fields dataset $\mathbf{T}$ of size $(N_{probes},N_t)$ is therefore available. The same proposed surrogate modeling can be used for the complete simulation, using $N_{probes}$ the total number of nodes available in the meshed domain.

The space domain reduced basis is constructed using a truncated singular value decomposition (SVD) of the thermal fields. For instance, we compute the following:
\begin{equation}\label{svdeq1}
    [\mathbf{V}_{full},\bm\lambda_{full},\mathbf{W}_{full}]=svd(\mathbf{T}),
\end{equation}

$\mathbf{V}_{full}$ being the left eigenvectors, $\mathbf{W}_{full}$ the right one, and $\bm\lambda_{full}$ the eigenvalues. The truncation is performed, such as
\begin{equation}\label{truncationeq}
\kappa= \max \left\{ i \in \{1, \dots, M\} \;:\; \frac{\lambda_i}{\lambda_1} > \epsilon_t\right\},
\end{equation}

with $\epsilon_t$ small enough truncation tolerance. Later, we define the space coordinates' reduced basis as $\mathbf{V}$, the truncated part of $\mathbf{V}_{full}$ considering only the first $\kappa$ vectors of $\mathbf{V}_{full}$. The reconstructed temperature fields can now be approximated using:
\begin{equation}
\mathbf{T}\approx\hat{\mathbf{T}}=\mathbf{V}\bm\xi,
\end{equation}

$\hat{\mathbf{T}}$ the approximated thermal fields, the reduced coordinates being $\bm\xi$, $\bm\lambda$ and $\mathbf{W}$ the truncated $\bm\lambda_{full}$ and $\mathbf{W}_{full}$ to the rank $\kappa$.

To predict future simulations, the reduced basis $\mathbf{V}$ is considered constant, while the time-dependent coefficients are computed in time.

Inspired by the Proper Orthogonal Decomposition  method (POD), our method will use the reduced basis as a fixed space basis with a time evolution scheme \citep{Tannous2025}. In fact, to revisit the POD method, the finite element formulation at any time step $i$ is always written in the form:
\begin{equation}
    \mathbf{K}_i T_i=f_i,
\end{equation}

with $T_i$ being the temperature vector at time step $i$ , $\mathbf{K}_i$ the corresponding stiffness matrix and $f_i$ the right hand side. When using the reduced coordinates, a typical POD method writes:
\begin{equation}
\begin{array}{c}
    \mathbf{K}_i \mathbf{V}\xi_i=f_i\\
    \mathbf{V}^T \mathbf{K}_i \mathbf{V}\xi_i=\mathbf{V}^T f_i\\
    \bm\xi_i=\left(\mathbf{V}^T \mathbf{K}_i \mathbf{V}\right)^{-1}\mathbf{V}^T f_i,
\end{array}
\end{equation}

with $\xi_i$ the reduced coordinates at time step $i$. The aforementioned equation requires access to the reconstructed stiffness and mass matrices from the finite element solver. To overcome this limitation, a surrogate model is built to predict the evolution of $\bm\xi$ as a function of time using a long-short-term memory (LSTM) surrogate \citep{10.1162/neco.1997.9.8.1735,7780459}. The LSTM network is built on a recurrent neural network structure but allows a long memory approach to take into consideration the time dependencies on historical values. Therefore, we construct the surrogate $\mathcal{H}$ using the following:
\begin{equation}
    \bm\xi_{i+1}=\mathcal{H}(\mathbf{p}_i)
\end{equation}

with $\mathbf{p}_i$ the input parameters of the model, mainly the ambient temperature $T^a$, the water flux $W$, and the Potential EvapoTranspiration $P^{et}$, modeling to the potential water absorption by the vegetation. The last $\tau$ historical values are considered. For instance, we use the following:
\begin{equation}
    \mathbf{p}_i=\left(\begin{array}{ccc}
          T^a_{i+1} & W_{i+1} & P^{et}_{i+1}\\
          T^a_{i} & W_{i} & P^{et}_{i}\\
          T^a_{i-1} & W_{i-1} & P^{et}_{i-1}\\
          \vdots &  \vdots & \vdots\\
          T^a_{i-\tau+1} & W_{i-\tau+1} & P^{et}_{i-\tau+1}\\
    \end{array}\right)
\end{equation}

The LSTM algorithm implemented $\mathcal{H}$ uses a look-back memory of $\tau=60$ days, with several other $\tau$ values being tested and their results evaluated, but the best results were obtained for $\tau=60$. The reconstructed thermal fields can be obtained using:

\begin{equation}
    T_{i+1}=\mathbf{V} \mathcal{H}(\mathbf{p}_i)
\end{equation}

The modeling is performed for both the north-facing slope and the south-facing slope, as the behavior differs between both slopes due to the differential sun exposure and water flows.

\subsection{Surrogate modeling training}\label{surrogatetrainingsec}
\begin{figure}[!ht]
    \centering
    \subfigure[Sample simulation results on the south-facing slope under SSP126]{\includegraphics[width=\linewidth]{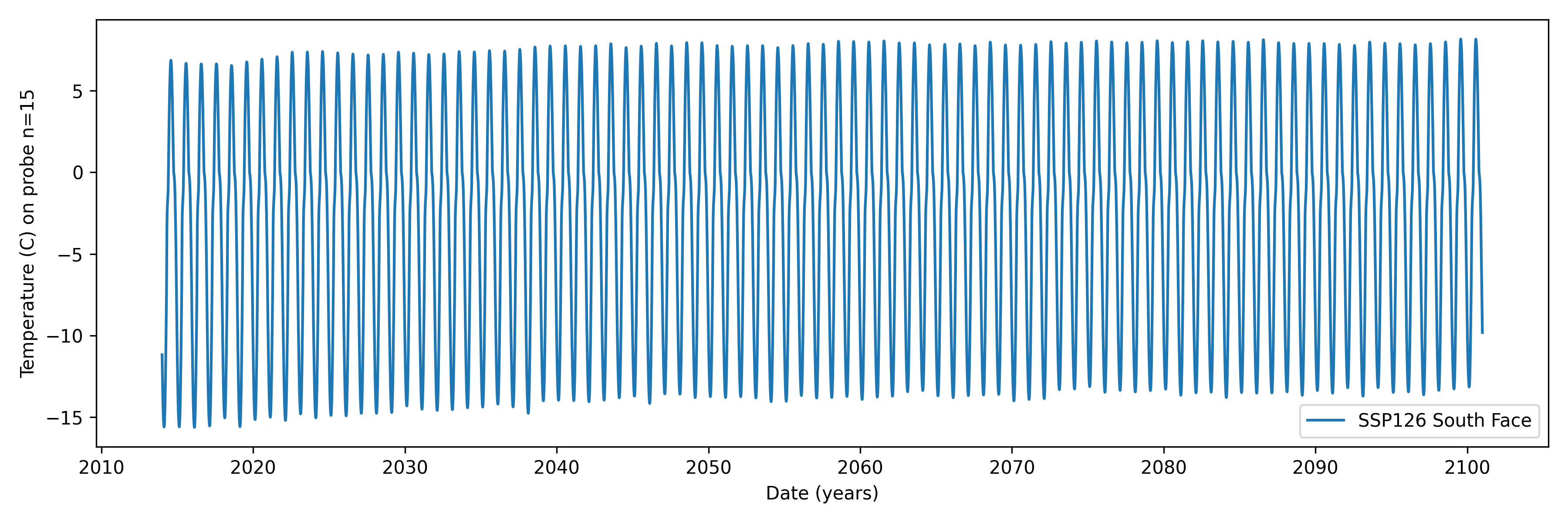}}
    \subfigure[Sample simulation results on the north-facing slope under SSP126]{\includegraphics[width=\linewidth]{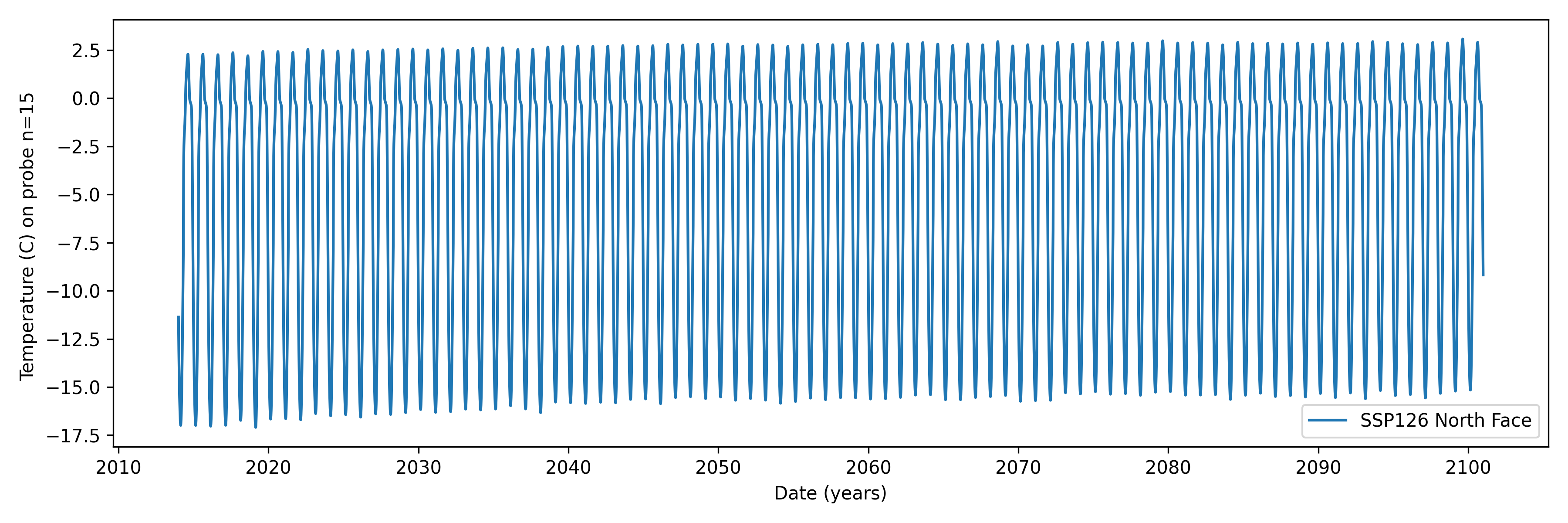}}
    \subfigure[Sample simulation results on the south-facing slope under SSP585]{\includegraphics[width=\linewidth]{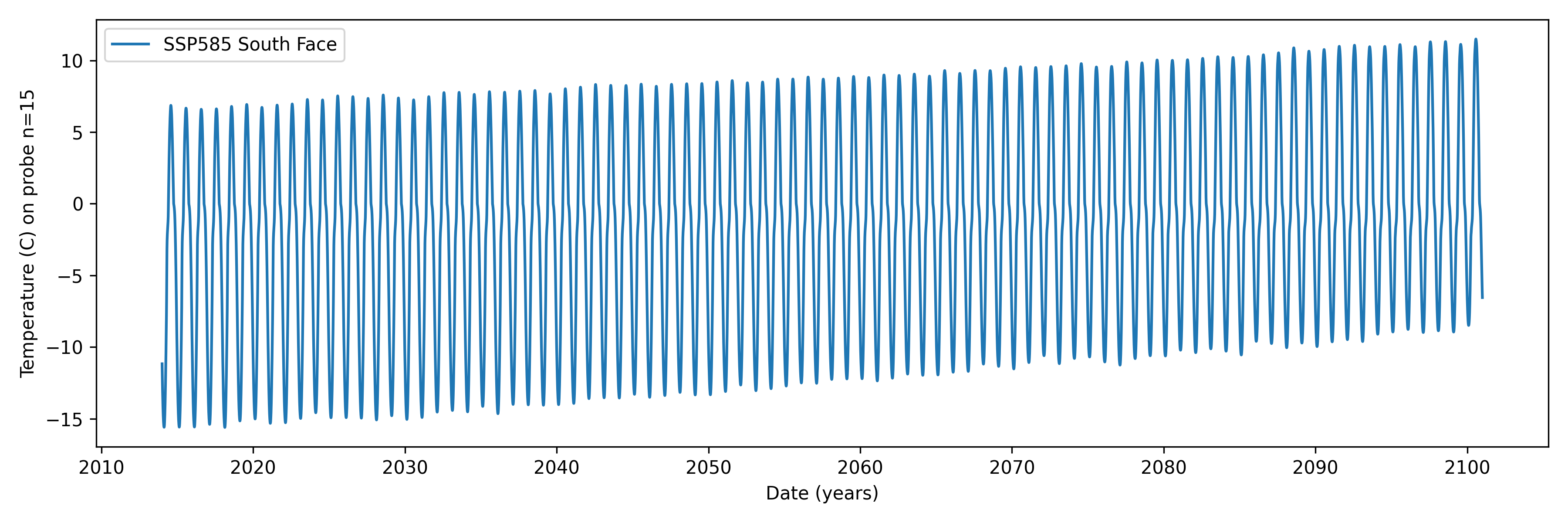}}
    \subfigure[Sample simulation results on the north-facing slope under SSP585]{\includegraphics[width=\linewidth]{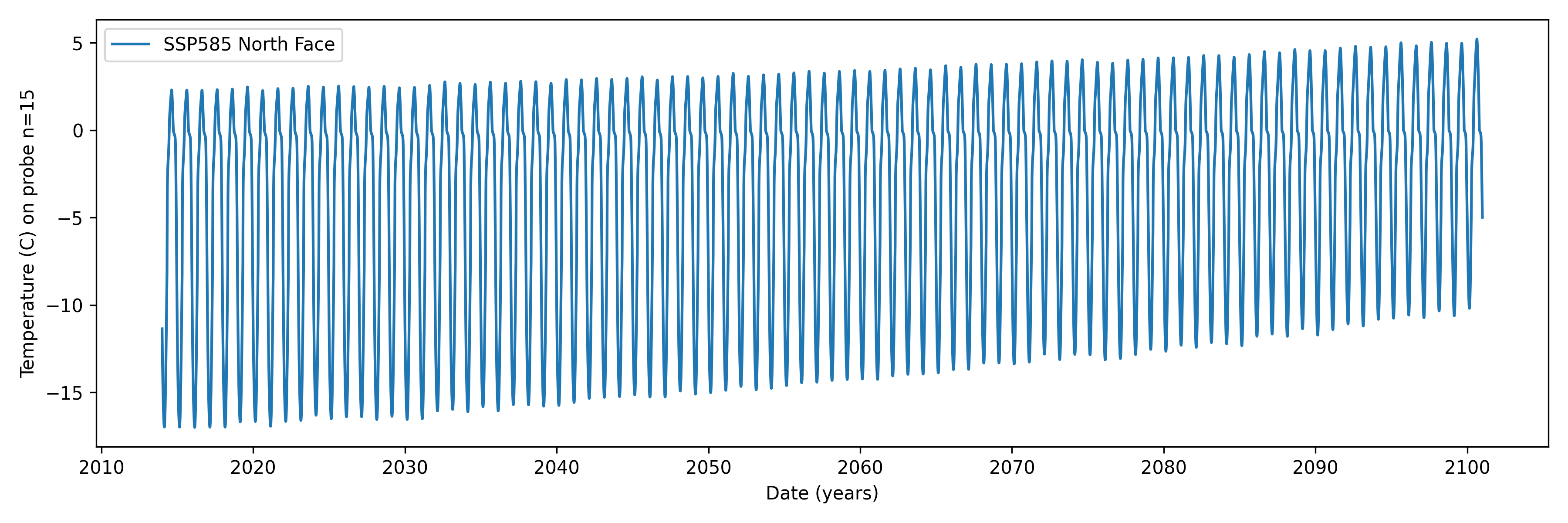}}
    \caption{Sample results from the PermaFOAM simulations for both north and south-facing slopes for a selected numerical probe. Source: Authors own work.}
    \label{samplesfigs}
\end{figure}

A generated dataset considering the four CMIP6 environmental pathway scenarios, SSP126, SSP245, SSP370, and SSP585 \citep{gmd-9-1937-2016,gmd-9-3461-2016,RIAHI2017153}, is simulated. The daily air temperature and precipitation projected until 2100 from these climatic scenarios are fed into PermaFOAM to generate the thermal evolution of the temperature within the permafrost in both south-facing slope and north-facing slope. The dataset consists of 61 selected numerical probes in each slope of the analyzed domain. A sample of the dataset is illustrated in Figure \ref{samplesfigs}.

To train the surrogate model, the scenarios SSP126 and SSP585 are considered as inputs. The scenarios SSP245 and SSP370 are kept out of the training to assess the predictive ability of the surrogate models for unseen datasets.

The dataset for each scenario consists of one temperature sample per day, sampled over 87 years. The training of the surrogate model $\mathcal{H}$ is performed to predict the reduced coordinates $\bm\xi$ for every time step $i$, after performing a linear model order reduction through singular value decomposition. The selected reduced basis size is $\kappa=9$ for both the north-facing slope and the south-facing slope. The truncation is performed as described in equation (\ref{truncationeq}), with a selected truncation tolerance $\epsilon_t=10^{-4}$.

To further establish the ability to perform interpolation in space, not all the 61 numerical probes are used to train the surrogate model $\mathcal{H}$. Instead, only $N=25$ sampled probes are used as target outputs to the surrogate model. For instance, the loss function $\mathcal{L}$ is computed as such:
\begin{equation}
\mathcal{L}=\frac{1}{m}\sum\limits_{i=1}^m \left(\mathbf{V}^{sampled}\mathcal{H}(\mathbf{p}_i)-T^{sampled}_i\right)^2,
\end{equation}
with $m$ being the total number of samples available in time, combined for both training scenarios, SSP126 and SSP585, and $\mathbf{V}^{sampled}$ the particularization of $\mathbf{V}$ on the sampled probed only. The target outputs $T_i^{sampled}$ are the PermaFOAM simulated temperatures at the sampled probes for time step $i$. The used input values $\mathbf{p}_i$ are the normalized input parameters, using a min-max scaler \citep{scikit-learn}. 

\begin{figure}[!ht]
    \centering
    \includegraphics[width=\linewidth]{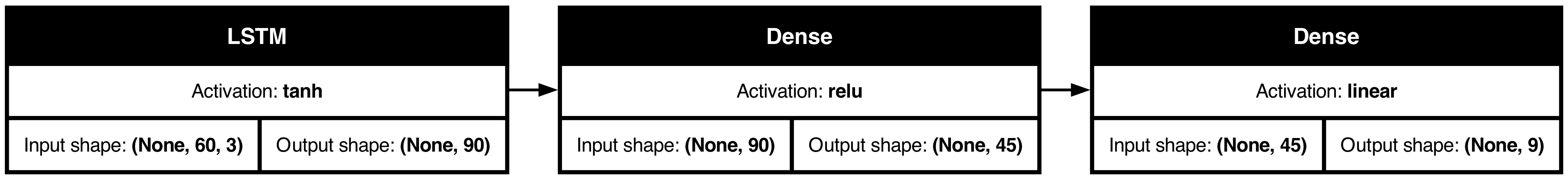}
    \caption{Surrogate model $\mathcal{H}$ architecture showing the neural network layers from left to right, their size and their activation functions. Source: Authors own work.}
    \label{fig:modelofH}
\end{figure}

The surrogate model is built using a TensorFlow deep dense neural network \citep{tensorflow2015-whitepaper}, with the structure depicted in Figure \ref{fig:modelofH}. The used model contains an LSTM layer, which is fed the last $\tau=60$ normalized inputs, and will output a set of values with the return sequence set to false. After flattening, the LSTM outputs are fed into a dense layer with linear activation to obtain the reduced coordinates $\bm\xi$ for every subsequent time step $i$. A total of 38,349 trainable parameters are available in the model.

The training is performed using the ADAM optimizer \citep{kingma2015adam}, along with a custom-built decreasing learning rate function. Batch gradient descent is used along with a total of 800 epochs performed during the training.

\subsection{Results of the surrogate model}\label{surrogateresultssec}
\begin{figure}[!ht]
    \centering
    \subfigure[Train set sample comparison of the surrogate model with PermaFOAM results on the north-facing slope under SSP126]{\includegraphics[width=\linewidth]{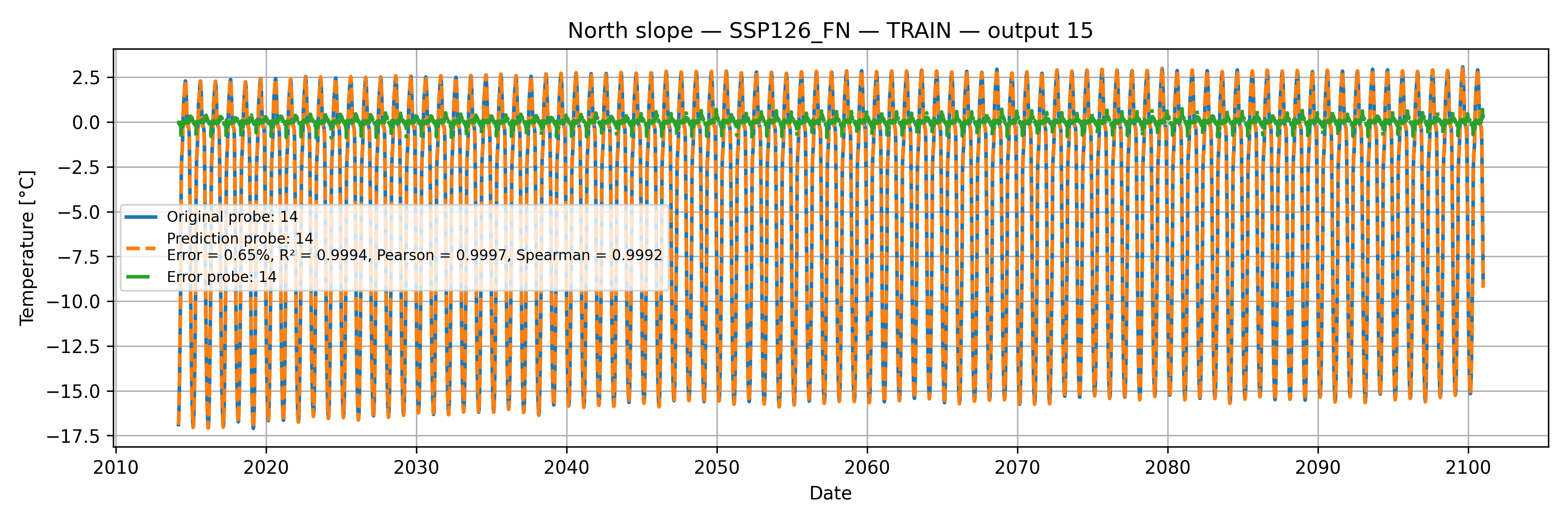}}
    \subfigure[Train set sample comparison of the surrogate model with PermaFOAM results on the north-facing slope under SSP585]{\includegraphics[width=\linewidth]{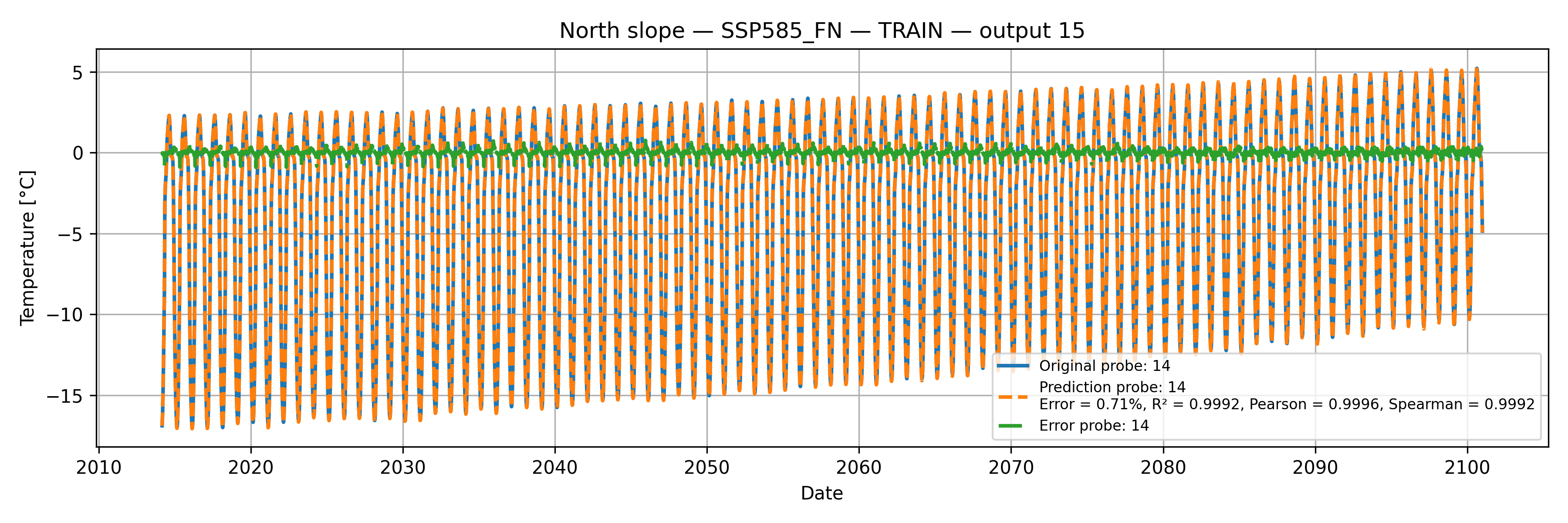}}
    \subfigure[Test set sample comparison of the surrogate model with PermaFOAM results on the north-facing slope under SSP245]{\includegraphics[width=\linewidth]{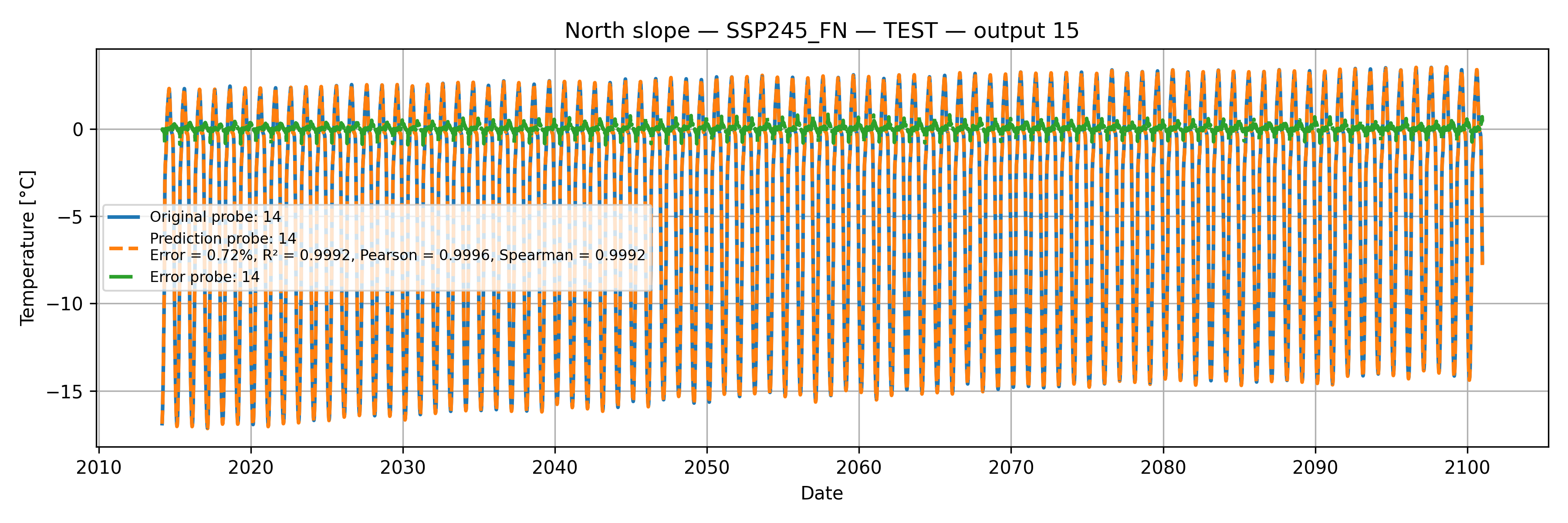}}
    \subfigure[Test set sample comparison of the surrogate model with PermaFOAM results on the north-facing slope under SSP370]{\includegraphics[width=\linewidth]{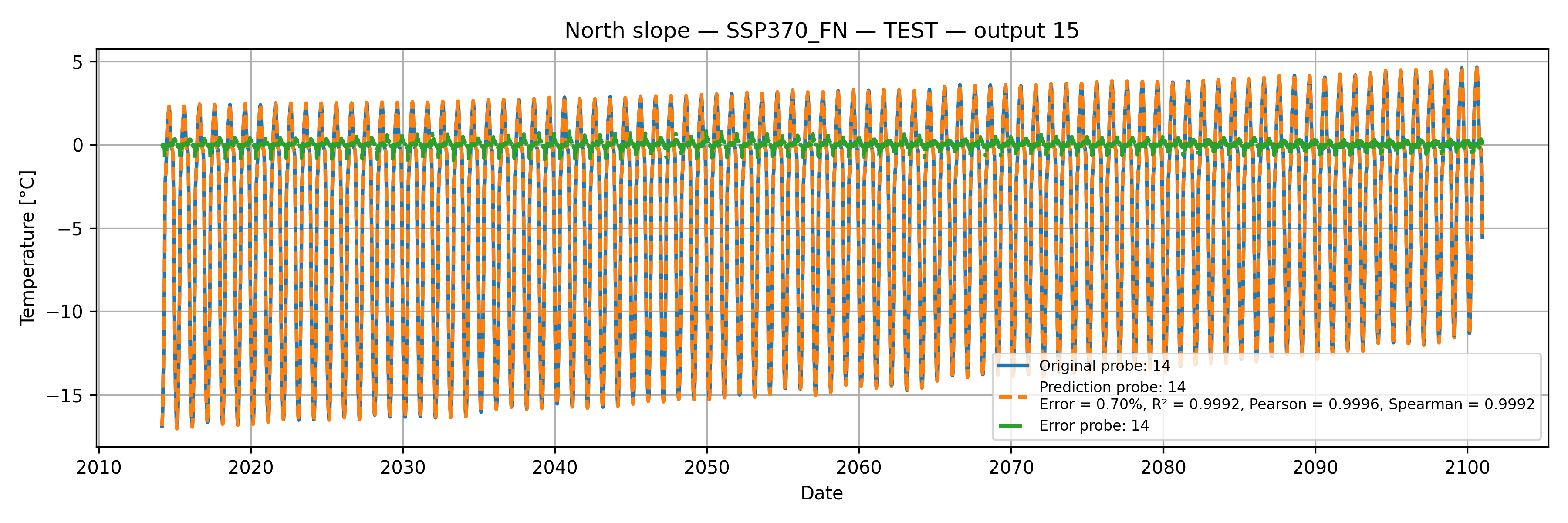}}
    \caption{Sample results from the surrogate model comparison to PermaFOAM simulations for both the train and test sets on the north-facing slope for the numerical probe 15, a sampled probe for the training of the surrogate model $\mathcal{H}$. The figures show that the error does not exceed a single degree Celsius. Source: Authors own work.}
    \label{surrogatenorthresults}
\end{figure}

\begin{figure}[!ht]
    \centering
    \subfigure[Train set sample comparison of the surrogate model with PermaFOAM results on the south-facing slope under SSP126]{\includegraphics[width=\linewidth]{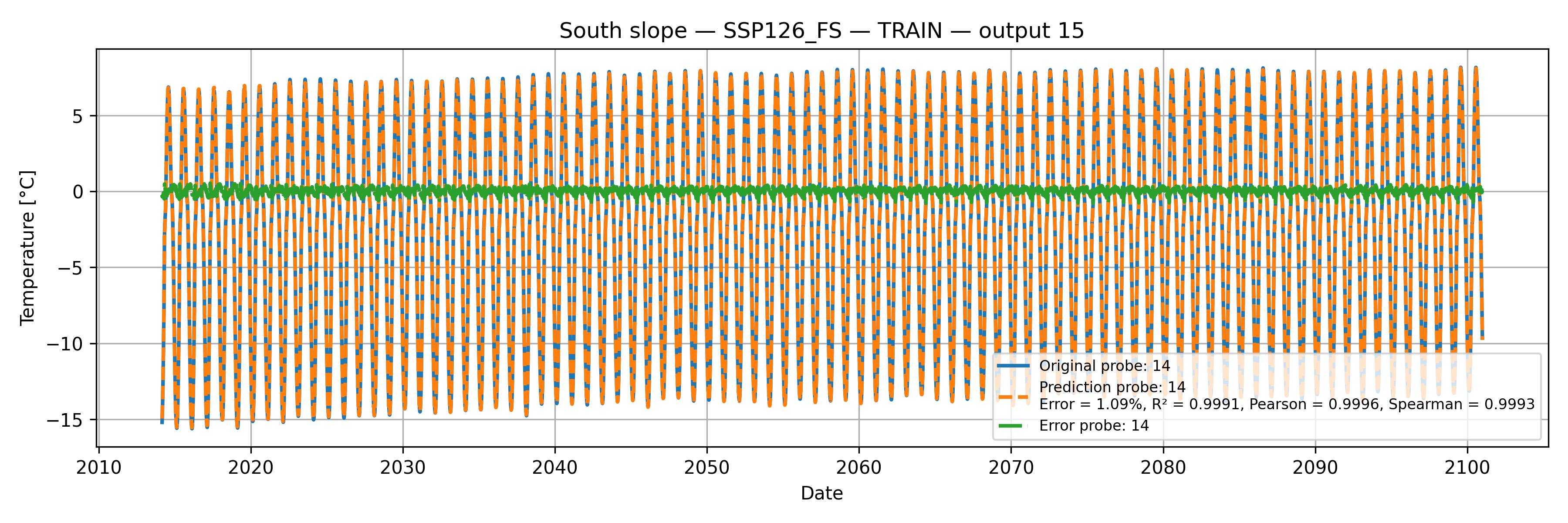}}
    \subfigure[Train set sample comparison of the surrogate model with PermaFOAM results on the south-facing slope under SSP585]{\includegraphics[width=\linewidth]{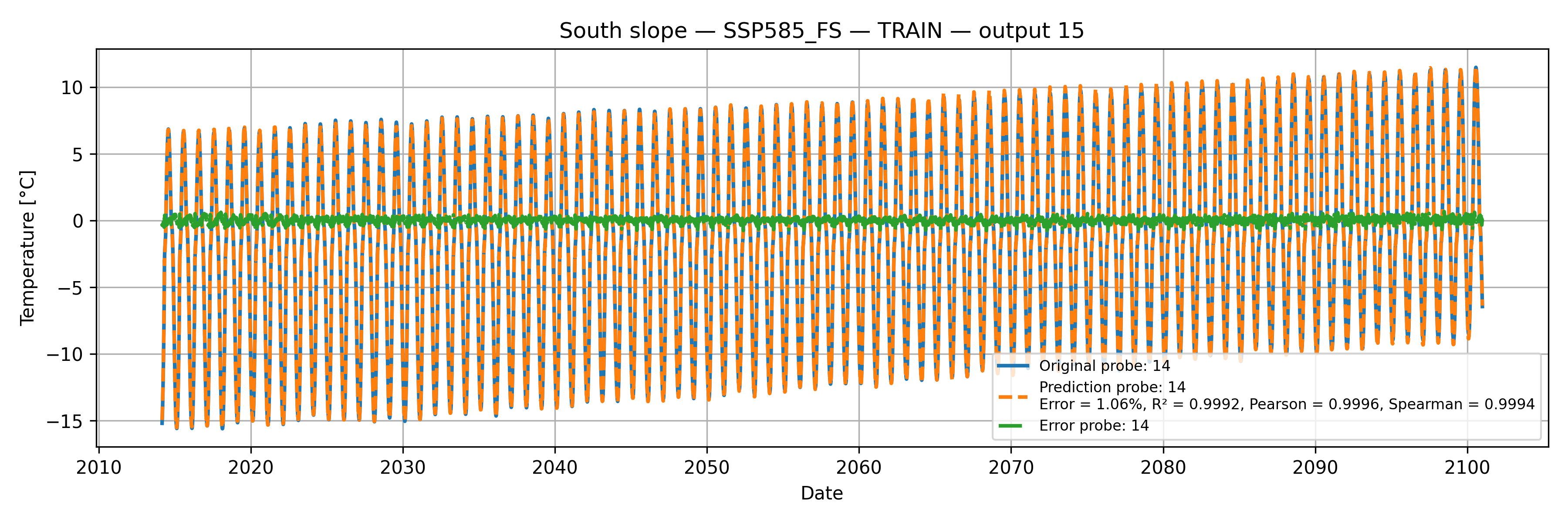}}
    \subfigure[Test set sample comparison of the surrogate model with PermaFOAM results on the south-facing slope under SSP245]{\includegraphics[width=\linewidth]{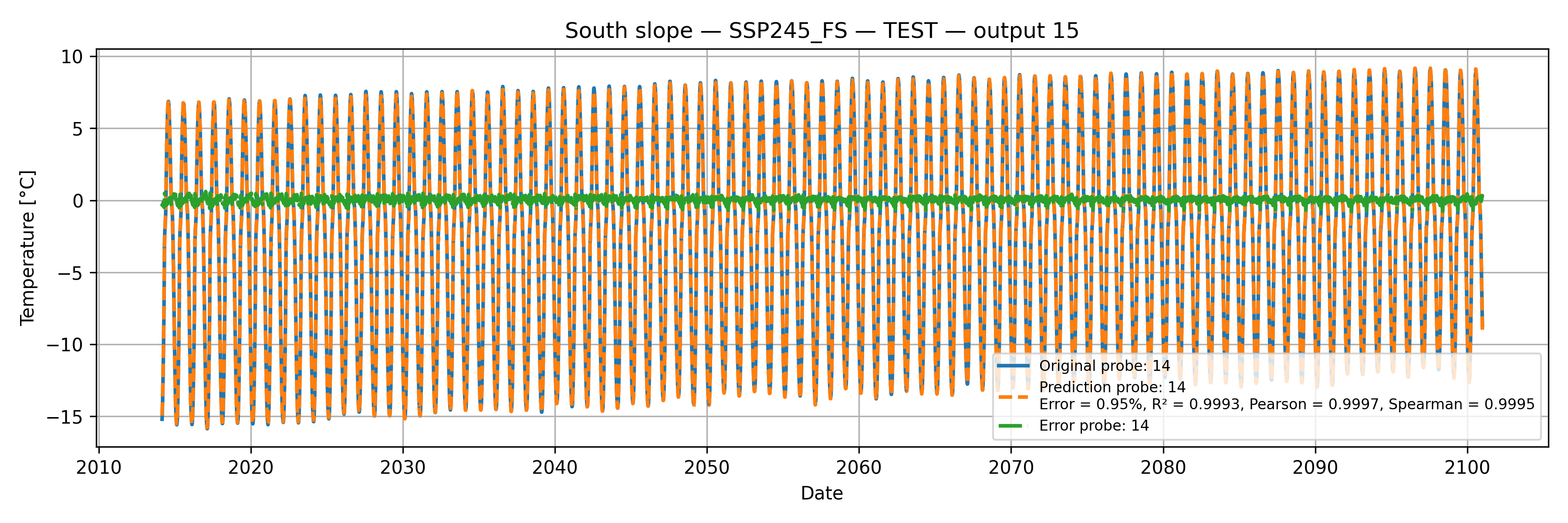}}
    \subfigure[Test set sample comparison of the surrogate model with PermaFOAM results on the south-facing slope under SSP370]{\includegraphics[width=\linewidth]{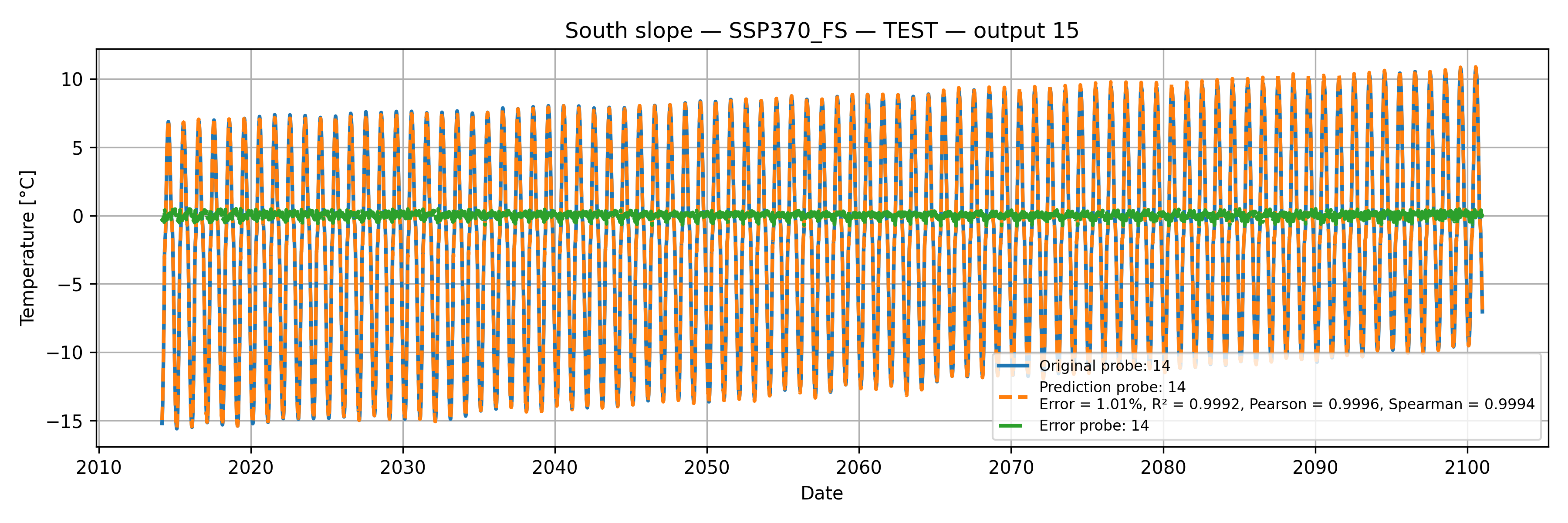}}
    \caption{Sample results from the surrogate model comparison to PermaFOAM simulations for both the train and test sets on the south-facing slope for the numerical probe 15, a sampled probe for the training of the surrogate model $\mathcal{H}$. The figures show that the error does not exceed a single degree Celsius. Source: Authors own work.}
    \label{surrogatesouthresults}
\end{figure}

The north-facing slope results on a sampled probe for the training are illustrated in Figure \ref{surrogatenorthresults} for both train and test sets. The south-facing slope results are illustrated in Figure \ref{surrogatesouthresults}. The relative error in percentage, the coefficient of determination $R^2$, as well as the Pearson and Spearman correlations are reported on the images.

To obtain a dimensionless error measure that is invariant to the chosen temperature scale and avoids point-wise division by zero, we compute the range-normalized mean absolute error $e_r$ with the temperature $T_{ij}$ being in Celsius for time step $i$ and location $j$ among the $N_p$ probes, such as
\begin{equation}
    e_r=\frac{100}{m\times N_p}\sum\limits_{j=1}^{N_p} \sum\limits_{i=1}^m \left(\frac{\left|\hat{T}_{ij}-T_{ij}\right|}{{T}_j^{max}-{T}_j^{min}}\right),
\end{equation}

with $\hat{T}_{ij}$ being the surrogate model predictions at time step $i$ location $j$, and $T_{ij}$ the corresponding exact measures when the experimental values are available, or the PermFOAM values in this section. ${T}_j^{max}$ and ${T}_j^{min}$ are the maximum and minimum temperatures observed in location $j$.

with $\mathbf{T}$ the sampled temperatures from the PermaFOAM-selected 25 probes, and the $\hat{\mathbf{T}}$ is the predicted temperatures on the same probes, such as, at each time step $i$:
\begin{equation}
    \hat{T}_i=\mathbf{V}\mathcal{H}(\mathbf{p}_i)
\end{equation}

\begin{figure}[!ht]
    \centering
    \subfigure[Train set sample comparison of the surrogate model with PermaFOAM results on the north-facing slope under SSP126]{\includegraphics[width=\linewidth]{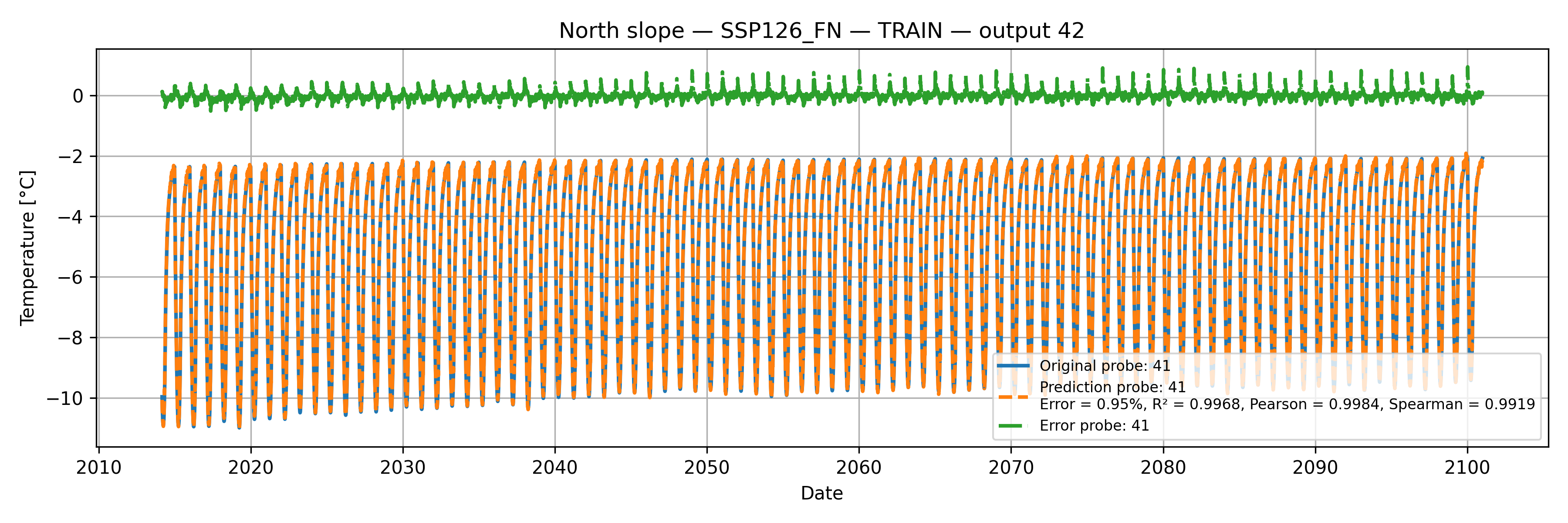}}
    \subfigure[Train set sample comparison of the surrogate model with PermaFOAM results on the north-facing slope under SSP585]{\includegraphics[width=\linewidth]{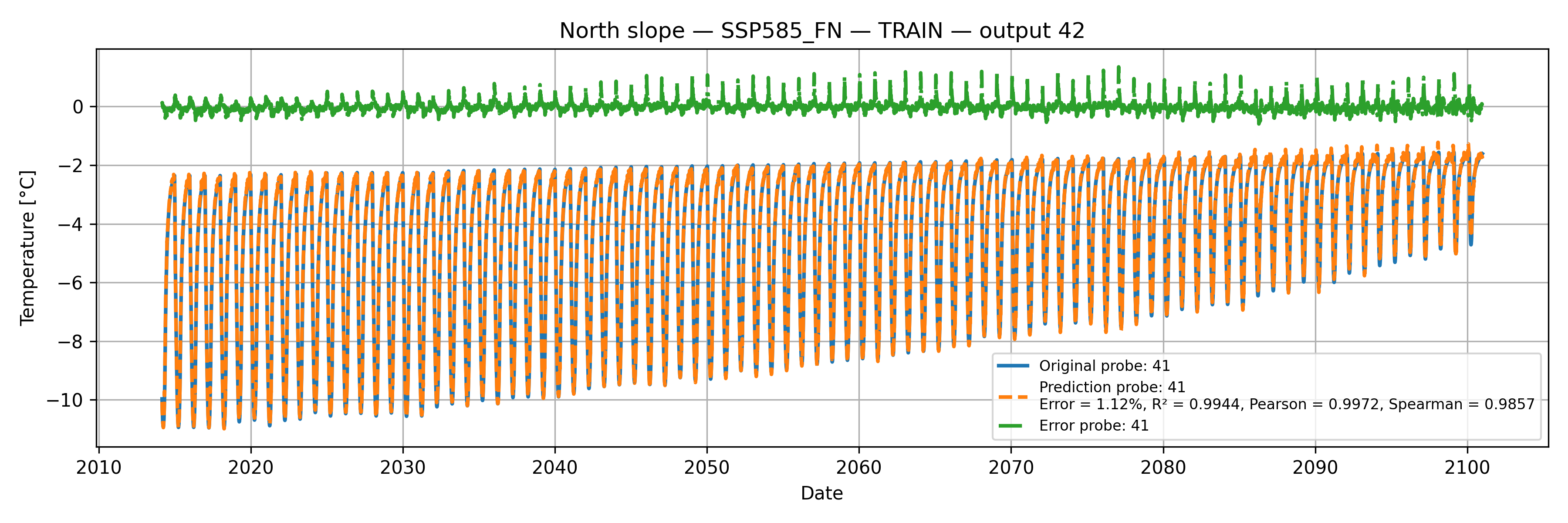}}
    \subfigure[Test set sample comparison of the surrogate model with PermaFOAM results on the north-facing slope under SSP245]{\includegraphics[width=\linewidth]{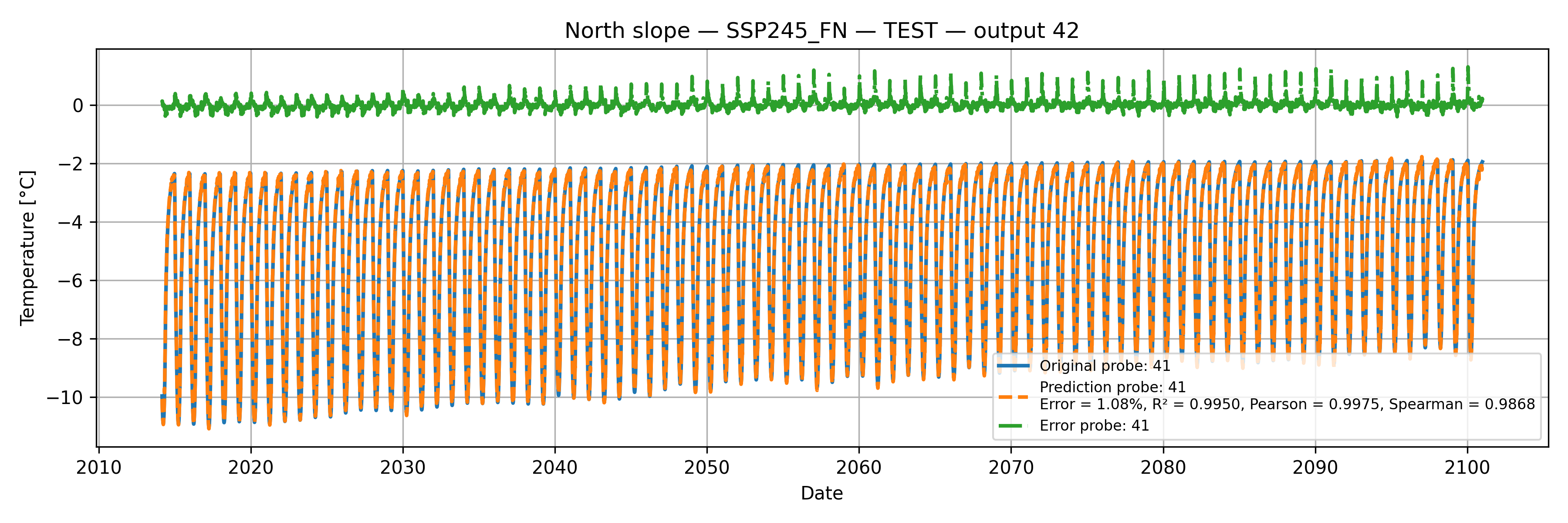}}
    \subfigure[Test set sample comparison of the surrogate model with PermaFOAM results on the north-facing slope under SSP370]{\includegraphics[width=\linewidth]{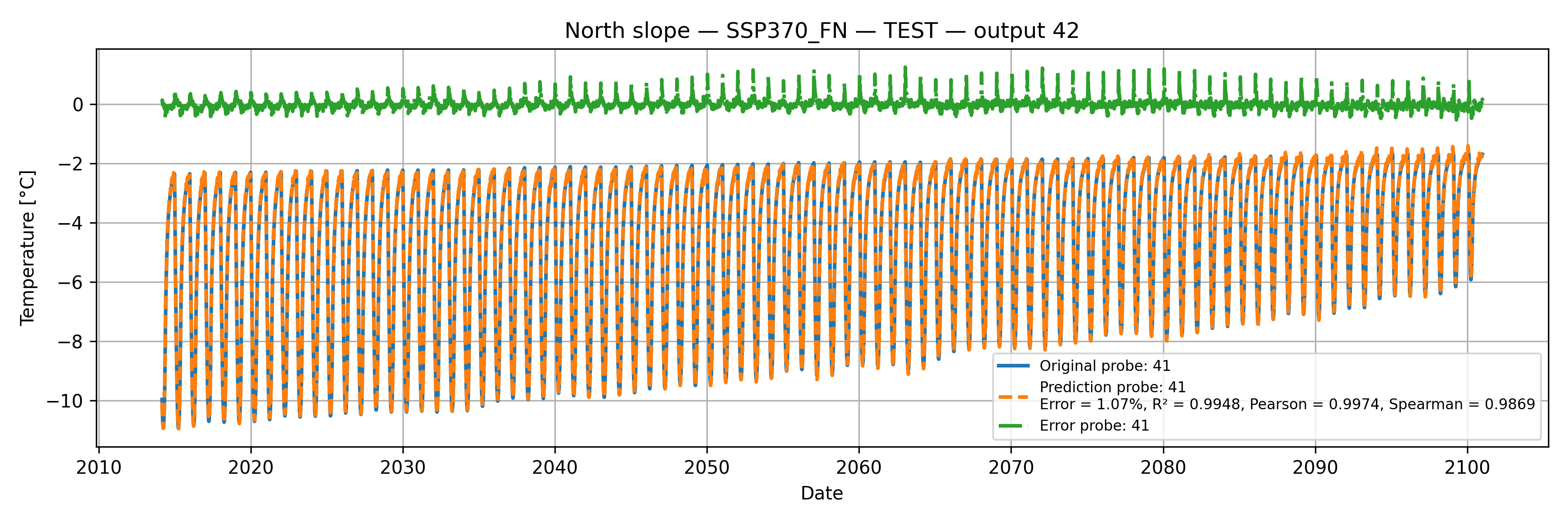}}
    \caption{Sample results from the surrogate model comparison to PermaFOAM simulations for both the train and test sets on the north-facing slope for the numerical probe 42, a probe not sampled during the training of the surrogate model $\mathcal{H}$. The figures show that the error does not exceed two degree Celsius. Source: Authors own work.}
    \label{surrogatenorthresultsunsampled}
\end{figure}

\begin{figure}[!ht]
    \centering
    \subfigure[Train set sample comparison of the surrogate model with PermaFOAM results on the south-facing slope under SSP126]{\includegraphics[width=\linewidth]{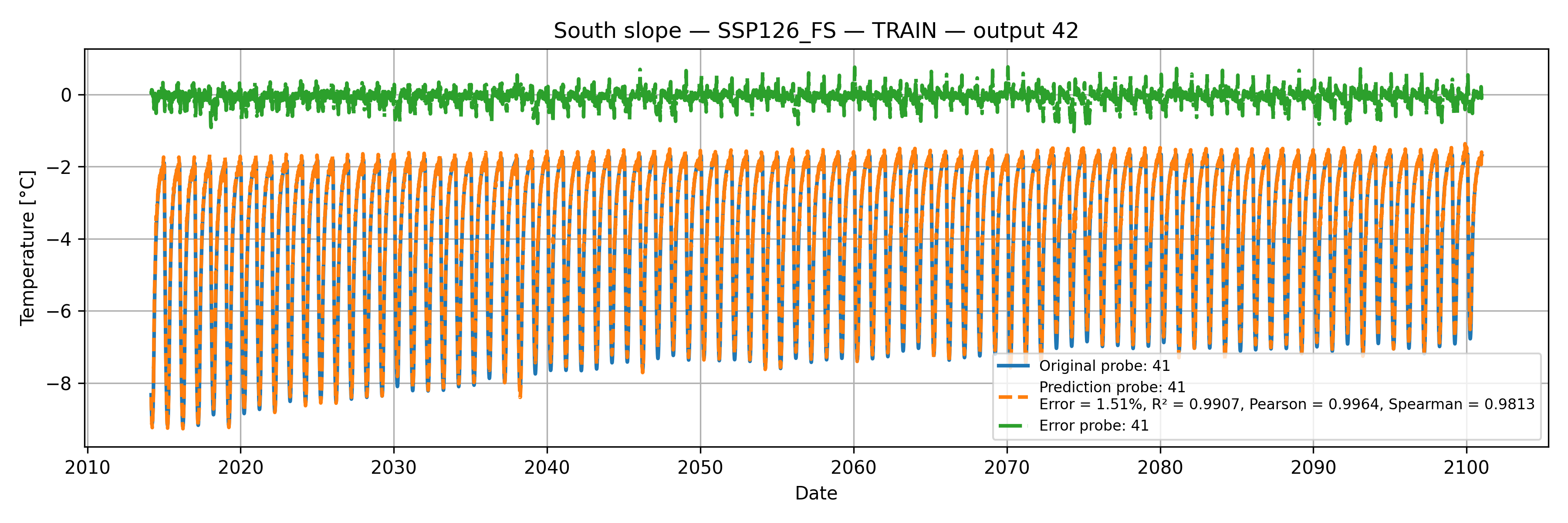}}
    \subfigure[Train set sample comparison of the surrogate model with PermaFOAM results on the south-facing slope under SSP585]{\includegraphics[width=\linewidth]{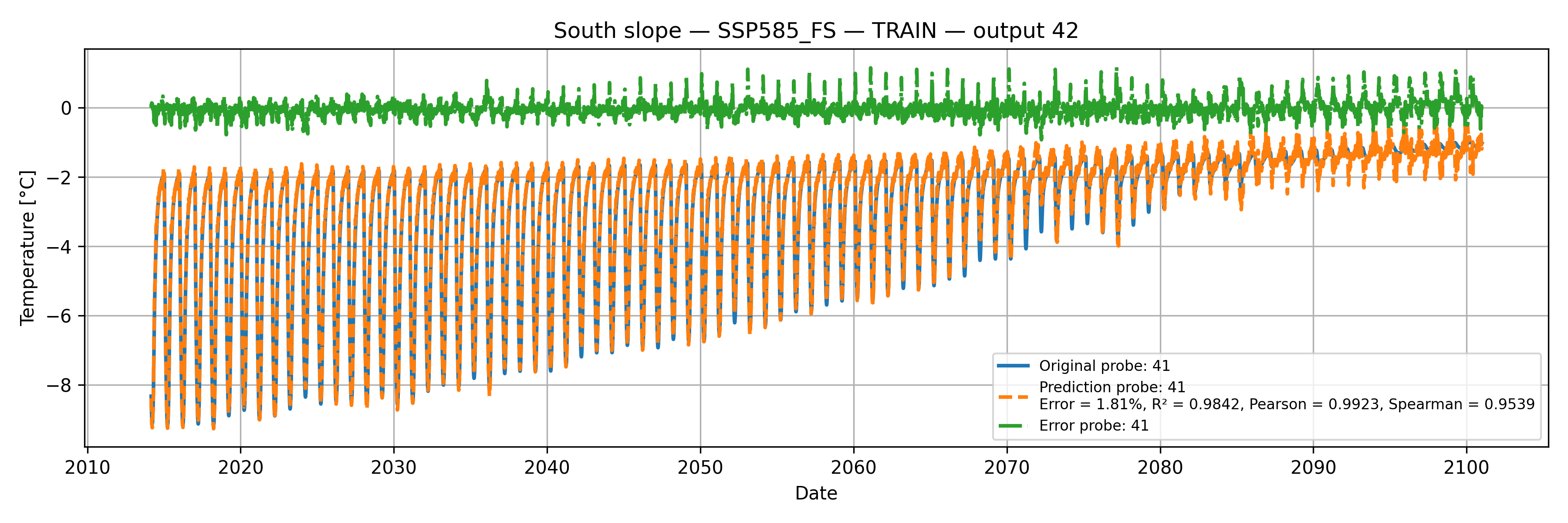}}
    \subfigure[Test set sample comparison of the surrogate model with PermaFOAM results on the south-facing slope under SSP245]{\includegraphics[width=\linewidth]{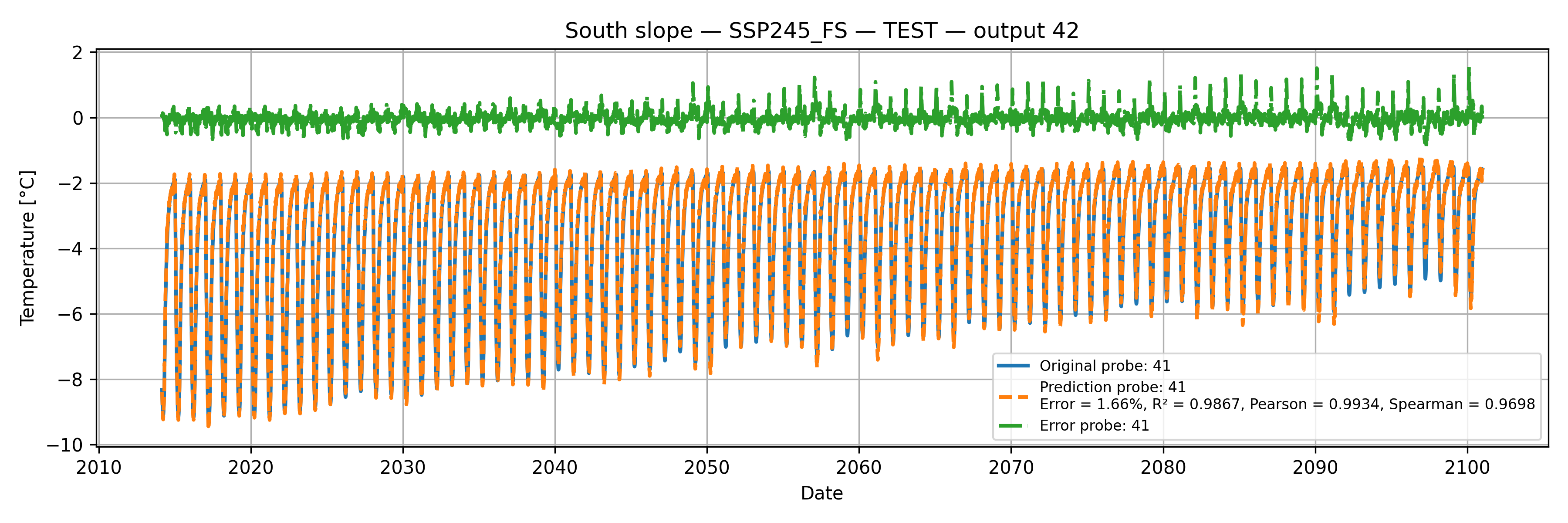}}
    \subfigure[Test set sample comparison of the surrogate model with PermaFOAM results on the south-facing slope under SSP370]{\includegraphics[width=\linewidth]{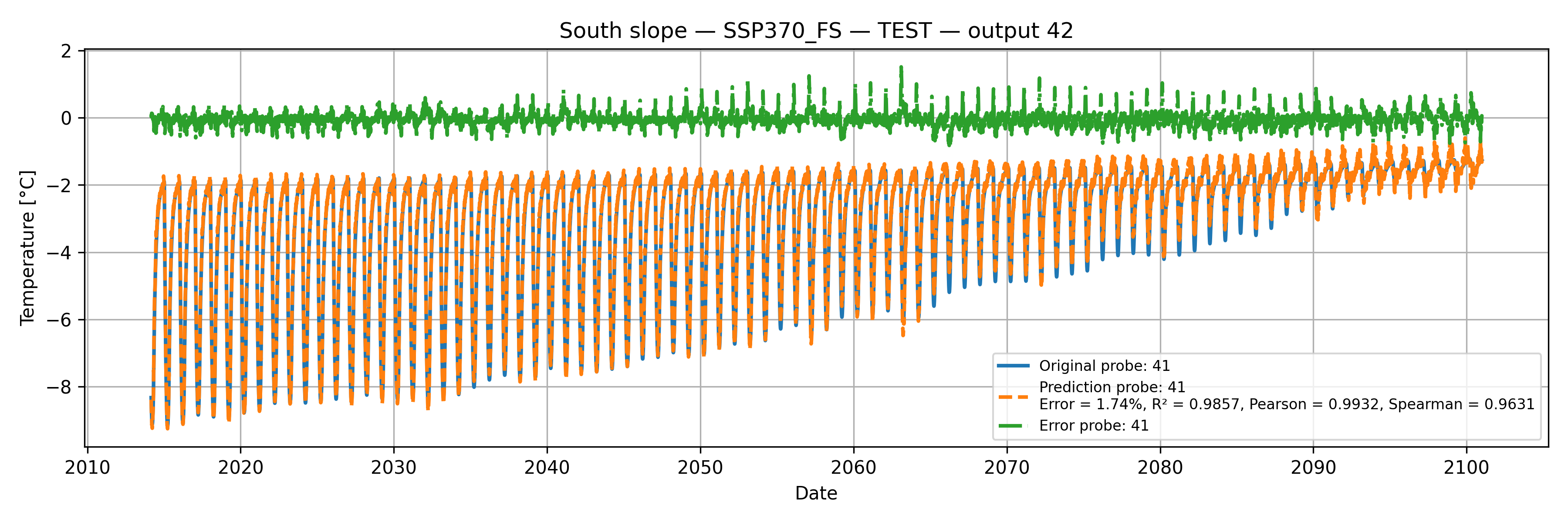}}
    \caption{Sample results from the surrogate model comparison to PermaFOAM simulations for both the train and test sets on the south-facing slope for the numerical probe 42, a probe not sampled during the training of the surrogate model $\mathcal{H}$. The figures show that the error does not exceed two degree Celsius. Source: Authors own work.}
    \label{surrogatesouthresultsunsampled}
\end{figure}

The results for a selected, not sampled, probe are illustrated in Figures \ref{surrogatenorthresultsunsampled} and \ref{surrogatesouthresultsunsampled} for both the north-facing slope and the south-facing slopes, respectively. A review of the average results for all the probes in both sampled and unsampled sets are illustrated in Table \ref{surrogatetablereview_north} for the north-facing slope and Table \ref{surrogatetablereview_south} for the south-facing slope.

\begin{table}[!ht]
\centering
\caption{Surrogate model $\mathcal{H}^{North}$ performance review on train sets and test sets for both sampled probes and unsampled probes on the north-facing slope. Source: Authors own work.}
\label{surrogatetablereview_north}
\small{
%\begin{tabular}{lcccc}
%\toprule
%Averaged values & Train sampled & Train not sampled & Test sampled & Test not sampled \\
%\midrule
%Relative error (RE) [\%] & 1.4806 & 1.3196 & 1.5472 & 1.3562 \\
%Absolute error (AE) [$^\circ$C] & 0.1484 & 0.1519 & 0.1528 & 0.1550 \\
%$R^2$ & 0.9810 & 0.9902 & 0.9812 & 0.9896 \\
%Pearson correlation & 0.9909 & 0.9954 & 0.9908 & 0.9950 \\
%Spearman correlation & 0.9867 & 0.9887 & 0.9865 & 0.9878 \\
%95\% CI-RE [\%] & 1.4806 $\pm$ 0.1650 & 1.3196 $\pm$ 0.0982 & 1.5472 $\pm$ 0.1823 & 1.3562 $\pm$ 0.1059 \\
%95\% CI-AE [$^\circ$C] & 0.1484 $\pm$ 0.0077 & 0.1519 $\pm$ 0.0060 & 0.1528 $\pm$ 0.0074 & 0.1550 $\pm$ 0.0056 \\
%\bottomrule
%\end{tabular}
\begin{tabular}{lcccc}
\toprule
Averaged values & Train sampled & Train not sampled & Test sampled & Test not sampled \\
\midrule
Relative error ($e_r$) [\%] & 0.258 & 0.241 & 0.275 & 0.255 \\
%Relative error ($e_r$) [\%] & 1.0143 & 0.9524 & 1.0571 & 0.9836 \\
%\textcolor{red}{Relative error ($e_r^K$) [\%]} & \textcolor{red}{0.0594} & \textcolor{red}{0.0636} & \textcolor{red}{0.0617} & \textcolor{red}{0.0664} \\
Absolute error (AE) [$^\circ$C] & 0.1181 & 0.1233 & 0.1220 & 0.1269 \\
$R^2$ & 0.9934 & 0.9963 & 0.9923 & 0.9957 \\
Pearson & 0.9968 & 0.9982 & 0.9964 & 0.9980 \\
Spearman & 0.9942 & 0.9941 & 0.9935 & 0.9935 \\
%95\% CI-RE [\%] & 1.0143 $\pm$ 0.0773 & 0.9524 $\pm$ 0.0455 & 1.0571 $\pm$ 0.0838 & 0.9836 $\pm$ 0.0473 \\
95\% CI-RE [\%] & 0.258 $\pm$ 0.02 & 0.241 $\pm$ 0.0115 & 0.275 $\pm$ 0.0218 & 0.255 $\pm$ 0.0123 \\
95\% CI-AE [$^\circ$C] & 0.1181 $\pm$ 0.0070 & 0.1233 $\pm$ 0.0050 & 0.1220 $\pm$ 0.0060 & 0.1269 $\pm$ 0.0045 \\
\bottomrule
\end{tabular}
}
\end{table}

\begin{table}[!ht]
\centering
\caption{Surrogate model $\mathcal{H}^{South}$ performance review on train sets and test sets for both sampled probes and unsampled probes on the south-facing slope. Source: Authors own work.}
\label{surrogatetablereview_south}
\small{
\begin{tabular}{lcccc}
\toprule
Averaged values & Train sampled & Train not sampled & Test sampled & Test not sampled \\
\midrule
Relative error ($e_r$) [\%] & 0.556 & 0.482 & 0.815 & 0.691 \\
%Relative error ($e_r$) [\%] & 1.9468 & 1.6868 & 2.0373 & 1.7287 \\
%\textcolor{red}{Relative error ($e_r^K$) [\%]} & \textcolor{red}{0.0849} & \textcolor{red}{0.0869} & \textcolor{red}{0.0875} & \textcolor{red}{0.0894} \\
Absolute error (AE) [$^\circ$C] & 0.1787 & 0.1804 & 0.1835 & 0.1831 \\
$R^2$ & 0.9685 & 0.9842 & 0.9701 & 0.9835 \\
Pearson & 0.9849 & 0.9926 & 0.9852 & 0.9920 \\
Spearman & 0.9792 & 0.9834 & 0.9794 & 0.9822 \\
%95\% CI-RE [\%] & 1.9468 $\pm$ 0.2645 & 1.6868 $\pm$ 0.1489 & 2.0373 $\pm$ 0.2993 & 1.7287 $\pm$ 0.1670 \\
95\% CI-RE [\%] & 0.556 $\pm$ 0.076 & 0.482 $\pm$ 0.043 & 0.815 $\pm$ 0.12 & 0.691 $\pm$ 0.067 \\
95\% CI-AE [$^\circ$C] & 0.1787 $\pm$ 0.0070 & 0.1804 $\pm$ 0.0056 & 0.1835 $\pm$ 0.0060 & 0.1831 $\pm$ 0.0044 \\
\bottomrule
\end{tabular}
}
\end{table}

The results showcased in Tables \ref{surrogatetablereview_north} and \ref{surrogatetablereview_south} showcase the ability of the surrogate model to reproduce PermaFOAM results for unseen datasets and, within the space domain, for unseen probe locations. In fact, only $25$ probes are sampled and included in the dataset, while $36$ other probes are not sampled and used to assess the ability to generalize along different locations in the analyzed permafrost. The reported error and the confidence intervals show the ability of the neural network to reproduce the results within a fraction of a degree.

When it comes to computation time, the prediction of a complete dataset over the course of 87 years requires only $0.696$ seconds on a typical laptop, instead of multiple months on a cluster with multiple cores, reported as half a million CPU hours. A speedup factor of $25.86\times 10^9$ times is therefore achieved, with a precision reduction of less than a single degree Celsius. 

\section{Hybridization methodology}\label{hybridsec}
When it comes to hybrid modeling, the modeling effort starts by acknowledging that the knowledge we have about the natural phenomena, modeled by partial differential equations, is not perfect. The discrepancy between the models and the measurements is partially epistemic and can be assimilated within the models \citep{CRMECA_2021__349_2_259_0}. The modeling supposes the presence of a possible data-driven correction such as:
\begin{equation}
    \mathbf{T}^{measured}=\mathbf{T}^{model}+\mathbf{T}^{correction}+\bm\eta,
\end{equation}

with $\mathbf{T}^{measured}$ the measured values,$\mathbf{T}^{model}$ the surrogate values, $\mathbf{T}^{correction}$ a correction term, and $\bm\eta$ a noise term such as $\bm\eta\sim \mathcal{N}(\mathbf{0}, \bm{\Sigma})$, $\mathcal{N}(\mathbf{0}, \bm{\Sigma})$ being a normal distribution of mean $0$ and covariance matrix $\bm{\Sigma}$. In our specific application, we aim to model the permafrost predictions, such as:
\begin{equation}\label{modelingeq}
    \mathbf{T}^{measured}=\mathbf{V}\mathcal{H}(\mathbf{p})+\mathbf{V}_g \mathcal{G}\left(\mathbf{p},\mathbf{V}\mathcal{H}(\mathbf{p})\right)+\bm\eta.
\end{equation}

In other words, $\mathbf{T}^{model}$ is approximated using surrogate modeling $\mathbf{V}\mathcal{H}(\mathbf{p})$, to avoid computational time and cost. At the same time, a correction model $\mathbf{V}_g \mathcal{G}$, predicting the epistemic error $\mathbf{T}^{correction}$ using a reduced basis $\mathbf{V}_g$ and a weight predictor $\mathcal{G}$, is constructed. The epistemic error is computed as a function of the normalized input parameters, but also as a function of the model temperatures $\mathbf{T}^{model}$, predicted using the available surrogate $\mathcal{H}$. 

\begin{figure}[!ht]
    \centering
    \subfigure[North slope measurements \label{northmeasurespic}]{\includegraphics[width=0.7\linewidth]{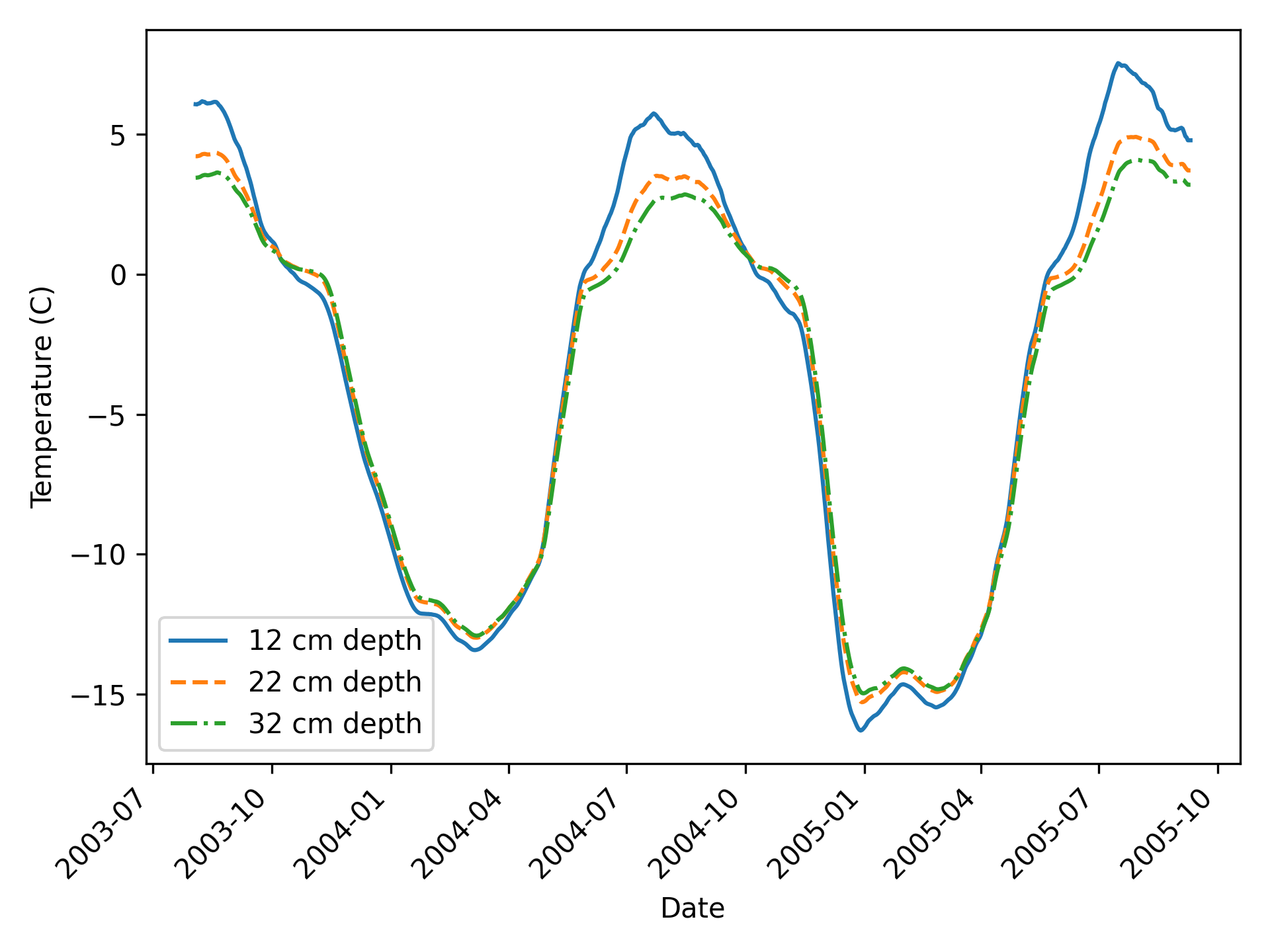}}
    \subfigure[South slope measurements \label{southmeasurespic}]{\includegraphics[width=0.7\linewidth]{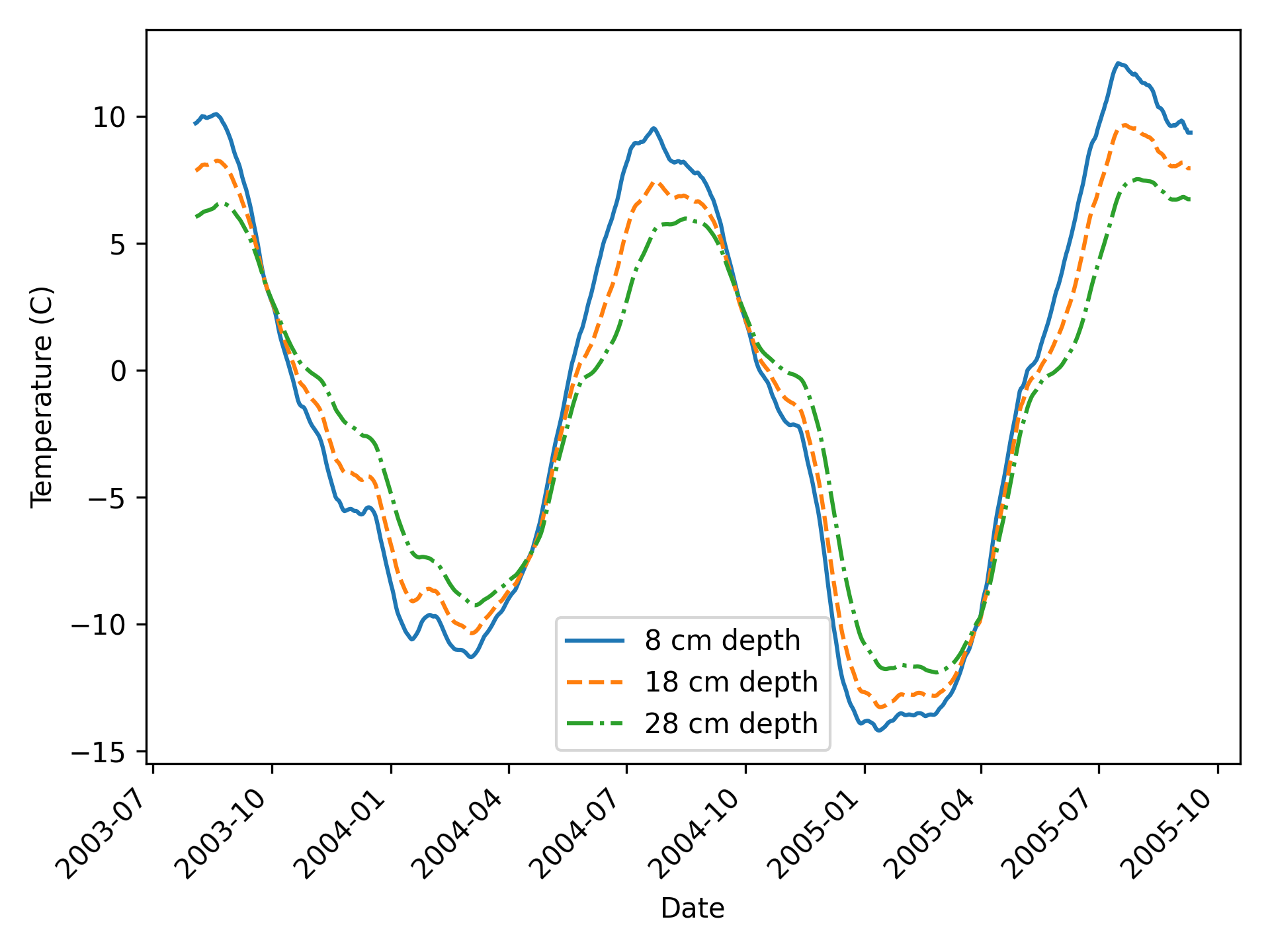}}
    \caption{Experimental measurements performed in the permafrost in the modeled site. Source: Authors own work.}
    \label{fig:experiementalmeasures}
\end{figure}

The modeling proposed in Eq. (\ref{modelingeq}) is proposed throughout the complete physical space domain $\mathbf{x}$. However, the measured temperature is restricted to several experimentally probed locations $\mathbf{x}^\star$. In fact, the available experimental data consist of three probes placed at $12cm$, $22cm$ and $32 cm$ depth from the ground surface on the north-facing slope (numerical probes 6, 10 and 14), and $8cm$, $18cm$ and $28 cm$ depth on the south-facing slope (numerical probes 4, 8 and 12). The measurements are performed daily over the course of about 2 years. The results are illustrated in Figure \ref{fig:experiementalmeasures} for the north-facing slope and the south-facing slope (personal communication from A.S. Prokushkin).

\subsection{Hybrid modeling formulation}
To construct the correction surrogate model $\mathcal{G}$, the predictive values from the previous surrogate are considered as additional inputs to the model. For instance, the modeling uses as input the concatenation of $(\mathbf{p},\mathbf{T}^{model})$ for the last $\tau=60$ time steps. The structure used for the prediction is based on a combined LSTM and dense networks operating on a moving window of size $\tau$. To allow space interpolation, another reduced basis $\mathbf{V}_g$ is used in this section. 

The reduced basis is constructed using the PermaFOAM temperatures as detailed in Eq. (\ref{svdeq1}) and Eq. (\ref{truncationeq}). However, the truncation is performed using $\epsilon_t=10^{-5}$ (instead of $\epsilon_t=10^{-4}$ when building the surrogate model in Section \ref{surrogatetrainingsec}), leading to $\kappa^{hybrid}_{north}=21$ vectors for the north-facing slope and $\kappa^{hybrid}_{south}=20$ for the south-facing slope. This reduced basis with a larger number of modes $\kappa$ allows more correction flexibility through space without compromising the physical shape of the solution. 

The loss function used to train $\mathcal{G}$ is computed, therefore, using
\begin{equation}\label{losscorrection}
    \mathcal{L}^{correction}=\frac{1}{m}\sum\limits_{i=1}^m \left(\mathbf{V}_g^\star\mathcal{G}\left(\mathbf{p}_i,\mathbf{V}^\star\mathcal{H}(\mathbf{p}_i)\right)-{T^\star}^{correction}_i\right)^2    
\end{equation}

The superscript $~^\star$ restricts the field to the available experimental probe locations, $\mathbf{x}^\star$, and ${T^\star}_i^{correction}$ the target discrepancies of correction values sampled only on $\mathbf{x}^\star$, defined using
\begin{equation}
    {T^\star}_i^{correction}={T^\star}_i^{measured}-{T^\star}_i^{model}
\end{equation}
with ${T^\star}_i^{measured}$ the measured temperatures experimentally at $\mathbf{x}^\star$, for a time step $i$, and ${T^\star}_i^{model}$ the probed model values at the same locations for the same time step $i$.

The definition of ${T}_i^{correction}$ where the measures are not available will not impact the results, as the loss only acts on the points $\mathbf{x}^\star$ where the experimental measures are available as defined in Eq. (\ref{losscorrection}). The predicted temperatures remain continuous thanks to the projection on the continuous reduced basis $\mathbf{V}_g$, which performs a physically grounded interpolation in between the measuring points. The final prediction using the hybrid modeling effort writes:
\begin{equation}
    \hat{T}_i^{hybrid}=\mathbf{V}\mathcal{H}(\mathbf{p}_i)+\mathbf{V}_g\mathcal{G}\left(\mathbf{p}_i,\mathbf{V}\mathcal{H}(\mathbf{p}_i)\right),
\end{equation}

\begin{figure}[!ht]
    \centering
    \subfigure[North-facing slope surrogate model of the epistemic error $\mathcal{G}_{north}$\label{surrogatecorrectnorth}]{\includegraphics[width=\linewidth]{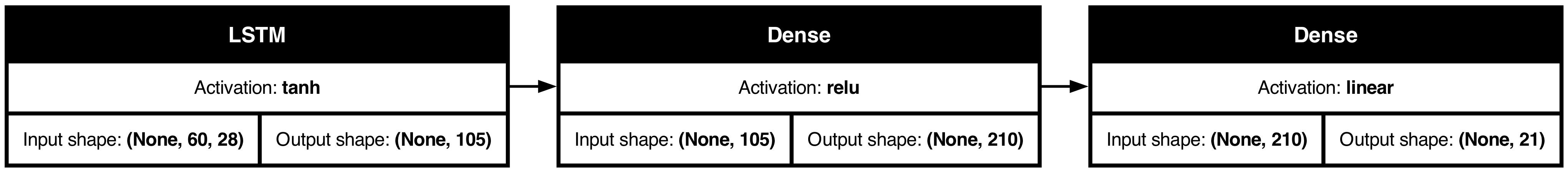}}
    \subfigure[South-facing slope surrogate model of the epistemic error $\mathcal{G}_{south}$ \label{surrogatecorrectsouth}]{\includegraphics[width=\linewidth]{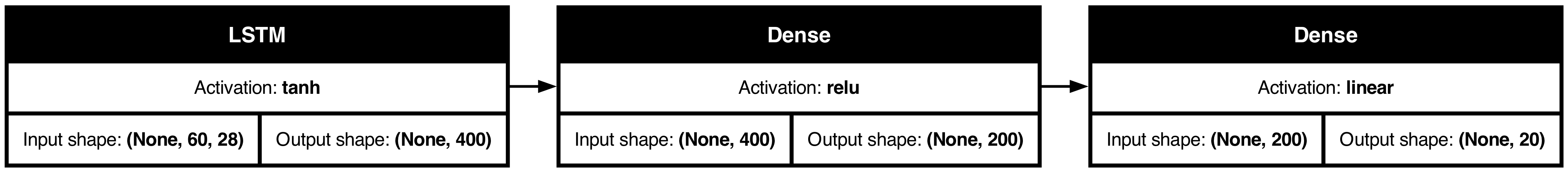}}
    \caption{Surrogate corrective models predicting the epistemic discrepancy between the model and the experimental measurements, showing the neural network layers from left to right, their size and their activation functions. Source: Authors own work.}
    \label{surrigatemodelhybrid}
\end{figure}

$\hat{T}_i^{hybrid}$ being the final hybrid model predictions. The surrogate model $\mathcal{H}$ is set to not trainable while training $\mathcal{G}$. The used surrogate model for $\mathcal{G}$ is illustrated in Figure \ref{surrigatemodelhybrid}. The number of parameters and layer sizes used are linked to the number of parameters to predict in the truncated reduced basis resulting from the SVD decomposition.

The training is performed using the ADAM optimizer \citep{kingma2015adam}, along with a custom-built decreasing learning rate function. Batch gradient descent is used along with a total of 800 epochs performed during the training and a batch size of 10 days.

\subsection{Hybrid modeling results}\label{hybridresultssec}
\begin{figure}[!ht]
    \centering
    \subfigure[North-facing slope predictions at $12 cm$ depth.]{\includegraphics[width=0.48\linewidth]{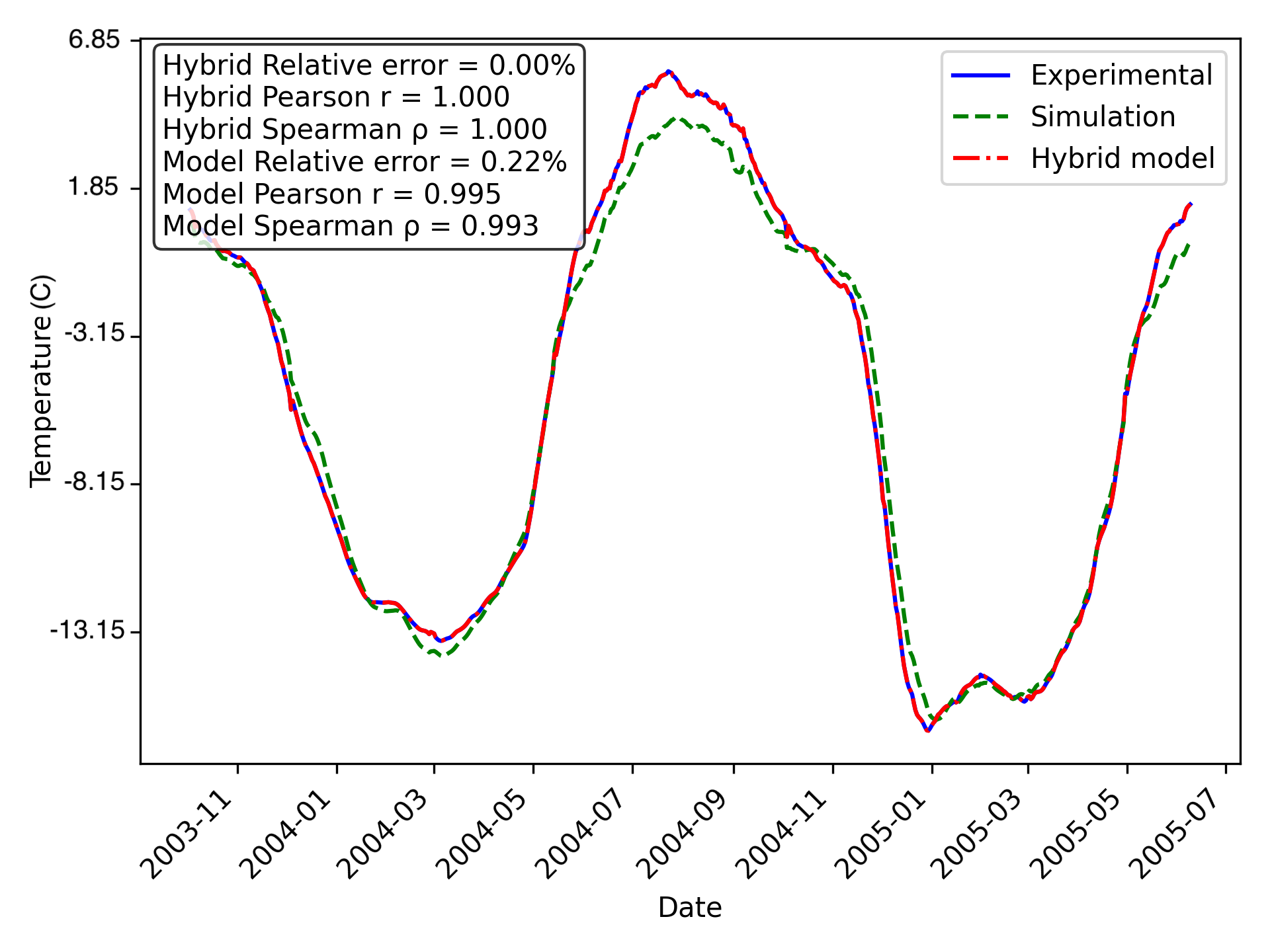}}
    \subfigure[North-facing slope predictions at $22 cm$ depth.]{\includegraphics[width=0.48\linewidth]{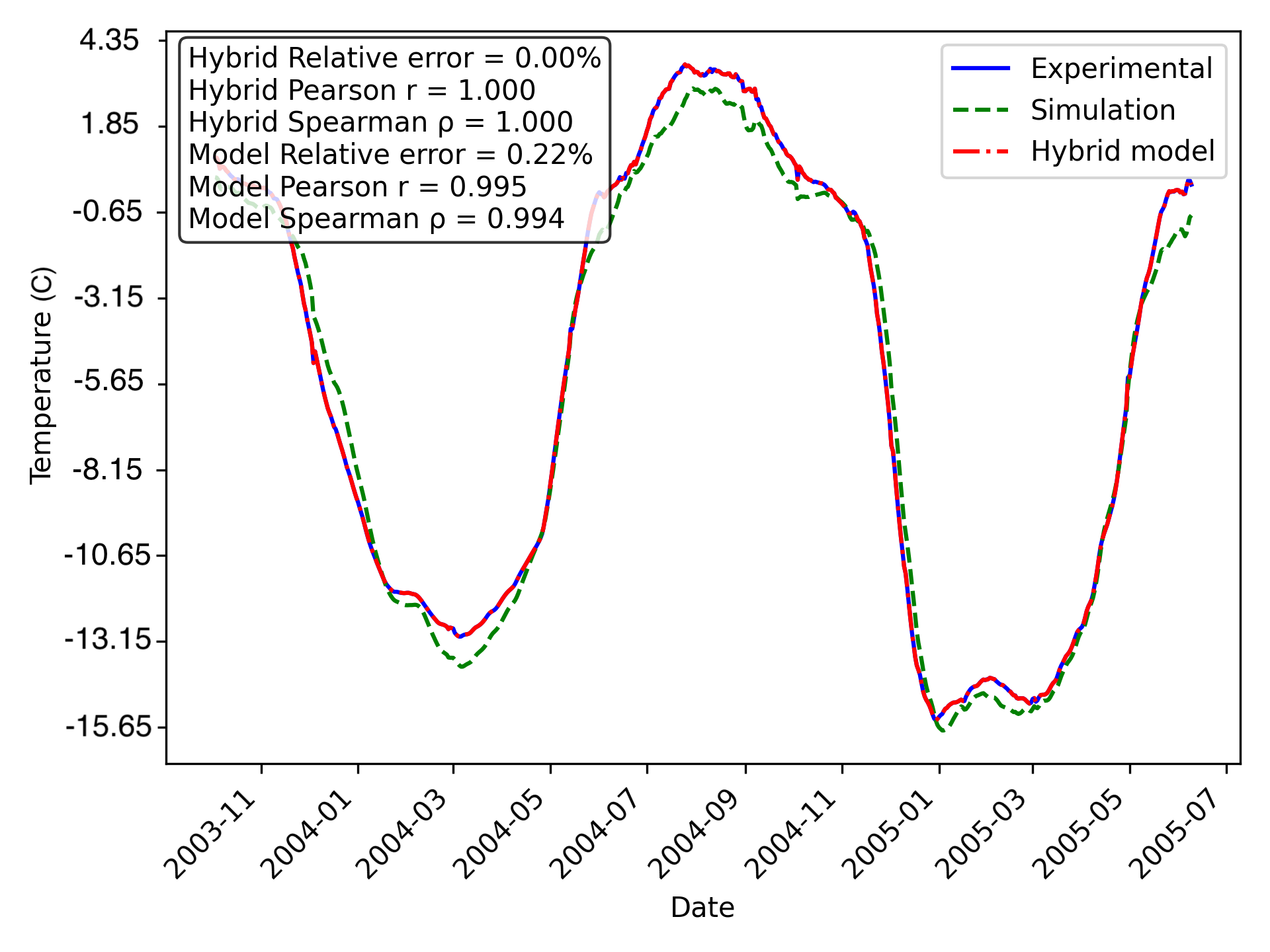}}
    \subfigure[North-facing slope predictions at $32 cm$ depth.]{\includegraphics[width=0.48\linewidth]{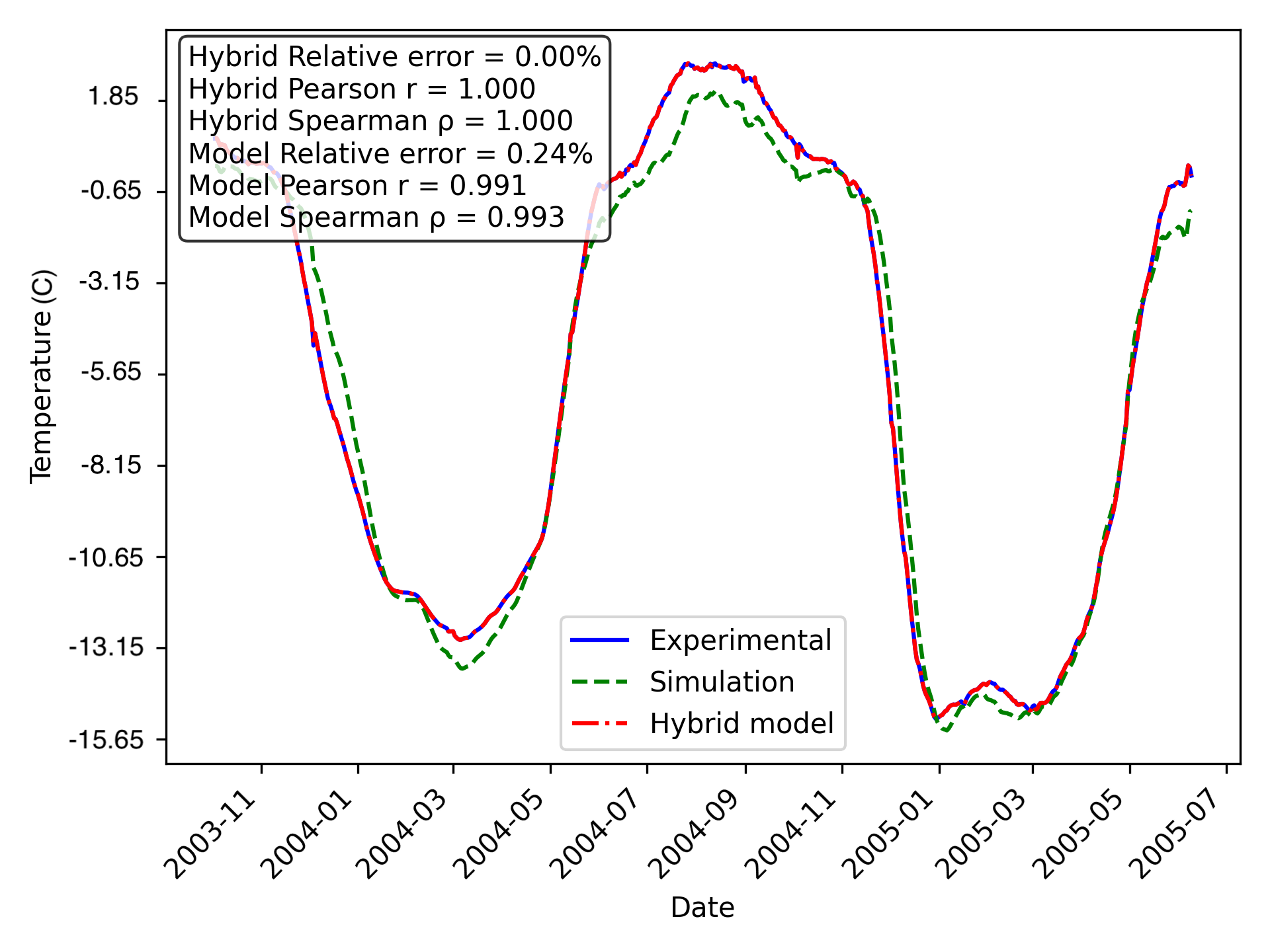}}
    \caption{North-facing slope comparison on the training set of the experimental measurements, the simulation obtained using the surrogate model $\mathbf{V}\mathcal{H}$, and the hybrid model $\mathbf{V}\mathcal{H}+\mathbf{V}_g\mathcal{G}$. Source: Authors own work.}
    \label{surrigatemodelhybridtrainingnorth}
\end{figure}

\begin{figure}[!ht]
    \centering
    \subfigure[South-facing slope predictions at $12 cm$ depth.]{\includegraphics[width=0.48\linewidth]{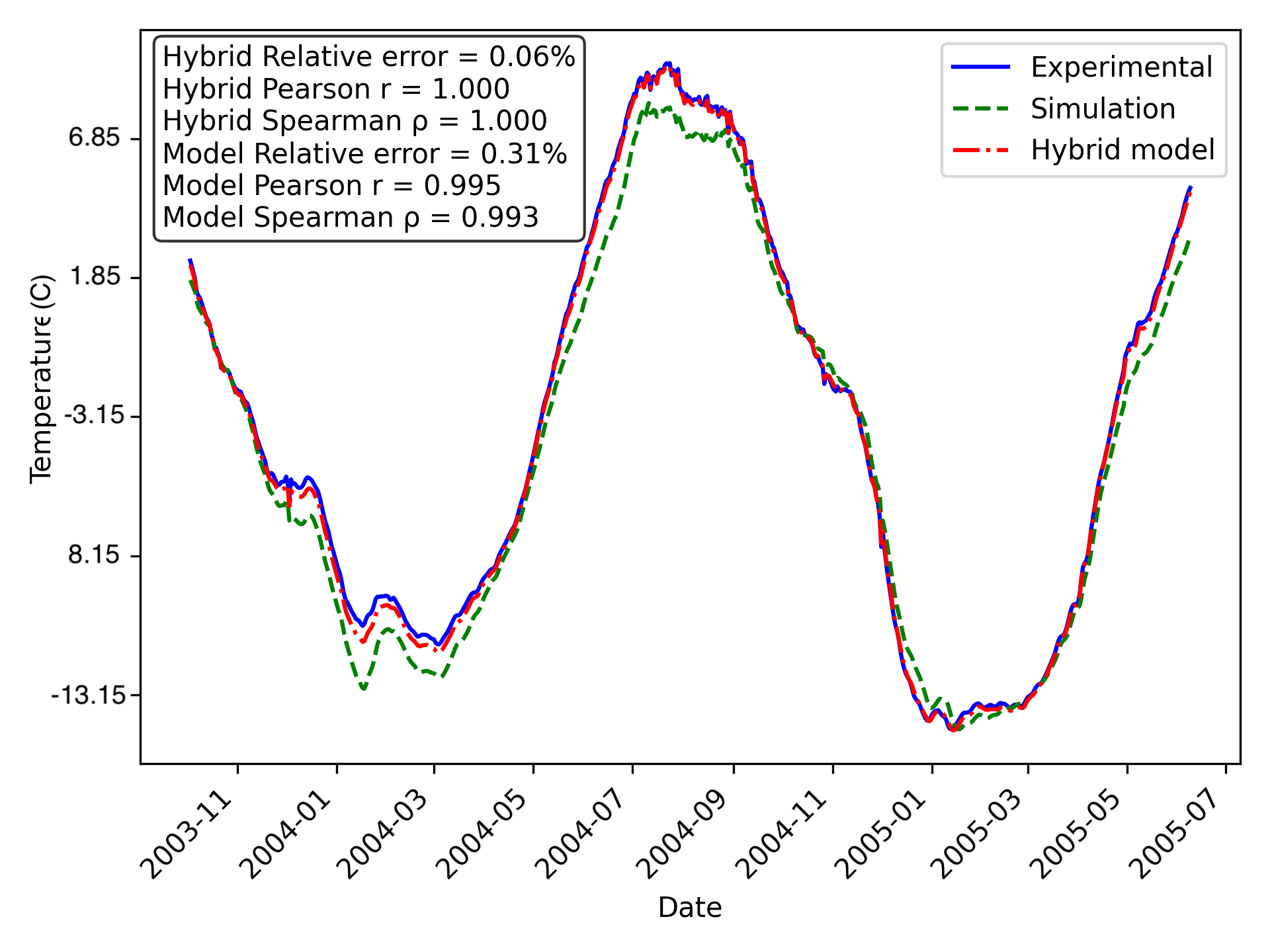}}
    \subfigure[South-facing slope predictions at $22 cm$ depth.]{\includegraphics[width=0.48\linewidth]{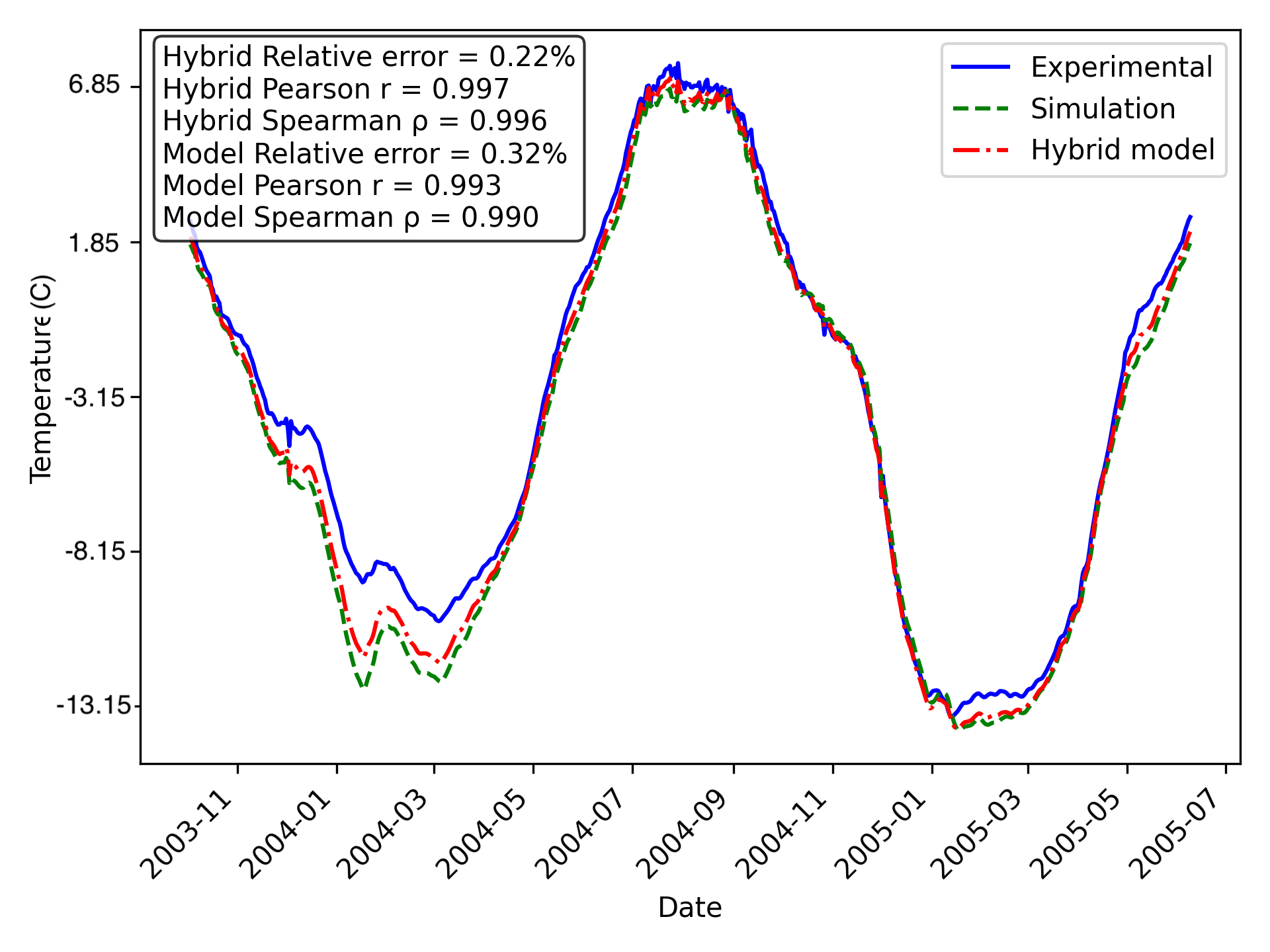}}
    \subfigure[South-facing slope predictions at $32 cm$ depth.]{\includegraphics[width=0.48\linewidth]{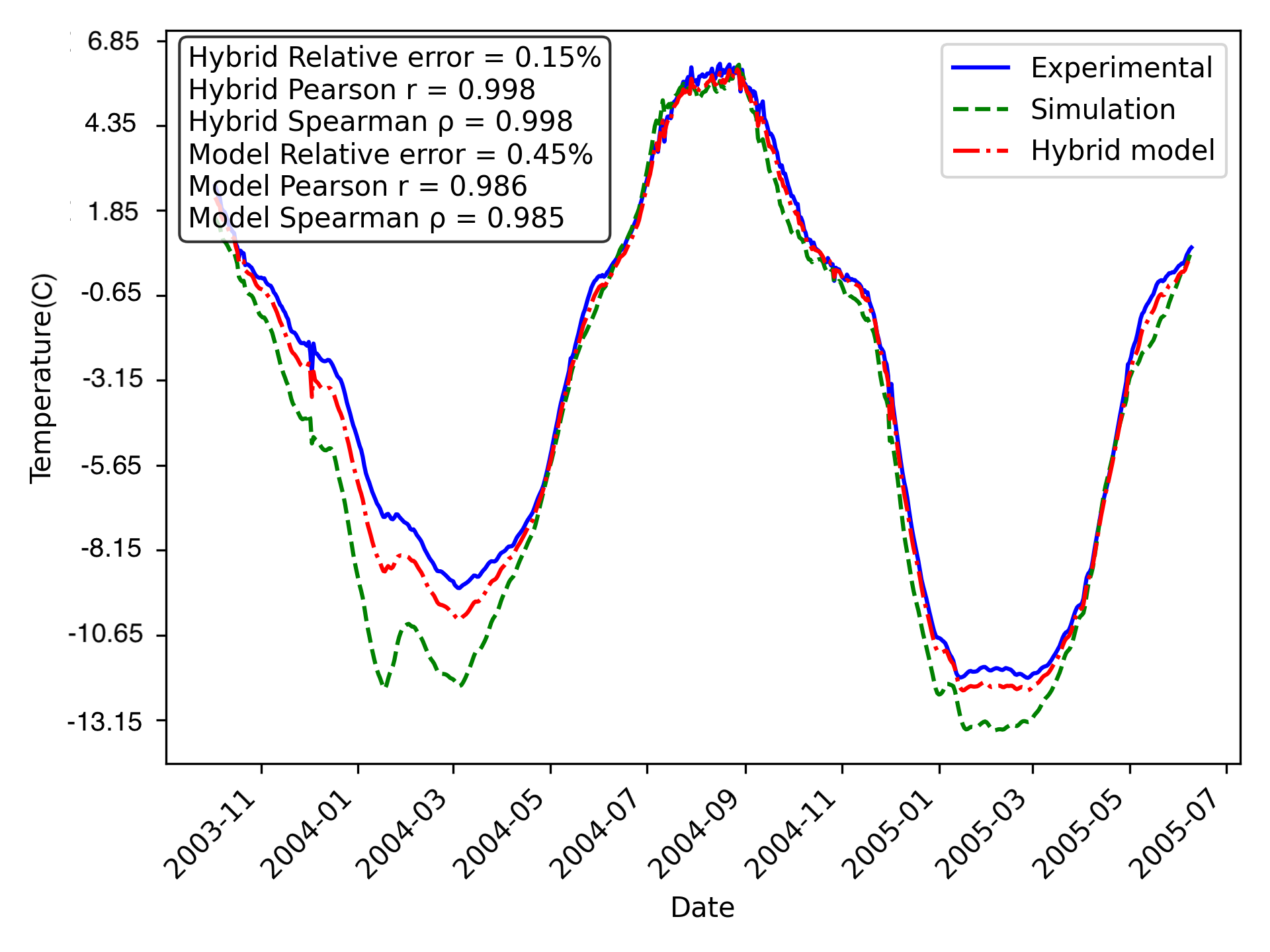}}
    \caption{South-facing slope comparison on the training set of the experimental measurements, the simulation obtained using the surrogate model $\mathbf{V}\mathcal{H}$, and the hybrid model $\mathbf{V}\mathcal{H}+\mathbf{V}_g\mathcal{G}$. Source: Authors own work.}
    \label{surrigatemodelhybridtrainingsouth}
\end{figure}

\begin{figure}[!ht]
    \centering
    \subfigure[North-facing slope predictions at $12 cm$ depth, test set results.]{\includegraphics[width=0.48\linewidth]{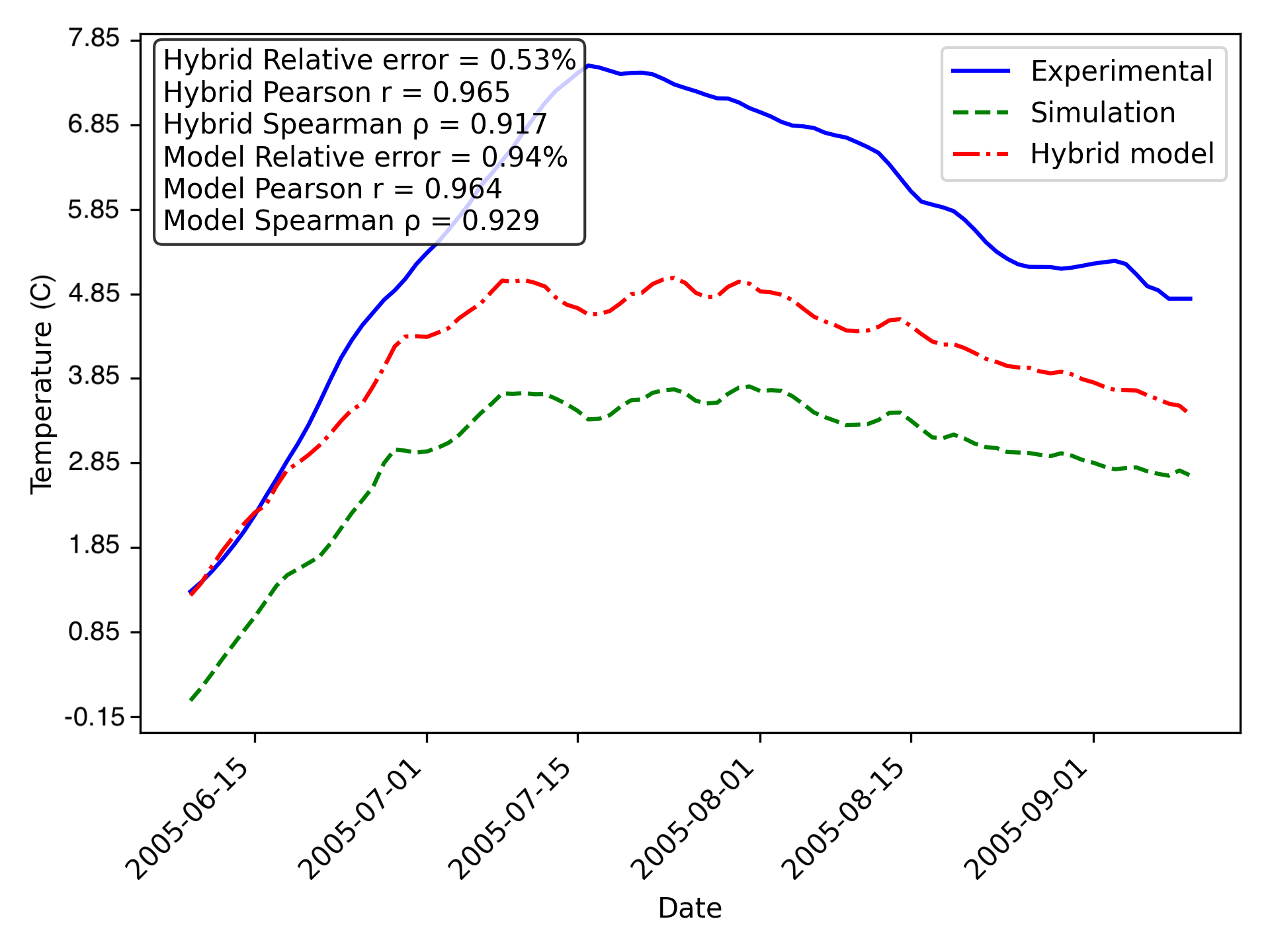}}
    \subfigure[North-facing slope predictions at $22 cm$ depth, test set results.]{\includegraphics[width=0.48\linewidth]{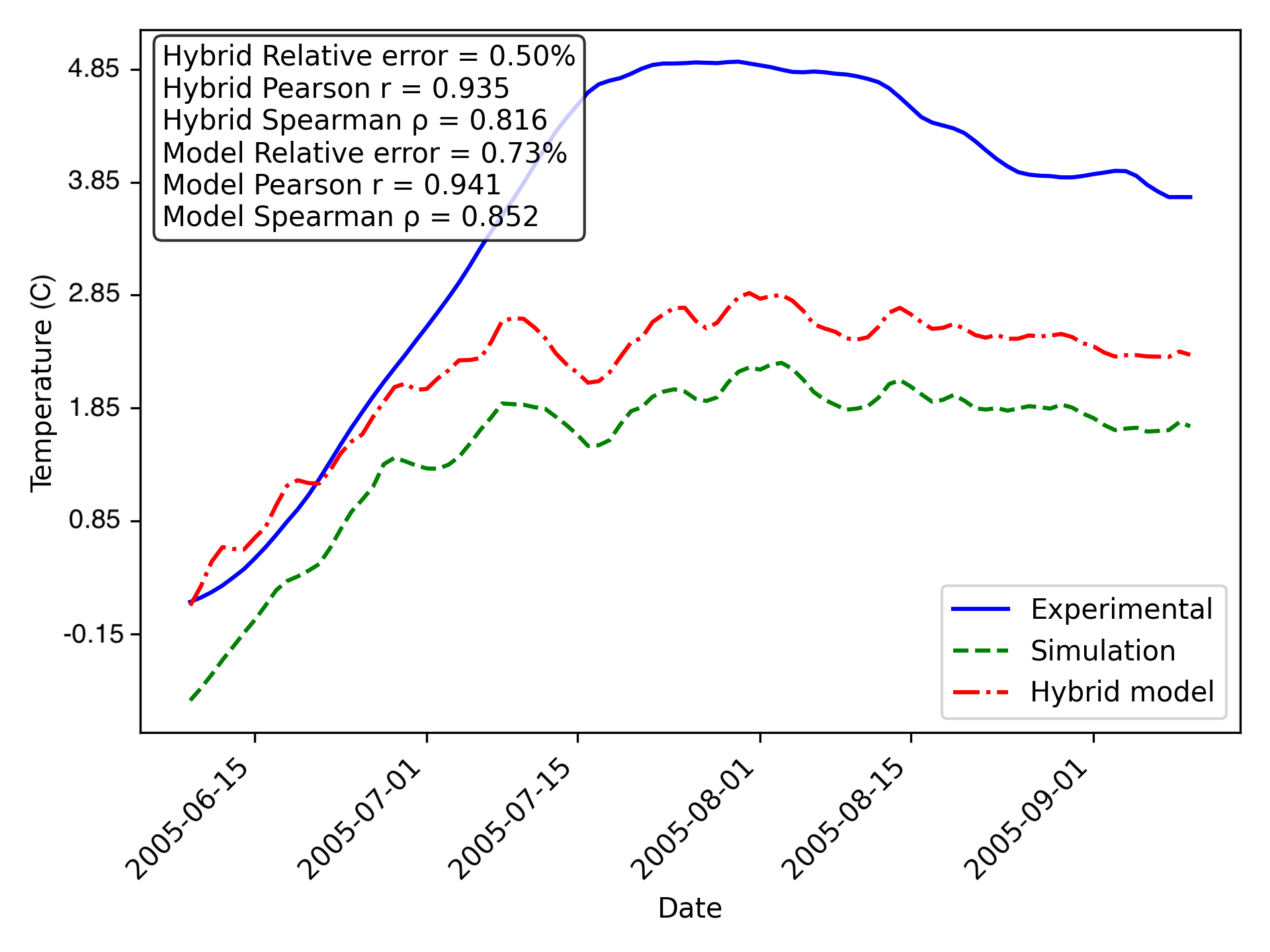}}
    \subfigure[North-facing slope predictions at $32 cm$ depth, test set results.]{\includegraphics[width=0.48\linewidth]{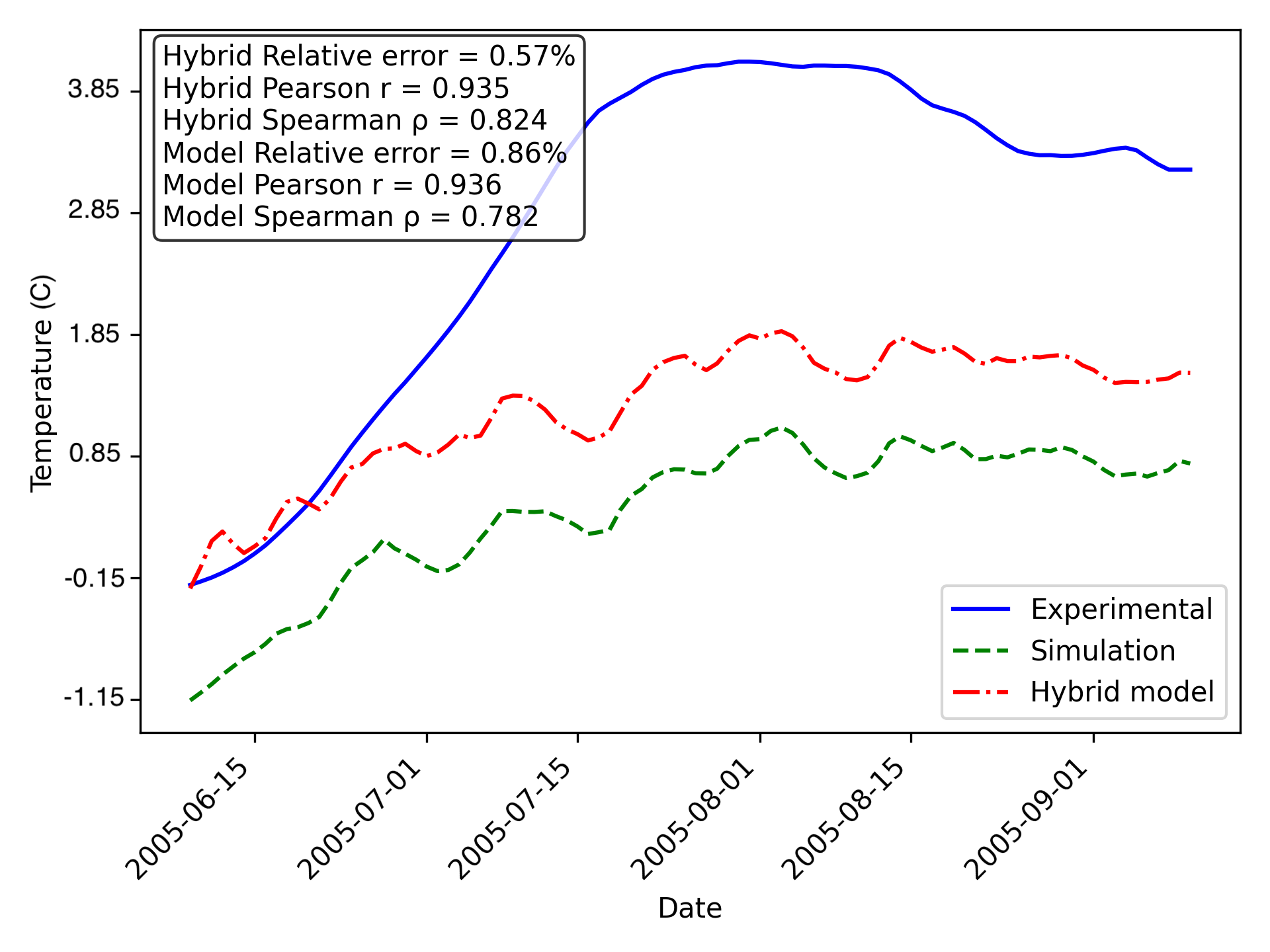}}
    \caption{North-facing slope comparison on the test set of the experimental measurements, the simulation obtained using the surrogate model $\mathbf{V}\mathcal{H}$, and the hybrid model $\mathbf{V}\mathcal{H}+\mathbf{V}_g\mathcal{G}$. Source: Authors own work.}
    \label{surrigatemodelhybridtestnorth}
\end{figure}

\begin{figure}[!ht]
    \centering
    \subfigure[South-facing slope predictions at $12 cm$ depth, test set results.]{\includegraphics[width=0.48\linewidth]{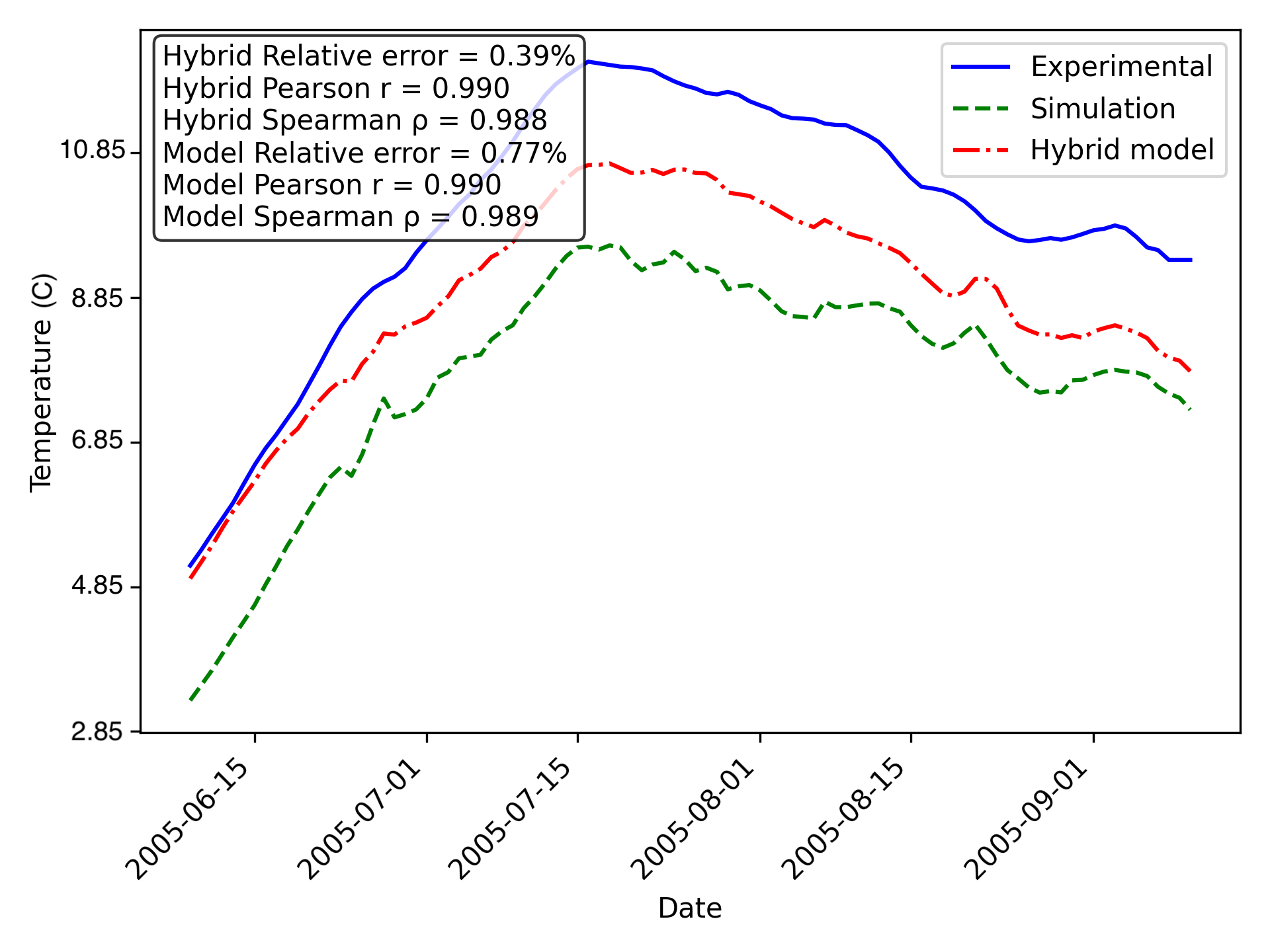}}
    \subfigure[South-facing slope predictions at $22 cm$ depth, test set results.]{\includegraphics[width=0.48\linewidth]{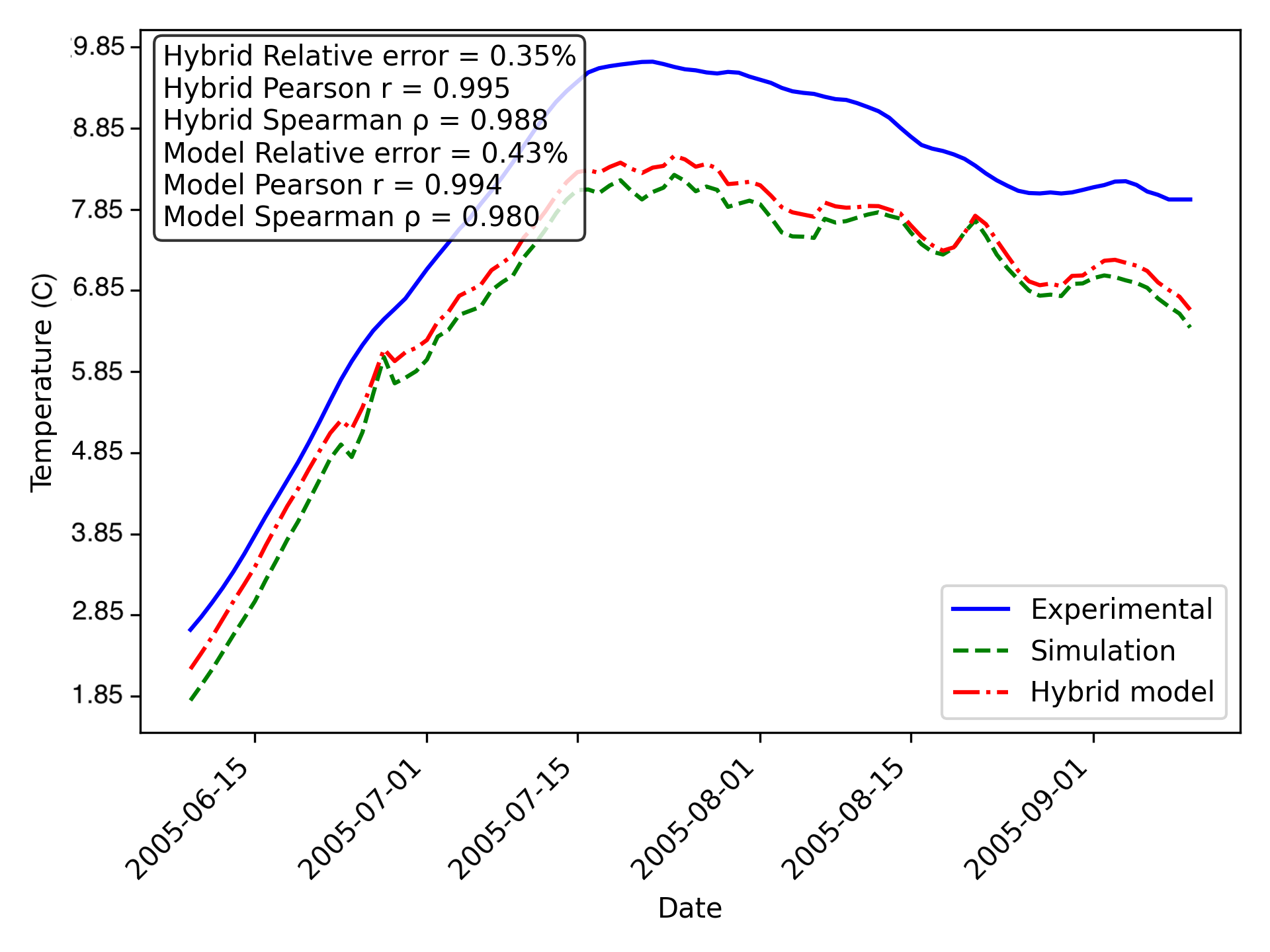}}
    \subfigure[South-facing slope predictions at $32 cm$ depth, test set results.]{\includegraphics[width=0.48\linewidth]{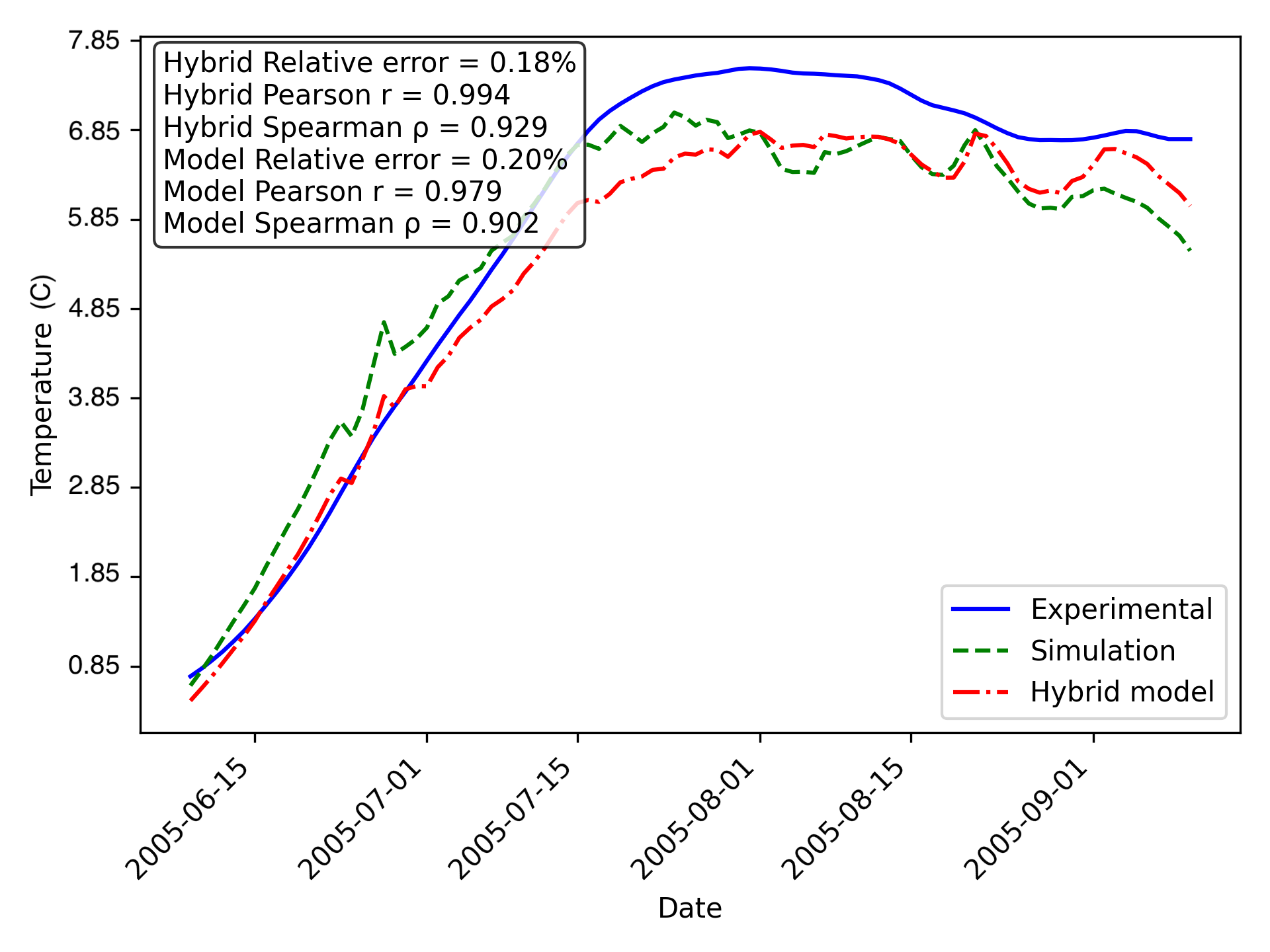}}
    \caption{South-facing slope comparison on the test set of the experimental measurements, the simulation obtained using the surrogate model $\mathbf{V}\mathcal{H}$, and the hybrid model $\mathbf{V}\mathcal{H}+\mathbf{V}_g\mathcal{G}$. Source: Authors own work.}
    \label{surrigatemodelhybridtestsouth}
\end{figure}

To evaluate the performance of the model, it is trained over 80\% of the data, and the last $20$\% is kept secret for testing and not shown for the surrogate model. The results consist, therefore, of an extrapolation in time for the surrogate model. The train results over the three measured probes are illustrated in Figures \ref{surrigatemodelhybridtrainingnorth} and \ref{surrigatemodelhybridtrainingsouth} for the training set on both north-facing and south-facing slopes, respectively. The test results are illustrated in Figures \ref{surrigatemodelhybridtestnorth} and \ref{surrigatemodelhybridtestsouth}. Table \ref{tab:hybrid_model_performance_north} shows average values through the test and train sets on the north-facing slope, while Table \ref{tab:hybrid_model_performance_south} shows the same statistics on the south-facing slope.

\begin{table}
\caption{Performance metrics for the baseline model approximating PermaFOAM and the hybrid model correcting the surrogate with $\mathcal{G}^{North}$ on the north-facing slope. Source: Authors own work.}
\label{tab:hybrid_model_performance_north}
\begin{tabular}{lllll}
\toprule
 Average values & Baseline train & Baseline test & Hybrid train & Hybrid test \\
\midrule
Absolute error (AE) [$^\circ$C]& 0.6317 & 2.3526 & 0.0023 & 1.4899 \\
%Relative error ($e_r$) [\%] & 0.3021 & 0.9059 & 0.0013 & 0.6133 \\
Relative error ($e_r$) [\%] & 0.061 & 0.181 & 0.0011 & 0.557 \\
Pearson & 0.9935 & 0.9248 & 1.0000 & 0.9075 \\
Spearman & 0.9929 & 0.9210 & 1.0000 & 0.9031 \\
95\% CI-AE [$^\circ$C]& [0.6087, 0.6539] & [2.2521, 2.4605] & [0.0022, 0.0025] & [1.3952, 1.5885] \\
%95\% CI-RE [\%] & [0.29, 0.31] & [0.87, 0.94] & [0.00, 0.00] & [0.59, 0.64] \\
95\% CI-RE [\%] & [0.059, 0.063] & [0.17, 0.19] & [0.00, 0.00] & [0.53, 0.58] \\
\bottomrule
\end{tabular}
\end{table}

\begin{table}
\caption{Performance metrics for the baseline model approximating PermaFOAM and the hybrid model correcting the surrogate with $\mathcal{G}^{South}$ on the south-facing slope. Source: Authors own work.}
\label{tab:hybrid_model_performance_south}
\begin{tabular}{lllll}
\toprule
  Average values & Baseline train & Baseline test & Hybrid train & Hybrid test \\
\midrule
Absolute error (AE) [$^\circ$C]& 1.0097 & 1.3154 & 0.3920 & 0.8609 \\
%Relative error ($e_r$) [\%] & 0.4871 & 0.5374 & 0.2076 & 0.3435 \\
Relative error ($e_r$) [\%] & 0.199 & 0.219 & 0.218 & 0.362 \\
Pearson & 0.9895 & 0.9703 & 0.9981 & 0.9953 \\
Spearman & 0.9882 & 0.9710 & 0.9975 & 0.9909 \\
95\% CI-AE [$^\circ$C]& [0.9717, 1.0471] & [1.2309, 1.4026] & [0.3735, 0.4100] & [0.8088, 0.9139] \\
%95\% CI-RE [\%] & [0.47, 0.51] & [0.51, 0.57] & [0.20, 0.22] & [0.33, 0.36] \\
95\% CI-RE [\%] & [0.192, 0.208] & [0.21, 0.233] & [0.21, 0.231] & [0.347, 0.38] \\
\bottomrule
\end{tabular}
\end{table}

The reported results show first the ability of the model to complement the north-facing slope surrogate much better than the performance obtained on the south-facing slope. This can be partially attributed to a higher fidelity in the PermaFOAM model when used on the north-facing slope. Moreover, the south-facing slope exhibits more errors in the deeper layers when it comes to the simulation results.

Moreover, the results showcase the ability of the model to extrapolate in time while maintaining stability on both north- and south-facing slopes. The results show a systematic improvement of the surrogate model when complemented with the data-driven correct.

\section{Projected permafrost dynamics until 2100}\label{forecastingsec}
In this section, we use the previously trained surrogate model $\mathcal{H}$ along with the epistemic correction model $\mathcal{G}$ to forecast the permafrost temperature and behavior up until year 2100, under the four climate scenarios considered in this work.

\begin{figure}[!ht]
    \centering
    \subfigure[North-facing slope predictions at $12 cm$ depth using SSP126.]{\includegraphics[width=\linewidth]{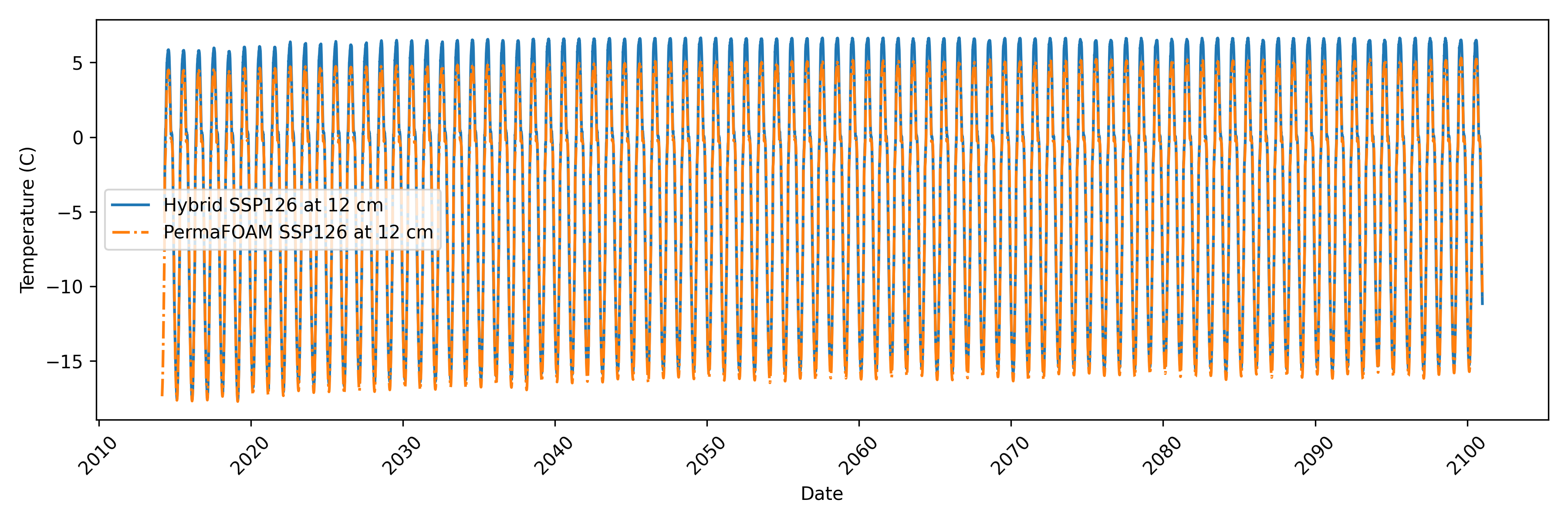}}
    \subfigure[North-facing slope predictions at $12 cm$ depth using SSP245.]{\includegraphics[width=\linewidth]{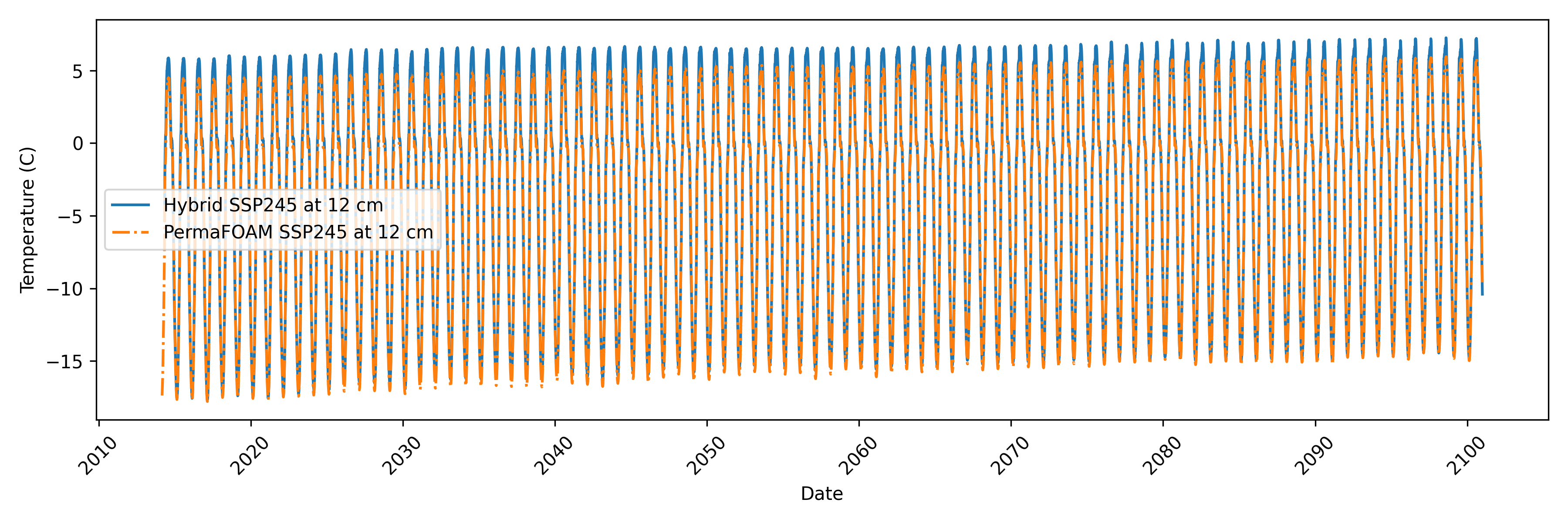}}
    \subfigure[North-facing slope predictions at $12 cm$ depth using SSP370.]{\includegraphics[width=\linewidth]{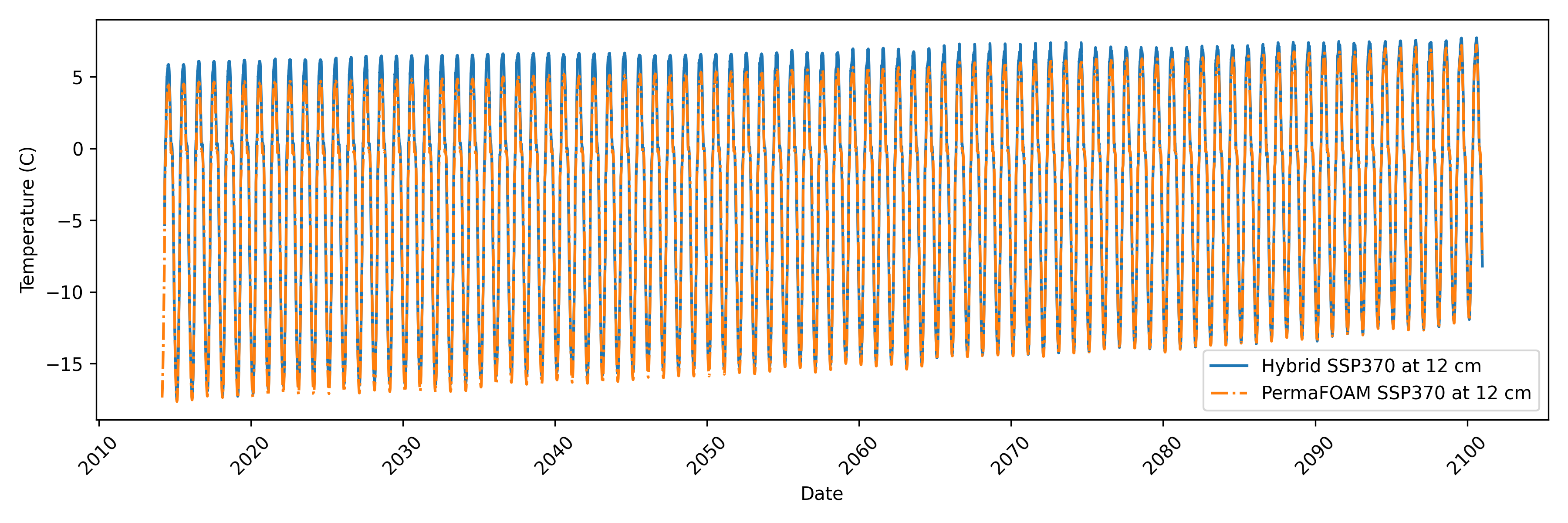}}
    \subfigure[North-facing slope predictions at $12 cm$ depth using SSP585.]{\includegraphics[width=\linewidth]{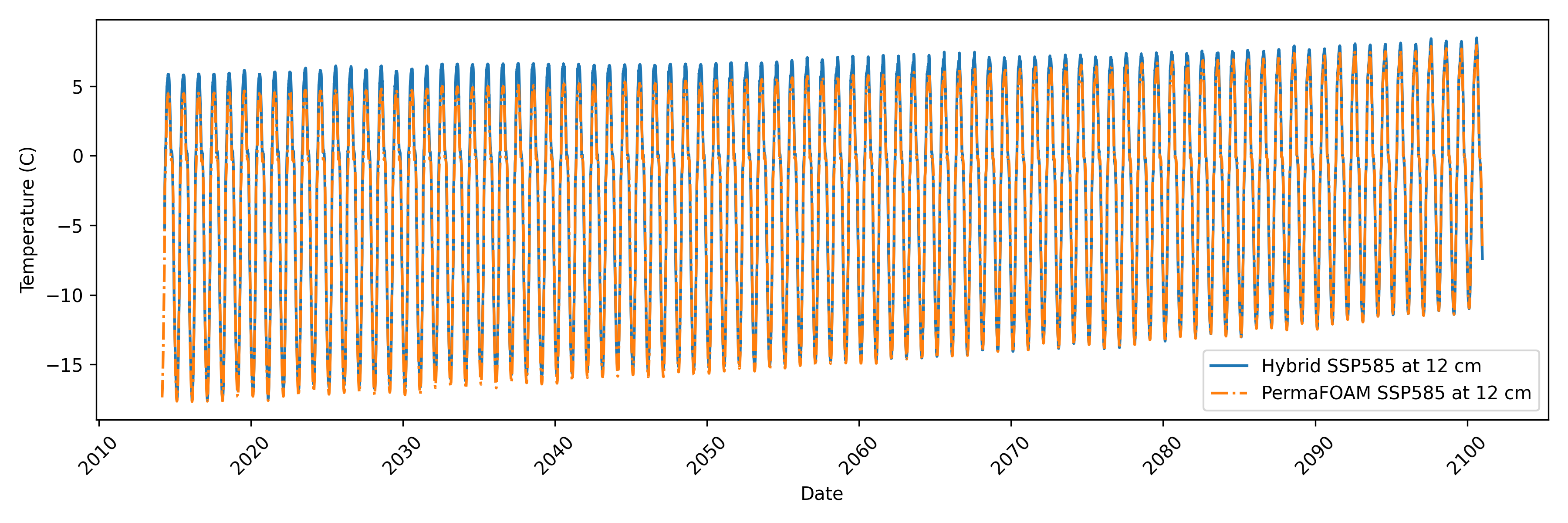}}
    \caption{North-facing slope comparison between PermaFOAM and the hybrid model at $12 cm$ depth. The temperature of the hybrid model are systemically higher than the PermaFOAM ones. Source: Authors own work.}
    \label{comparepermahybrid12north}
\end{figure}

\begin{figure}[!ht]
    \centering
    \subfigure[North-facing slope predictions at $32 cm$ depth using SSP126.]{\includegraphics[width=\linewidth]{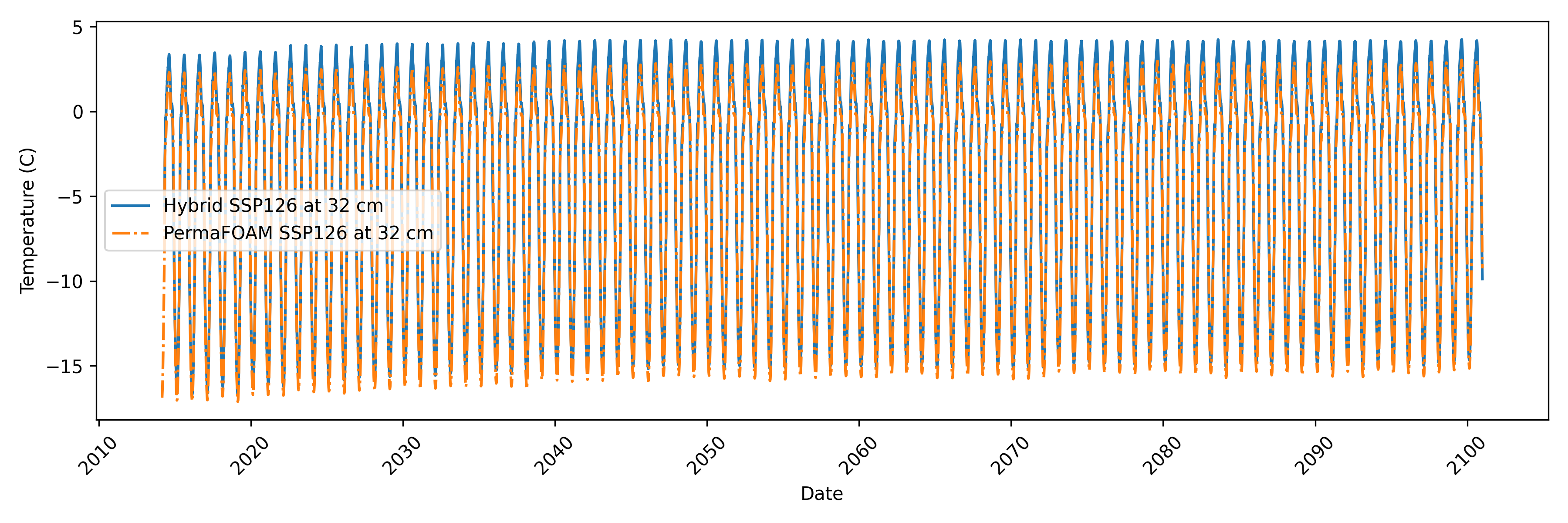}}
    \subfigure[North-facing slope predictions at $32 cm$ depth using SSP245.]{\includegraphics[width=\linewidth]{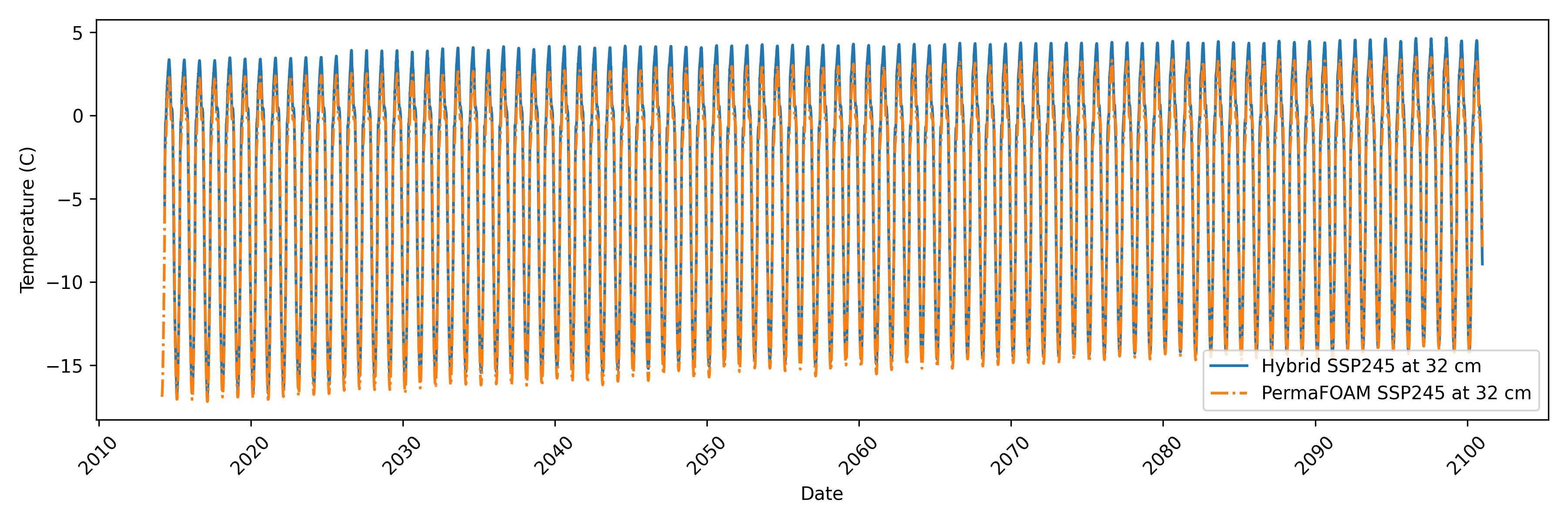}}
    \subfigure[North-facing slope predictions at $32 cm$ depth using SSP370.]{\includegraphics[width=\linewidth]{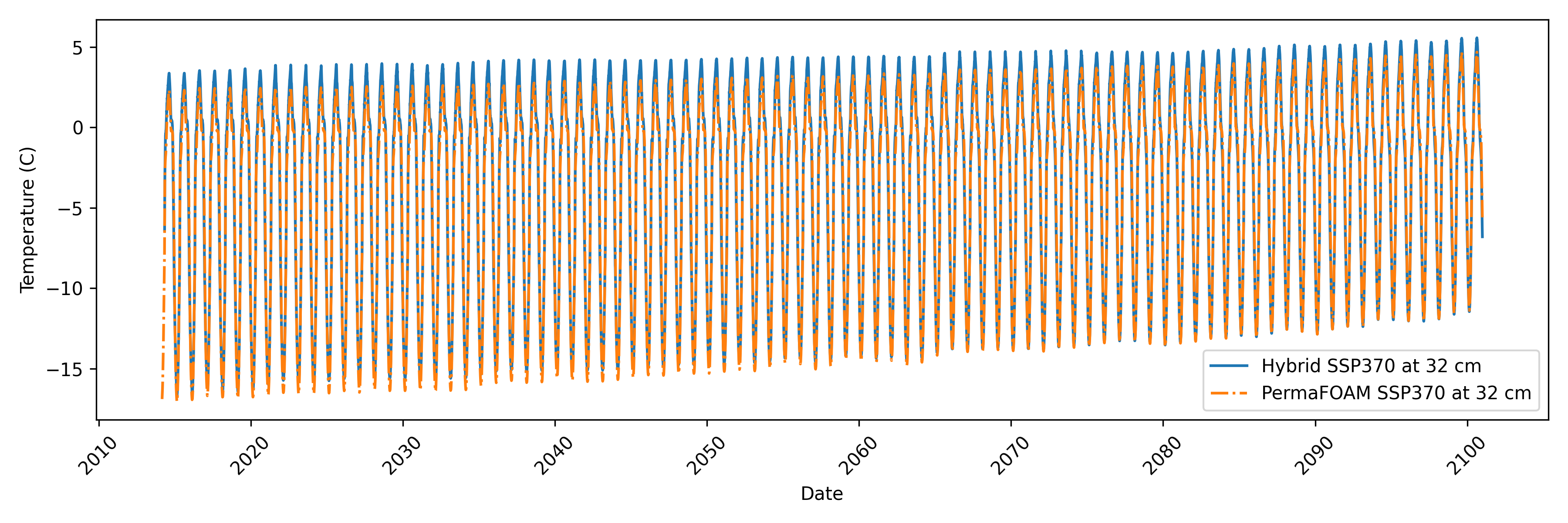}}
    \subfigure[North-facing slope predictions at $32 cm$ depth using SSP585.]{\includegraphics[width=\linewidth]{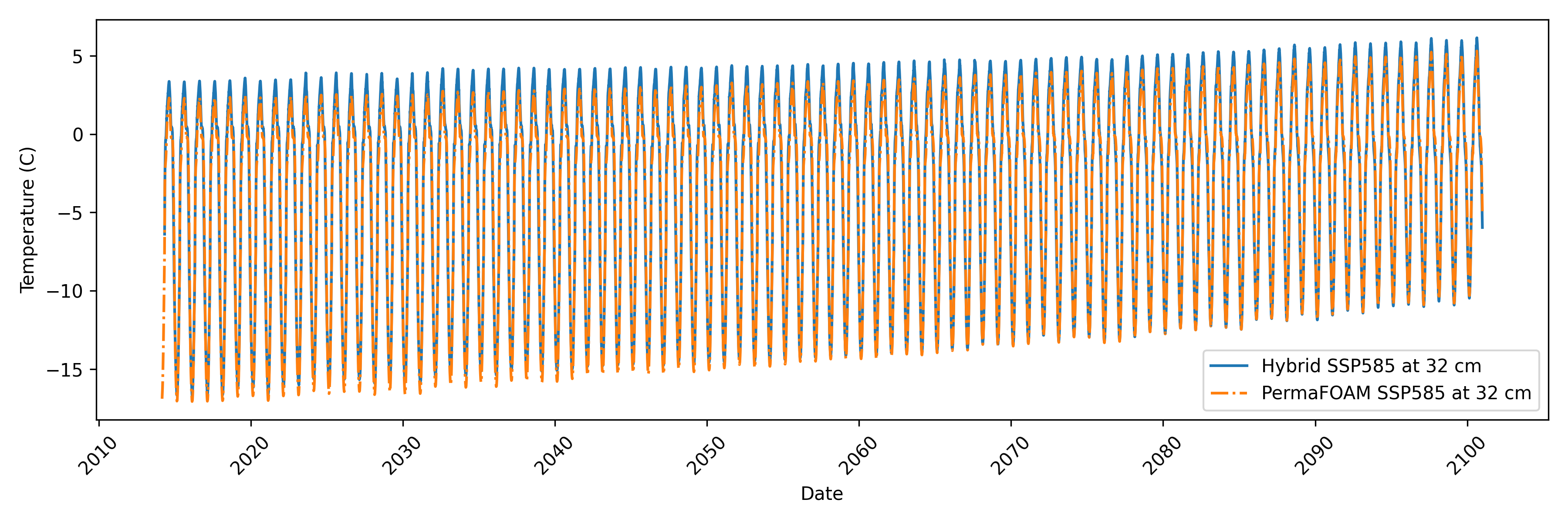}}
    \caption{North-facing slope comparison between PermaFOAM and the hybrid model at $32 cm$ depth. The temperature of the hybrid model are systemically higher than the PermaFOAM ones. Source: Authors own work.}
    \label{comparepermahybrid32north}
\end{figure}

The results of the extrapolation in time of the north-facing slope are illustrated in figures \ref{comparepermahybrid12north} and \ref{comparepermahybrid32north} for the probes at $12$ and $32 cm$ depth, respectively, for the four considered climate scenarios. Table \ref{tab:average_error_scenarios_north} reviews a comparison of the correction amplitude between the hybrid model and the PermaFOAM simulation results.

\begin{table}
\centering
\caption{Average correction predicted $\hat{T}_i^{hybrid}-T_i^{PermaFOAM}$ (the correction is reported in signed value and is positive) by the hybrid model for each depth and climate scenario on the north-facing slope. Source: Authors own work.}
\label{tab:average_error_scenarios_north}
\begin{tabular}{lrrrr}
\toprule
Depth & SSP126 & SSP245 & SSP370 & SSP585 \\
\midrule
12 cm & 0.4127 & 0.3427 & 0.2802 & 0.2472 \\
22 cm & 0.3641 & 0.3147 & 0.2868 & 0.2730 \\
32 cm & 0.4196 & 0.3642 & 0.3180 & 0.2883 \\
\bottomrule
\end{tabular}
\end{table}

\begin{figure}[!ht]
    \centering
    \subfigure[South-facing slope predictions at $8 cm$ depth using SSP126.]{\includegraphics[width=\linewidth]{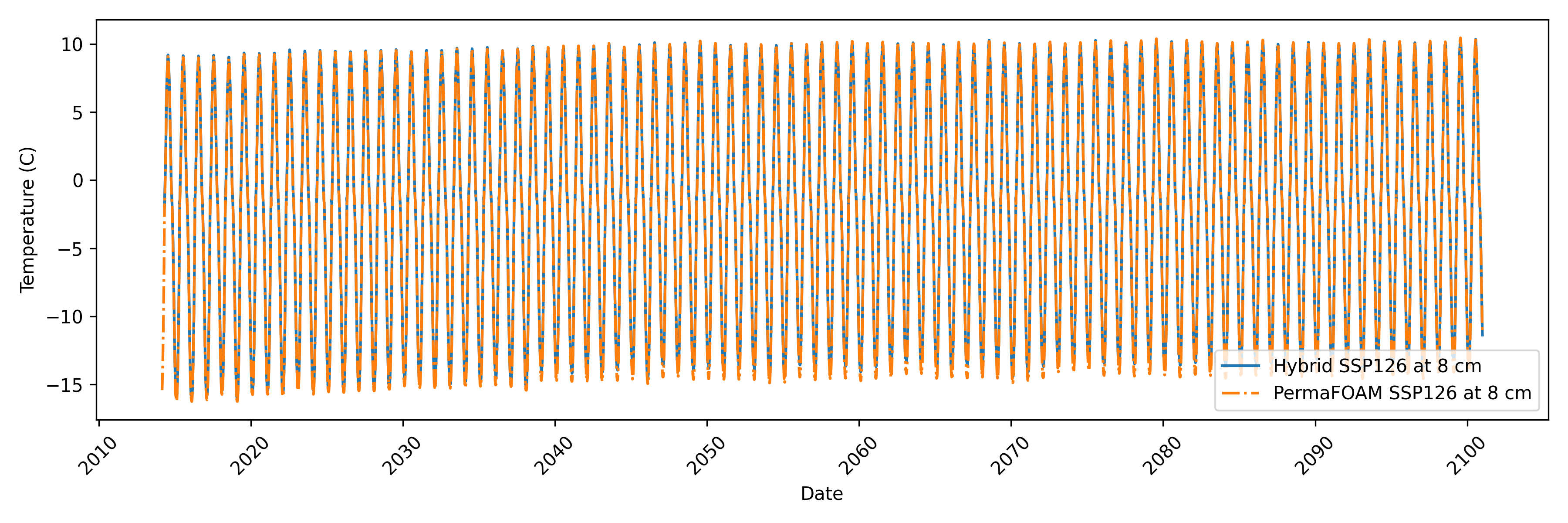}}
    \subfigure[South-facing slope predictions at $8 cm$ depth using SSP245.]{\includegraphics[width=\linewidth]{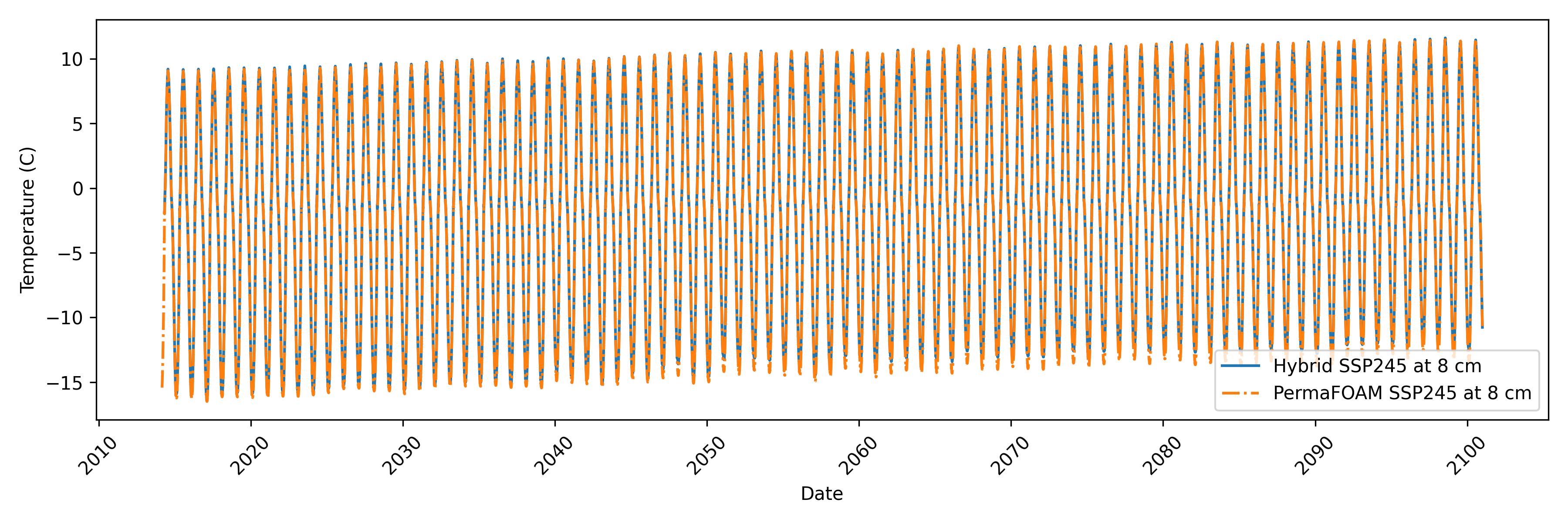}}
    \subfigure[South-facing slope predictions at $8 cm$ depth using SSP370.]{\includegraphics[width=\linewidth]{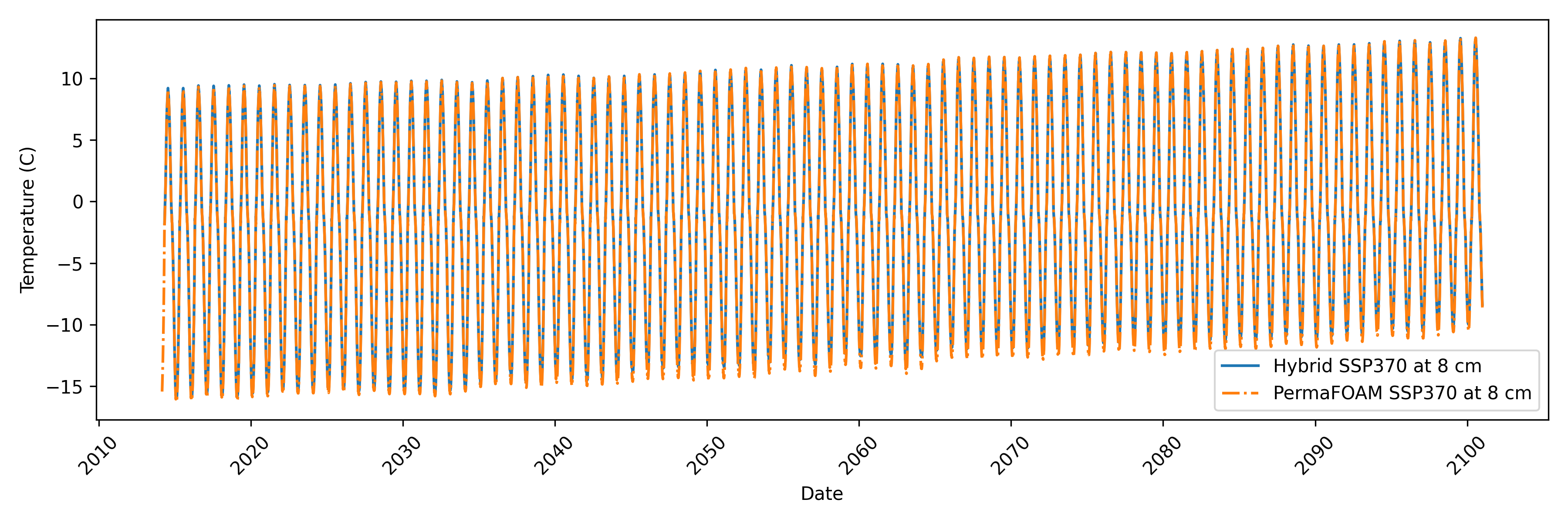}}
    \subfigure[South-facing slope predictions at $8 cm$ depth using SSP585.]{\includegraphics[width=\linewidth]{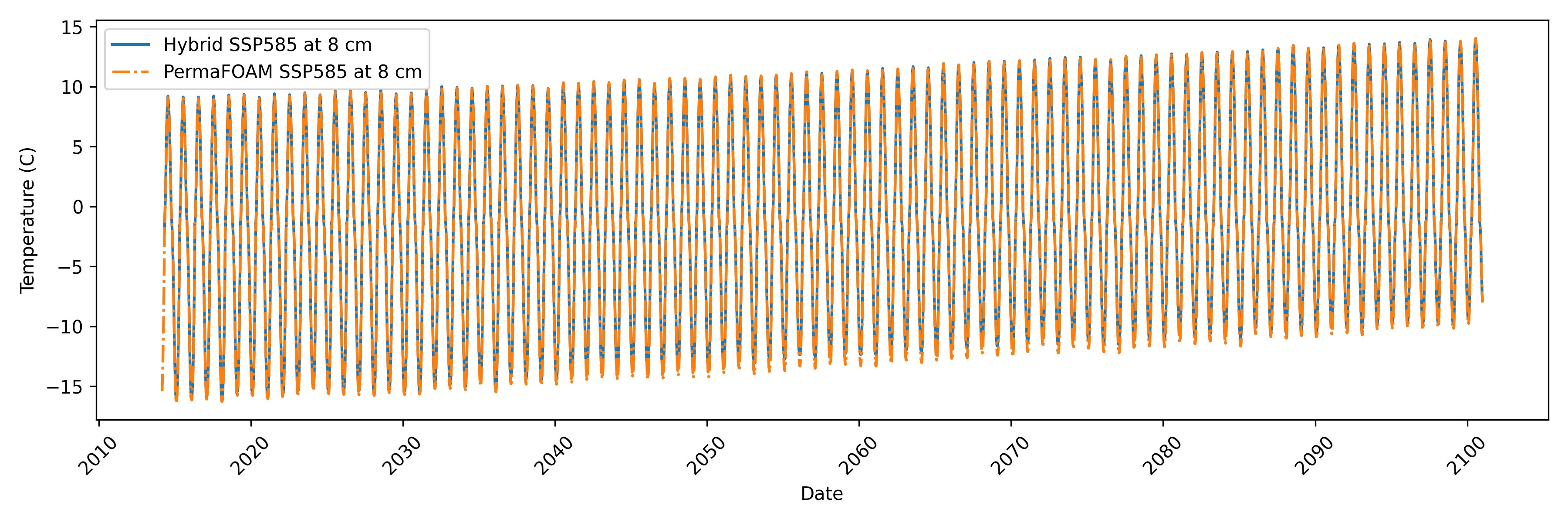}}
    \caption{South-facing slope comparison between PermaFOAM and the hybrid model at $8 cm$ depth. The temperature of the hybrid model are systemically higher than the PermaFOAM ones, with a less flagrant difference than for the north-facing slope. Source: Authors own work.}
    \label{comparepermahybrid8south}
\end{figure}

\begin{figure}[!ht]
    \centering
    \subfigure[South-facing slope predictions at $28 cm$ depth using SSP126.]{\includegraphics[width=\linewidth]{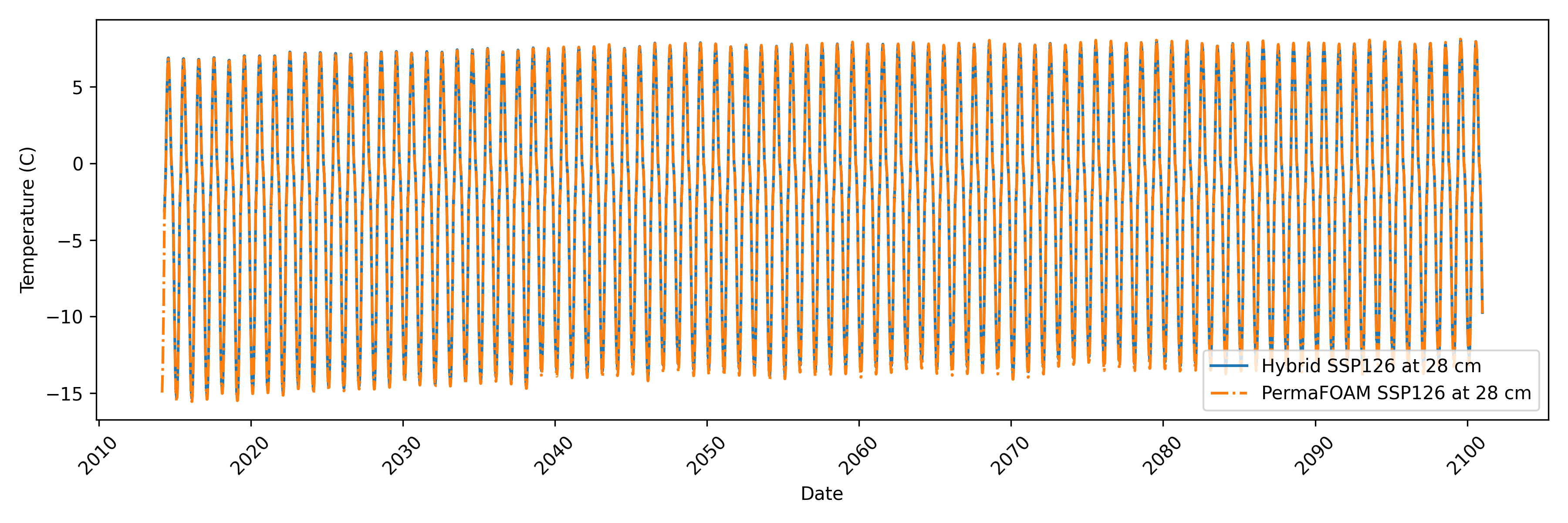}}
    \subfigure[South-facing slope predictions at $28 cm$ depth using SSP245.]{\includegraphics[width=\linewidth]{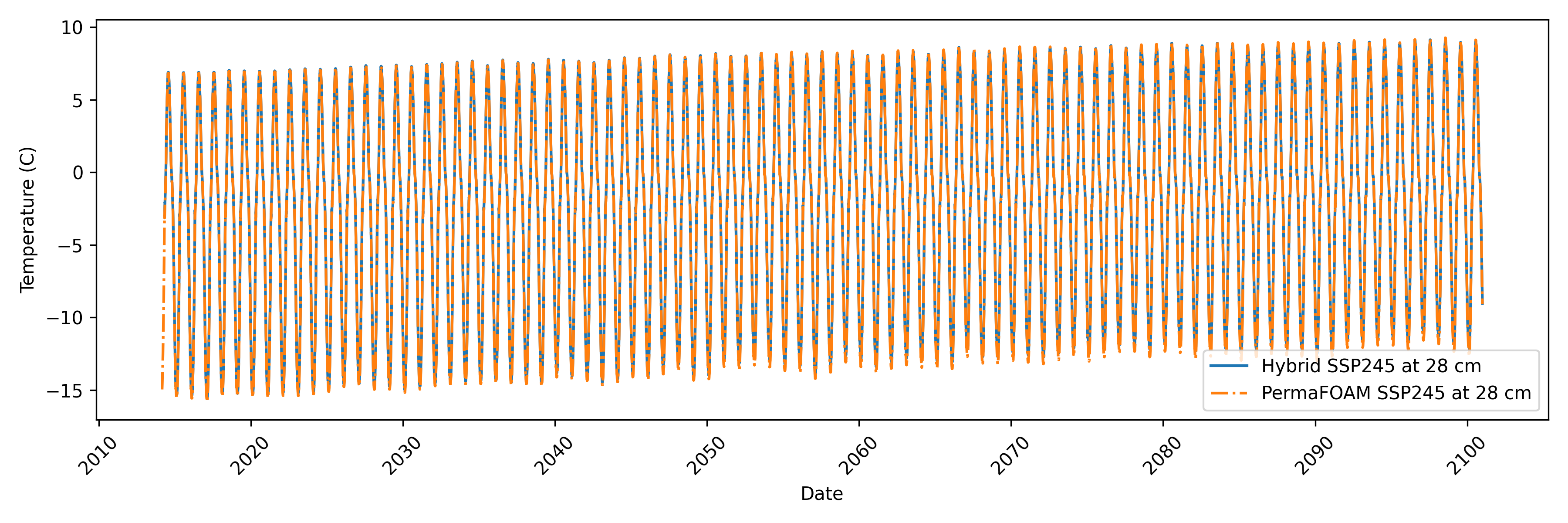}}
    \subfigure[South-facing slope predictions at $28 cm$ depth using SSP370.]{\includegraphics[width=\linewidth]{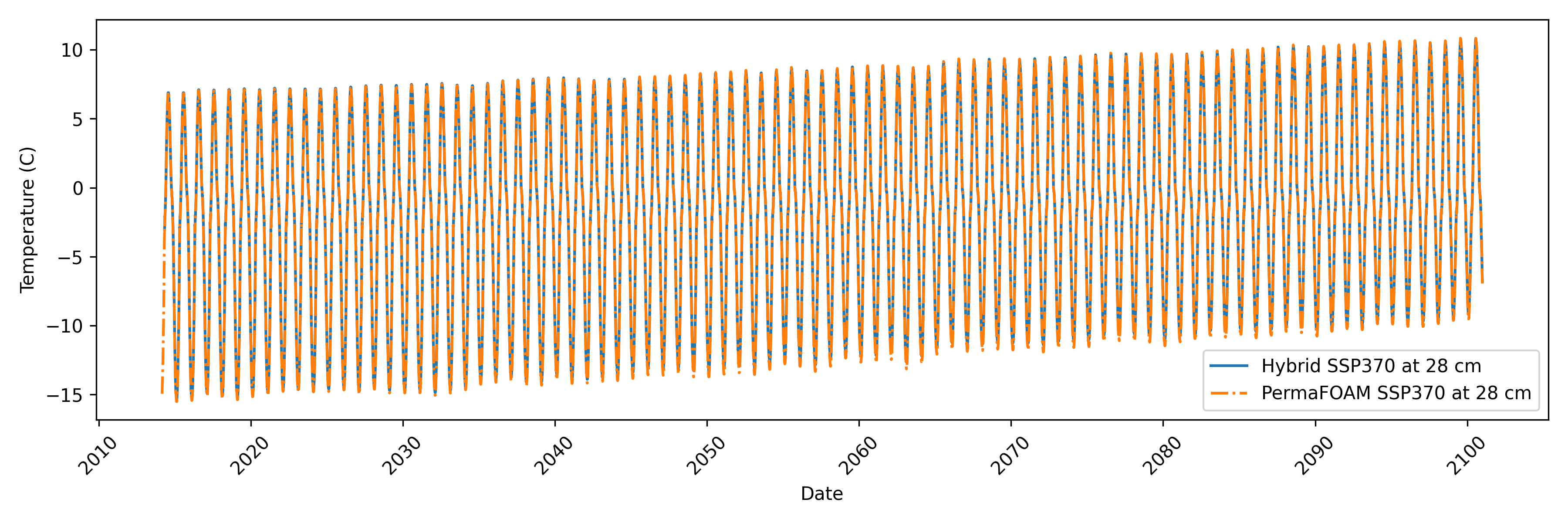}}
    \subfigure[South-facing slope predictions at $28 cm$ depth using SSP585.]{\includegraphics[width=\linewidth]{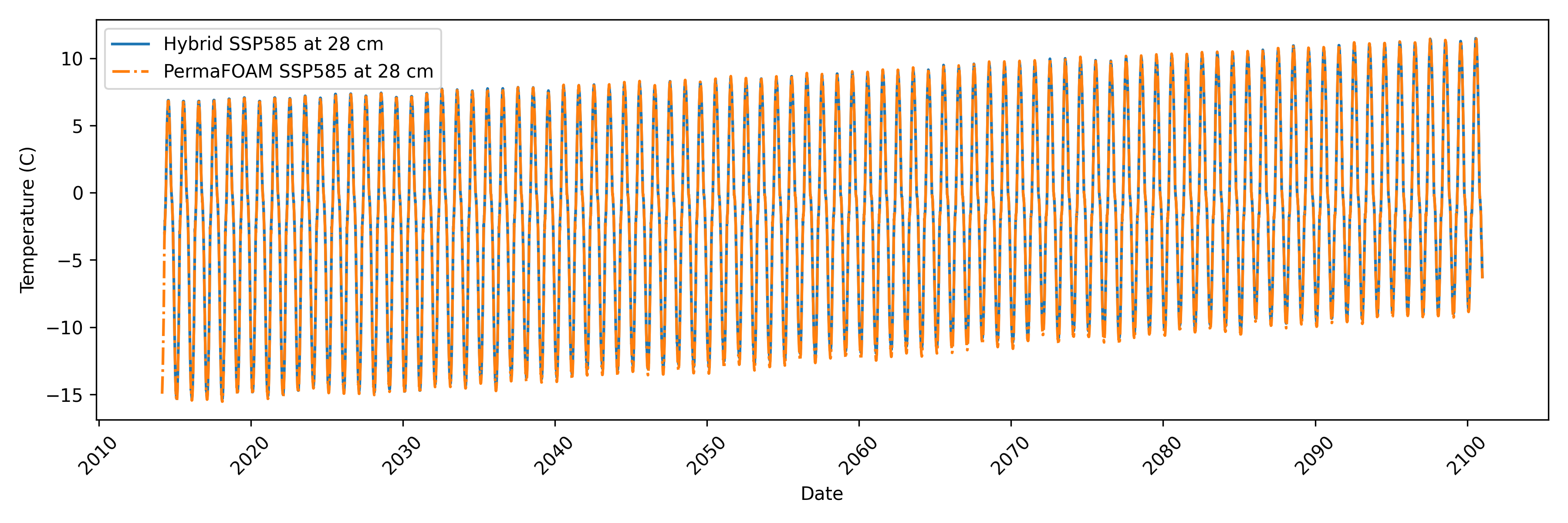}}
    \caption{South-facing slope comparison between PermaFOAM and the hybrid model at $28 cm$ depth. The temperature of the hybrid model are systemically higher than the PermaFOAM ones, with a less flagrant difference than for the north-facing slope. Source: Authors own work.}
    \label{comparepermahybrid28southth}
\end{figure}

When it comes to the south-facing slope, the results of the extrapolation in time are illustrated in figures \ref{comparepermahybrid8south} and \ref{comparepermahybrid28southth} for the probes at $8$ and $28 cm$ depth, respectively, for the four considered climate scenarios. Table \ref{tab:average_error_scenarios_south} reviews a comparison of the correction amplitude between the hybrid model and the PermaFOAM simulation results on the south-facing slope at different considered depths.

\begin{table}
\centering
\caption{Average correction predicted $\hat{T}_i^{hybrid}-T_i^{PermaFOAM}$ (the correction is reported in signed value and is positive) by the hybrid model for each depth and climate scenario on the south-facing slope. Source: Authors own work.}
\label{tab:average_error_scenarios_south}
\begin{tabular}{lrrrr}
\toprule
Depth & SSP126 & SSP245 & SSP370 & SSP585 \\
\midrule
8 cm & 0.1928 & 0.1834 & 0.1615 & 0.1578 \\
18 cm & 0.4699 & 0.4591 & 0.4436 & 0.4458 \\
28 cm & 0.1084 & 0.1040 & 0.1002 & 0.1006 \\
\bottomrule
\end{tabular}
\end{table}

\begin{figure}[!ht]
    \centering
    \subfigure[North-facing slope predictions at $12 cm$ depth using SSP126.]{\includegraphics[width=\linewidth]{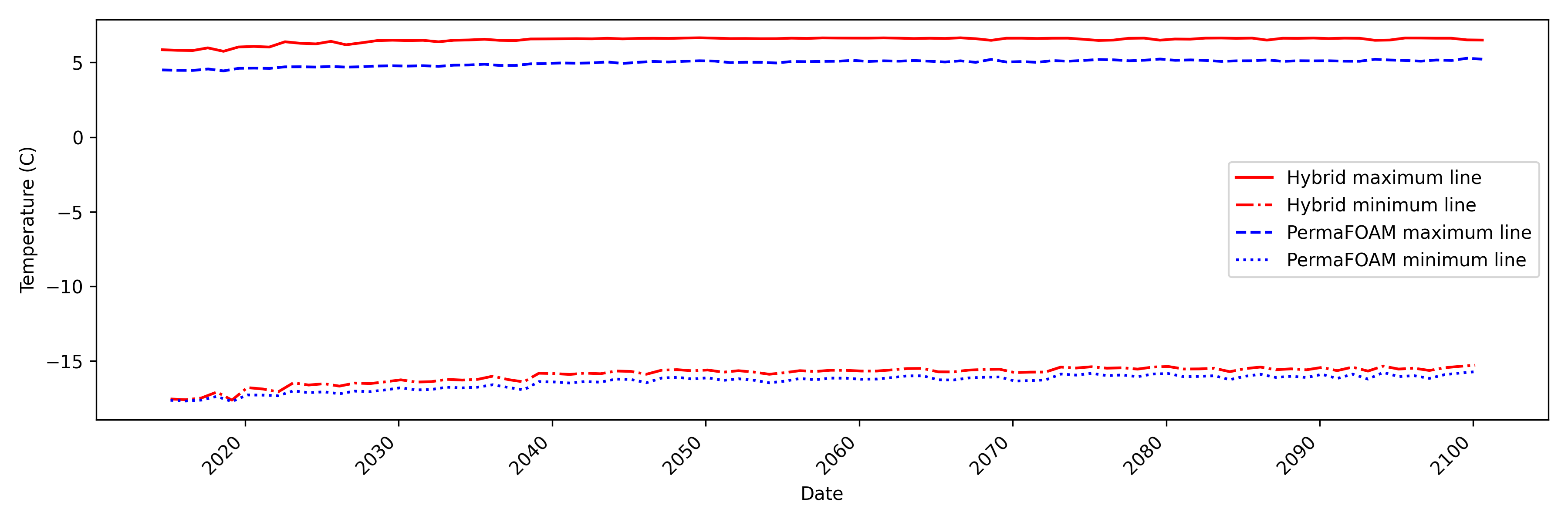}}
    \subfigure[North-facing slope predictions at $12 cm$ depth using SSP245.]{\includegraphics[width=\linewidth]{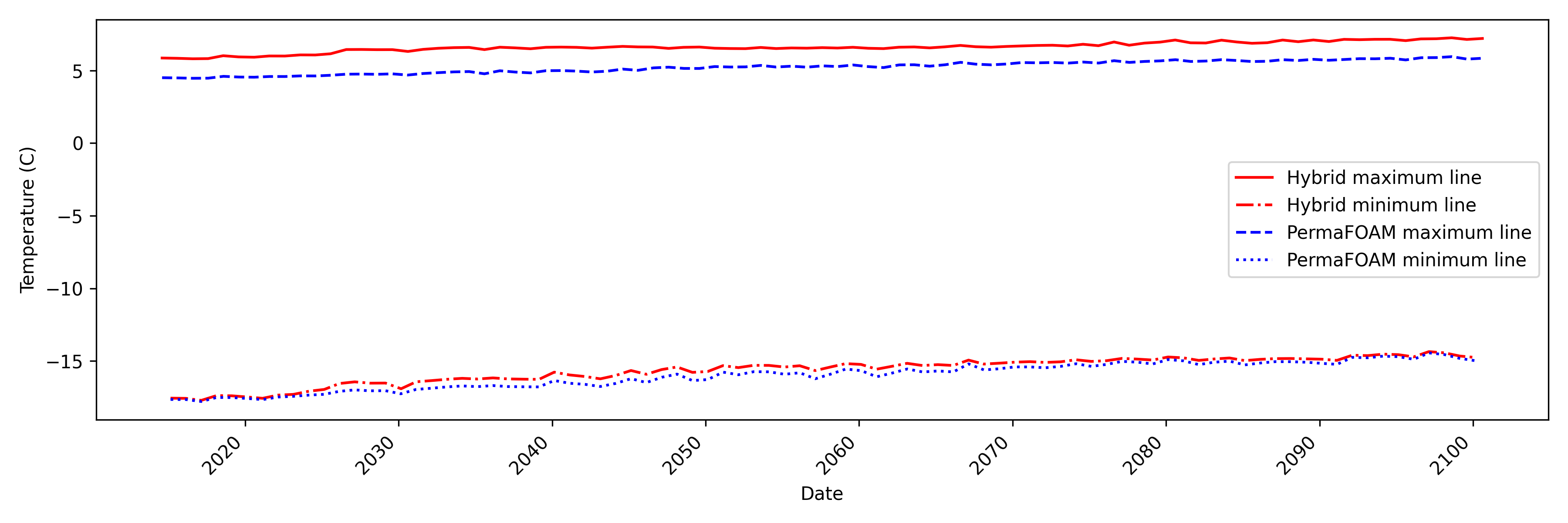}}
    \subfigure[North-facing slope predictions at $12 cm$ depth using SSP370.]{\includegraphics[width=\linewidth]{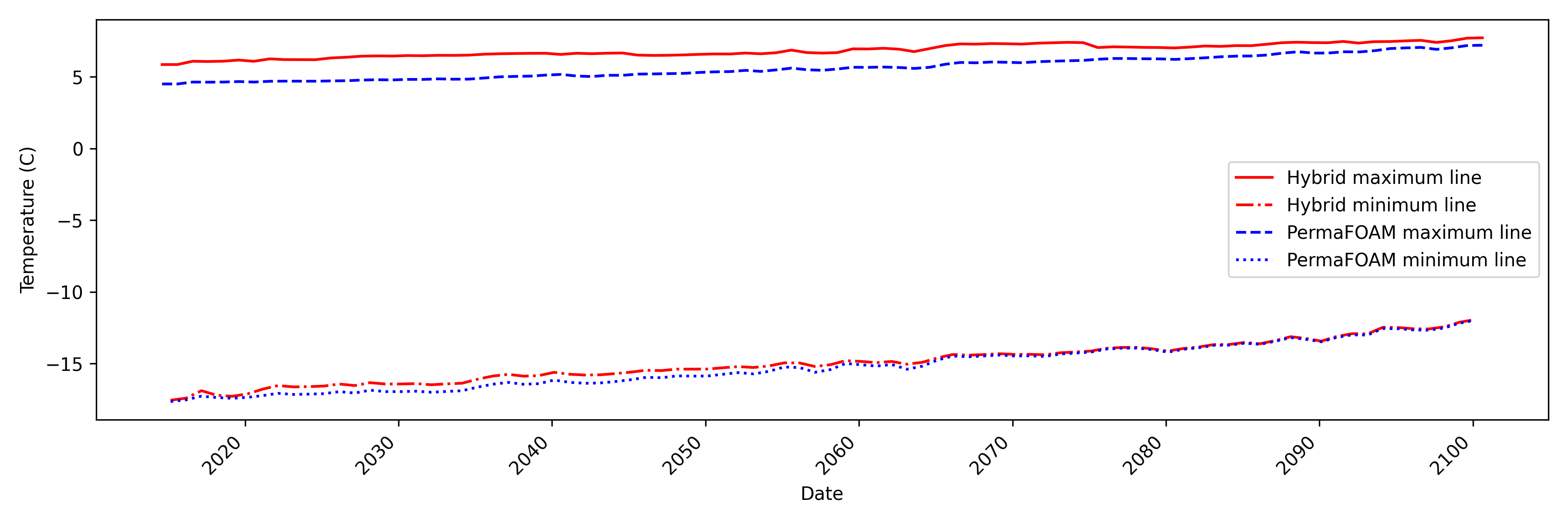}}
    \subfigure[North-facing slope predictions at $12 cm$ depth using SSP585.]{\includegraphics[width=\linewidth]{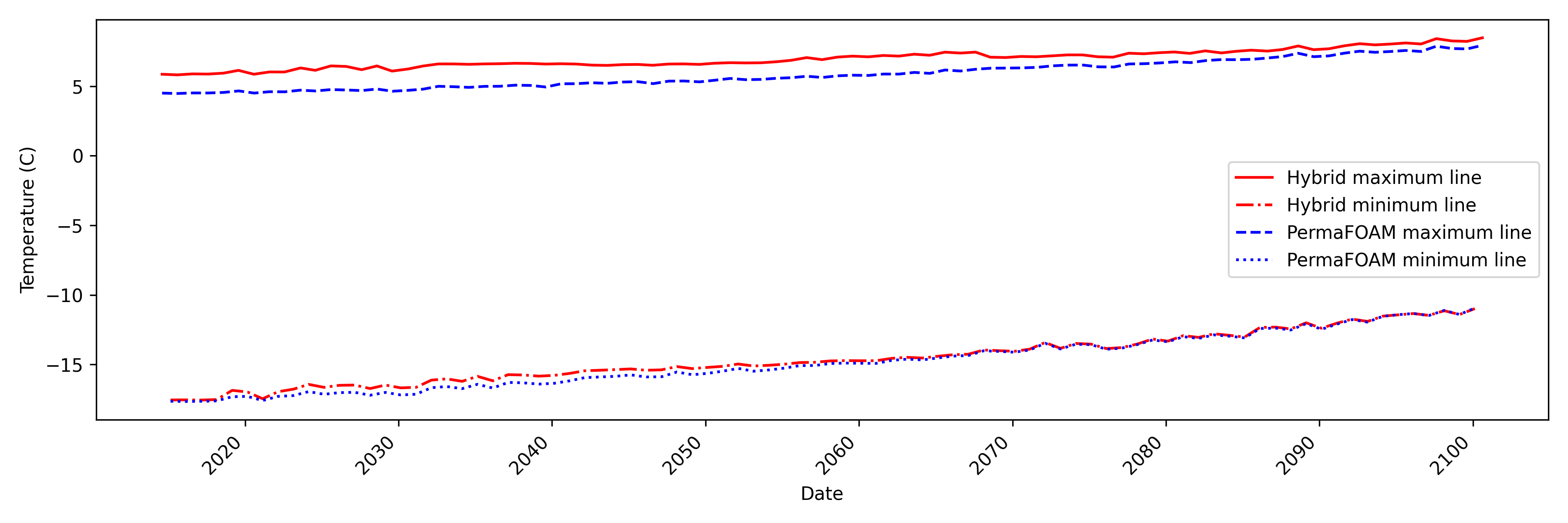}}
    \caption{North-facing slope comparison of the extrema lines between PermaFOAM and the hybrid model at $12 cm$ depth. Source: Authors own work.}
    \label{comparepermahybrid12northextreme}
\end{figure}

\begin{figure}[!ht]
    \centering
    \subfigure[North-facing slope predictions at $32 cm$ depth using SSP126.]{\includegraphics[width=\linewidth]{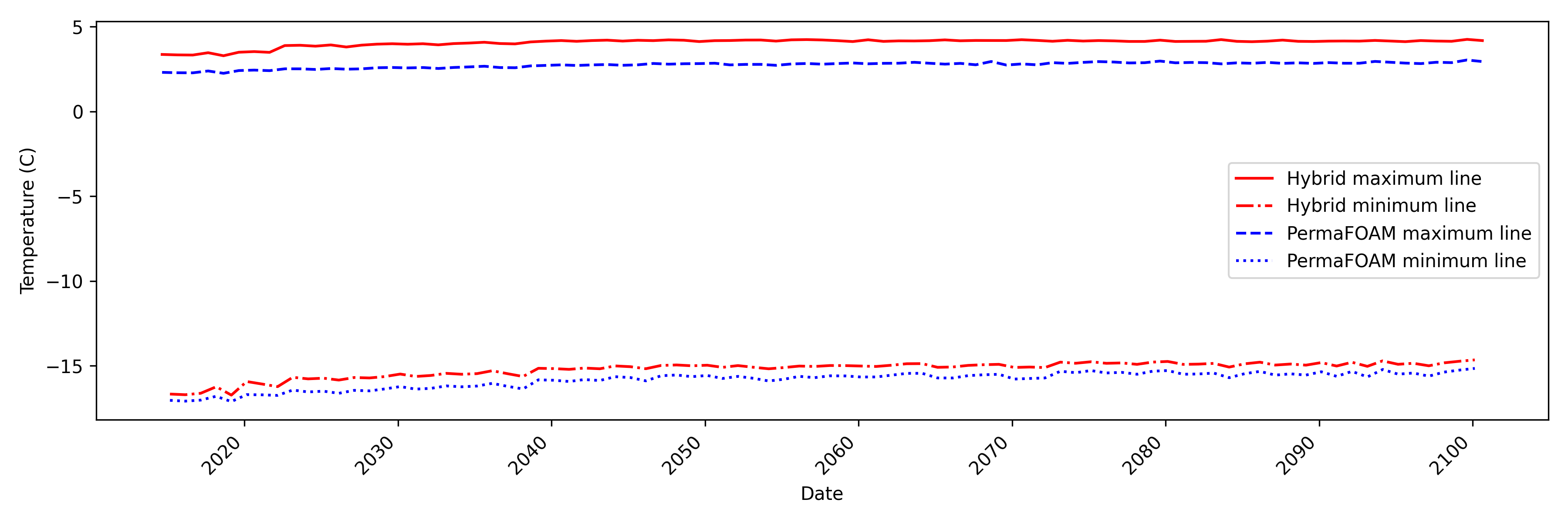}}
    \subfigure[North-facing slope predictions at $32 cm$ depth using SSP245.]{\includegraphics[width=\linewidth]{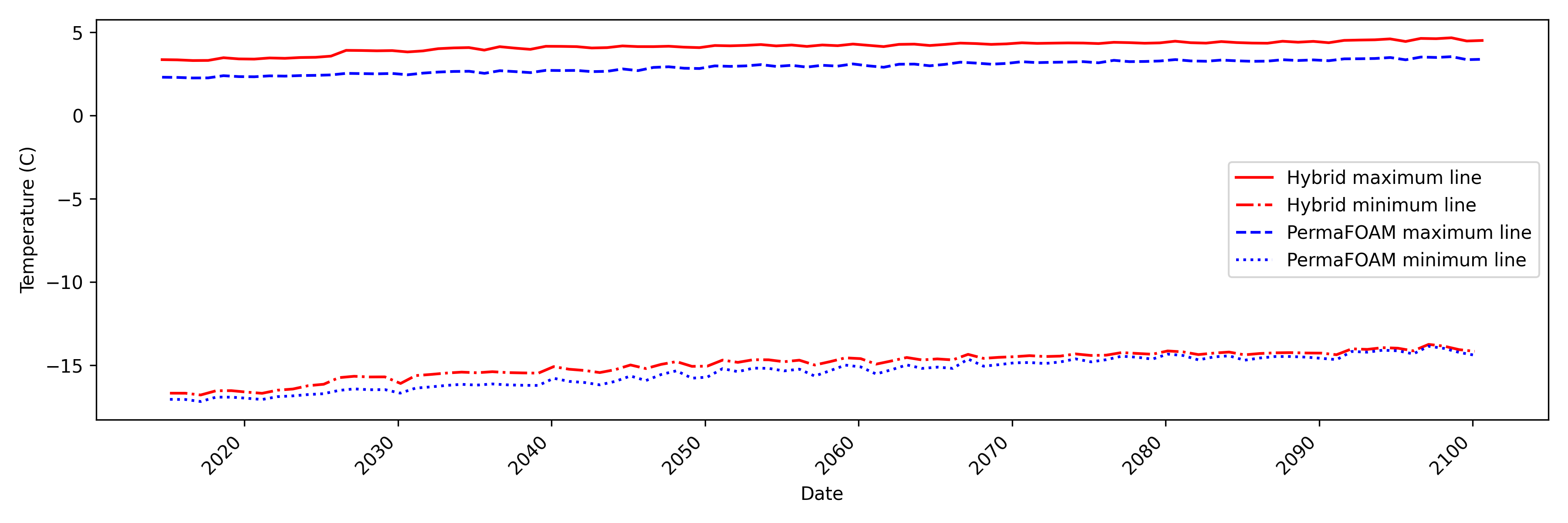}}
    \subfigure[North-facing slope predictions at $32 cm$ depth using SSP370.]{\includegraphics[width=\linewidth]{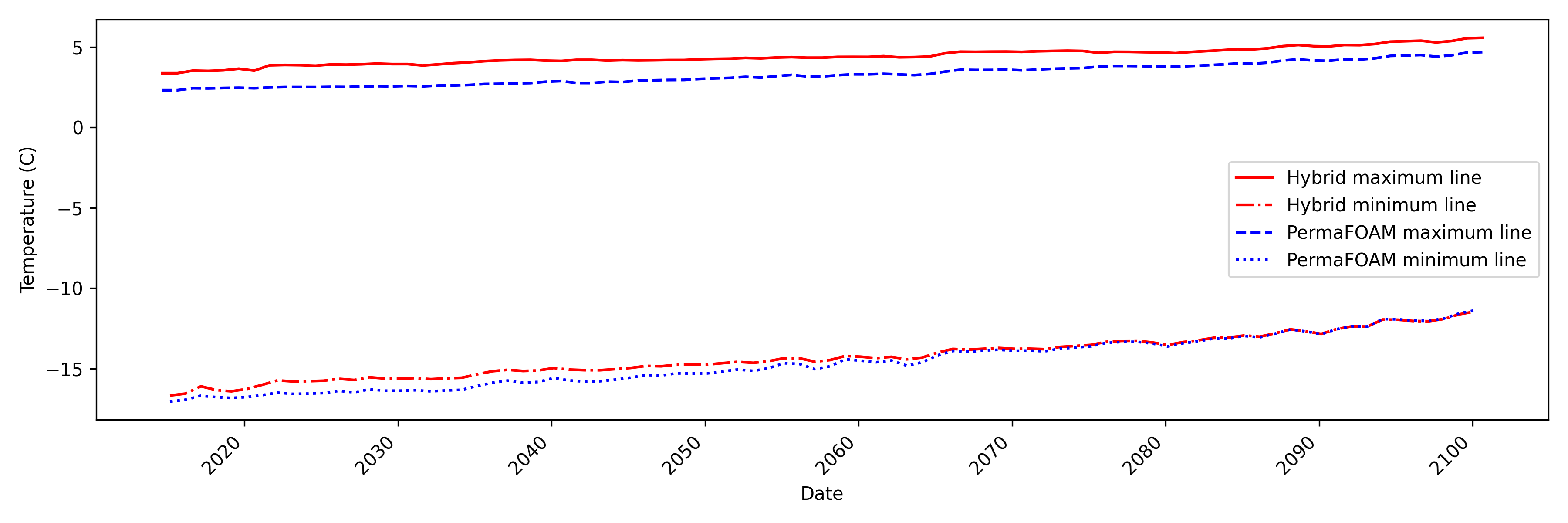}}
    \subfigure[North-facing slope predictions at $32 cm$ depth using SSP585.]{\includegraphics[width=\linewidth]{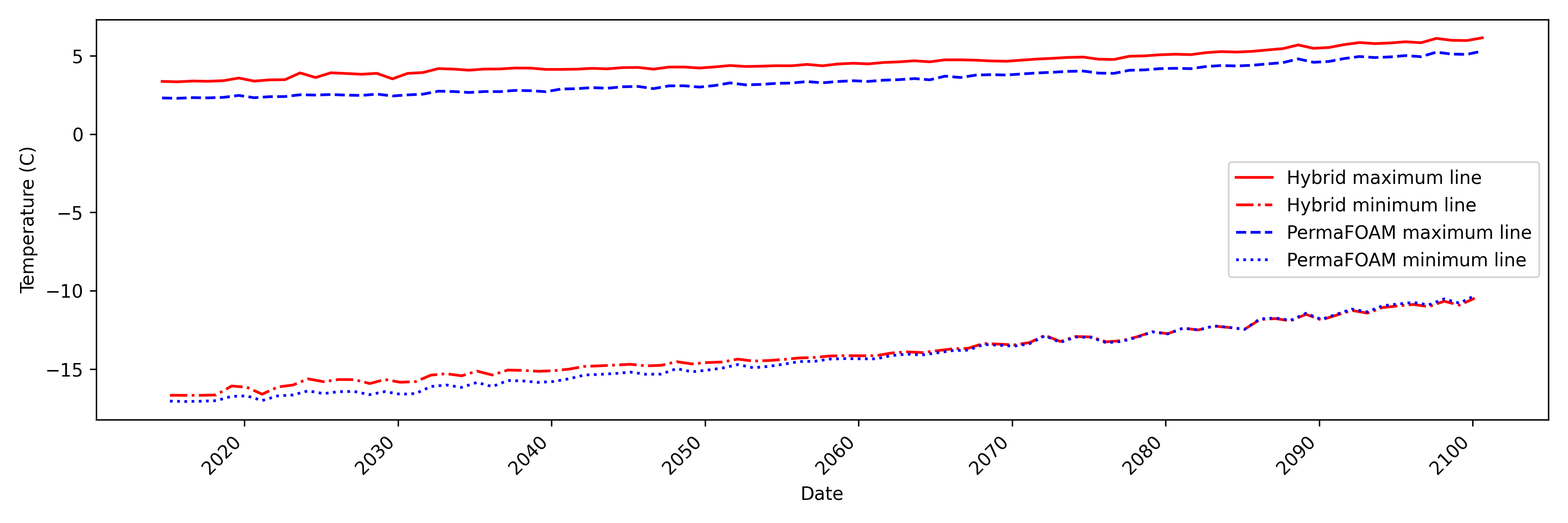}}
    \caption{North-facing slope comparison of the extrema lines between PermaFOAM and the hybrid model at $32 cm$ depth. Source: Authors own work.}
    \label{comparepermahybrid32northextreme}
\end{figure}

\begin{figure}[!ht]
    \centering
    \subfigure[South-facing slope predictions at $8 cm$ depth using SSP126.]{\includegraphics[width=\linewidth]{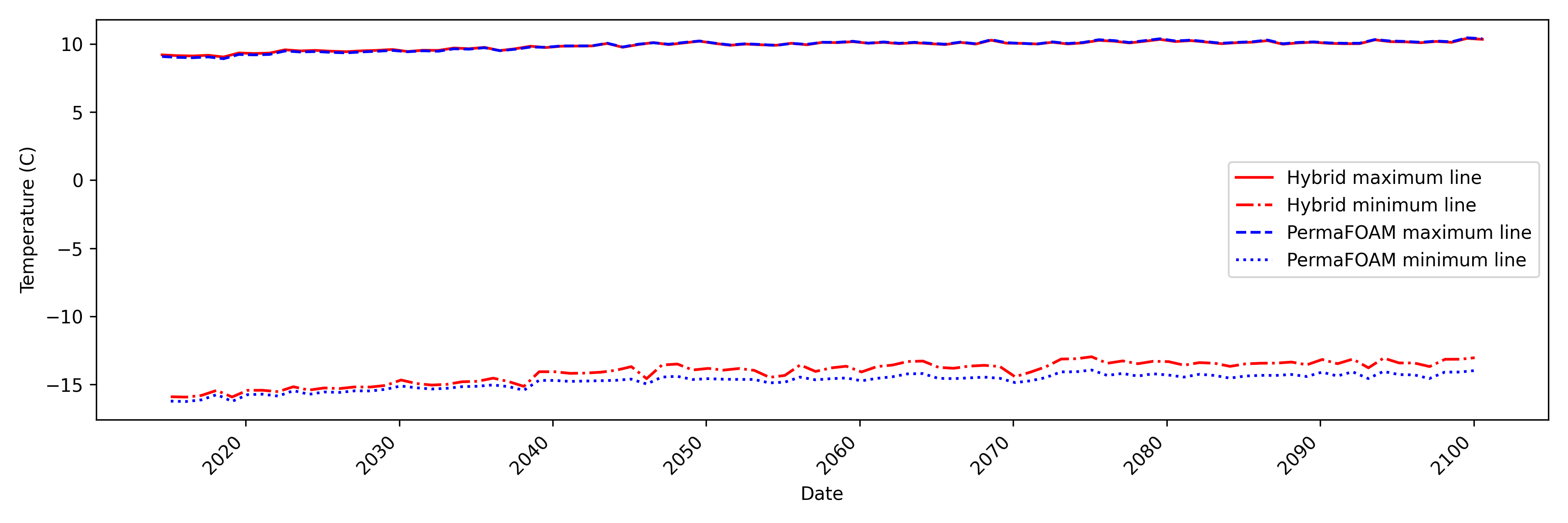}}
    \subfigure[South-facing slope predictions at $8 cm$ depth using SSP245.]{\includegraphics[width=\linewidth]{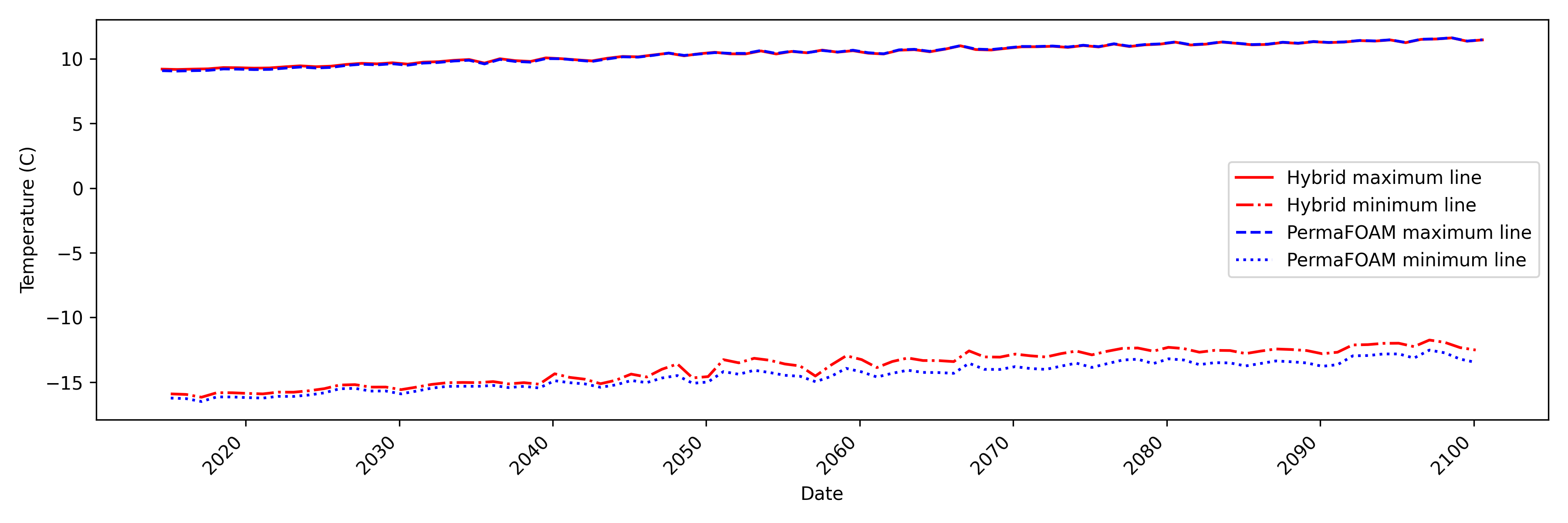}}
    \subfigure[South-facing slope predictions at $8 cm$ depth using SSP370.]{\includegraphics[width=\linewidth]{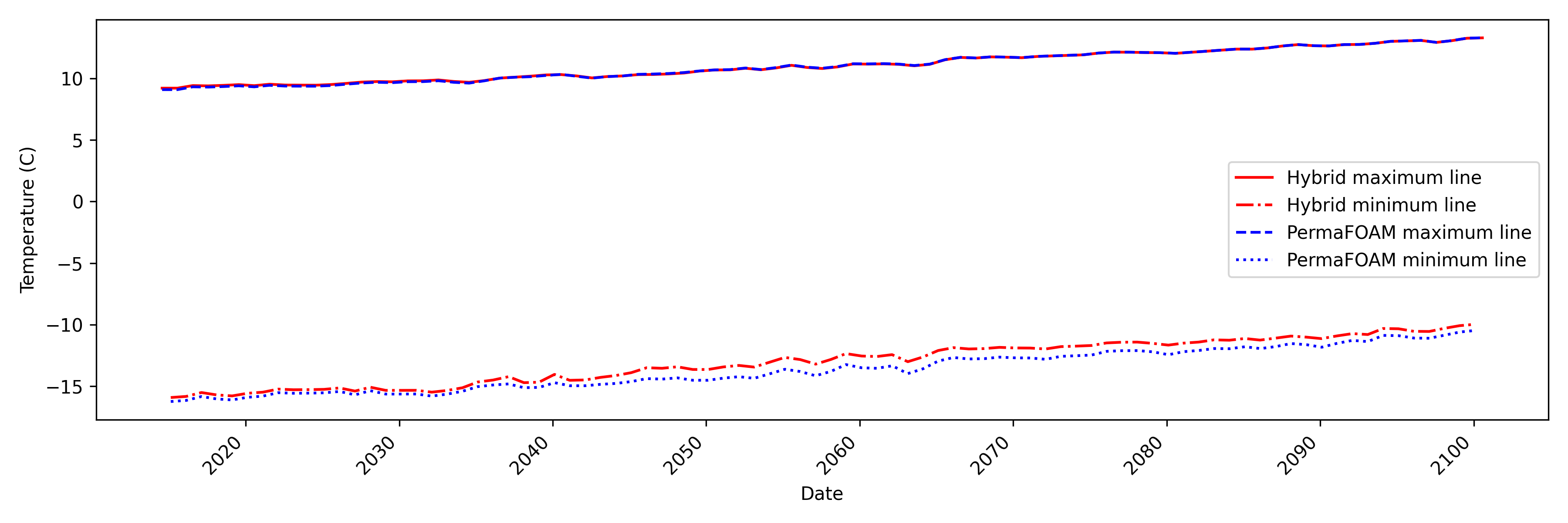}}
    \subfigure[South-facing slope predictions at $8 cm$ depth using SSP585.]{\includegraphics[width=\linewidth]{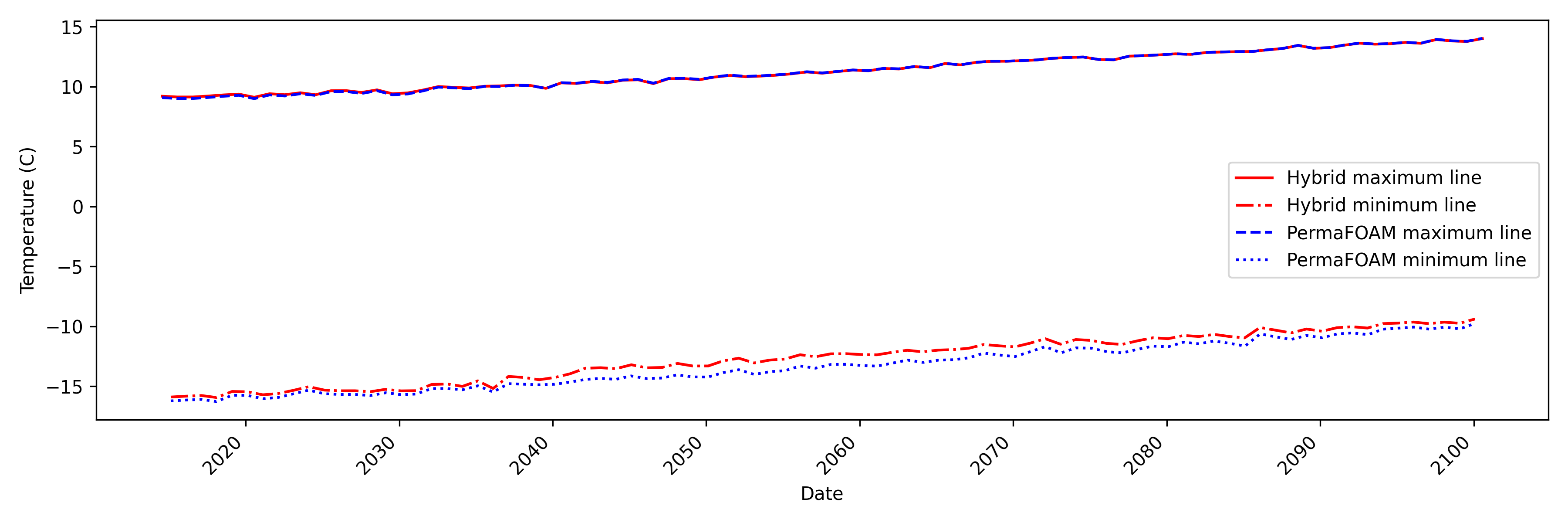}}
    \caption{South-facing slope comparison of the extrema lines between PermaFOAM and the hybrid model at $8 cm$ depth. Source: Authors own work.}
    \label{comparepermahybrid8southextr}
\end{figure}

\begin{figure}[!ht]
    \centering
    \subfigure[South-facing slope predictions at $28 cm$ depth using SSP126.]{\includegraphics[width=\linewidth]{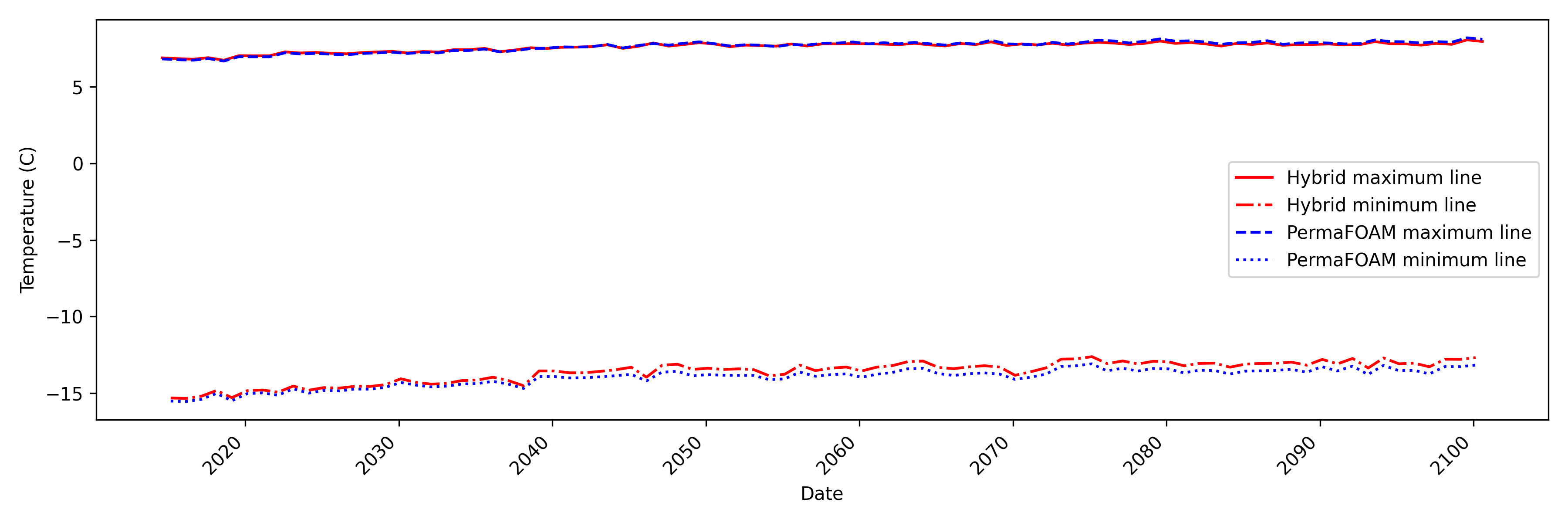}}
    \subfigure[South-facing slope predictions at $28 cm$ depth using SSP245.]{\includegraphics[width=\linewidth]{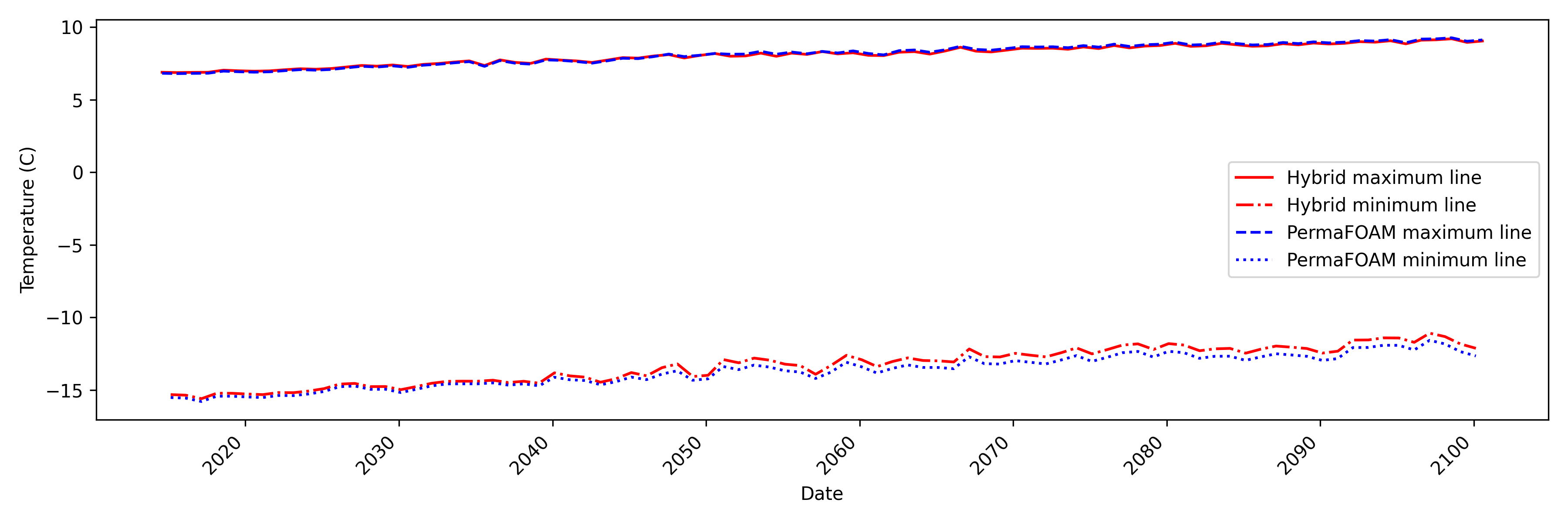}}
    \subfigure[South-facing slope predictions at $28 cm$ depth using SSP370.]{\includegraphics[width=\linewidth]{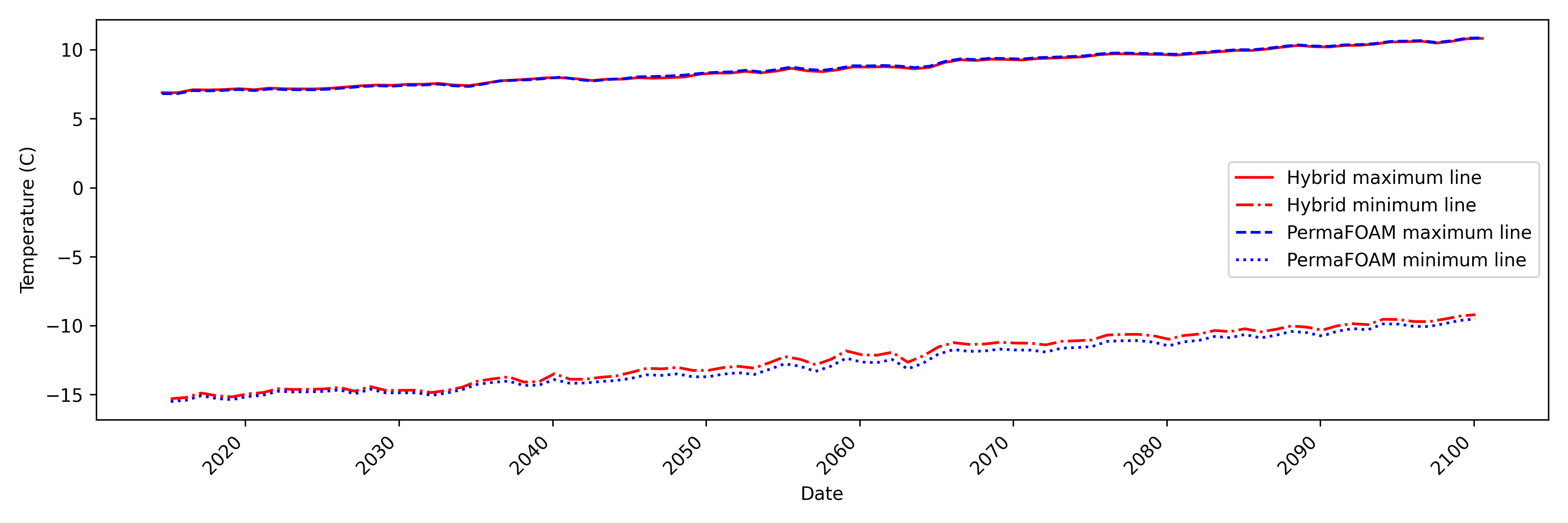}}
    \subfigure[South-facing slope predictions at $28 cm$ depth using SSP585.]{\includegraphics[width=\linewidth]{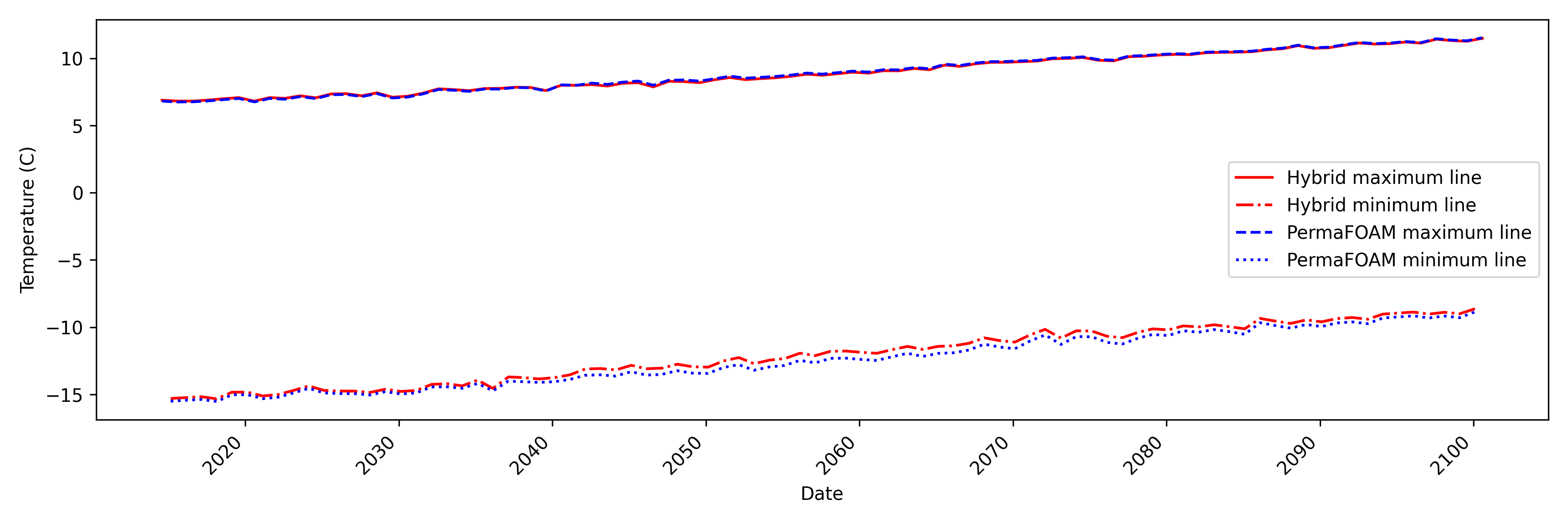}}
    \caption{South-facing slope comparison of the extrema lines between PermaFOAM and the hybrid model at $28 cm$ depth. Source: Authors own work.}
    \label{comparepermahybrid28southextr}
\end{figure}

To further investigate the effects of the correction, we plot the extrema lines along time, such as the lines of absolute maximum and absolute minimum temperature per year. The results are illustrates in Figures \ref{comparepermahybrid12northextreme} and \ref{comparepermahybrid32northextreme} for the north-facing slope and Figures \ref{comparepermahybrid8southextr} and \ref{comparepermahybrid28southextr} for the south-facing slope. A general remark we may note is that the corrected hybrid model is systematically generating as high or higher temperatures than the PermaFOAM simulations.

\section{Discussion}\label{discussionsec}
The hybrid modeling (surrogate model plus correction) of the permafrost shows a high correlation with the experimental measurements on both train and test sets. The relative error and the statistical metrics both show an improvement with respect to the PermaFOAM model alone. 

When it comes to the extrapolation to 2100, the hybrid modeling shows an increased temperature with respect to the PermaFOAM simulations, in both lowest and highest annual temperatures per year, as illustrated in Figures \ref{comparepermahybrid12northextreme}-\ref{comparepermahybrid28southextr}. The difference is marginal in the beginning of the century and increases systematically until the end of the century, reaching around $1^\textrm{o}$C by the end of the century for some climate scenarios.

The results show that the corrective effects of the hybrid modeling may highlight a faster-than-expected increase in temperature of the permafrost. This effect can potentially lead to faster degradation and more aggressive effects of the climate change on the permafrost layers. 

\section{Conclusion}\label{conclusionsec}
In this work we build a real-time surrogate model of the PermaFOAM simulation framework, using a combination of LSTM and dense neural networks, aiming to accelerate the simulation while preserving the computational fidelity. The simulation showed a high ability of the surrogate model to reproduce the PermaFOAM simulation with a mean relative error lower than 1\% across train and test sets. This relative error corresponds to less than that $0.19^\textrm{o}$C on average. The benefit in computation time spans several orders of magnitude, as the training time took a few hours on a typical laptop, while the inception time is less than 0.7 seconds on the same device, compared to more than a month on HPC facility with multi-core access for the original PermaFOAM simulation.

The availability of the surrogate model allows testing multiple climate scenarios within few seconds, allowing to gain further insights into the climate change effect on the permafrost without the time, energy and monetary cost involved in HPC computing.

Moreover, the surrogate model is complemented by a hybrid correction surrogate to improve the simulation accuracy with respect to the experimental measurements. The hybrid model improves the prediction results on both train and test sets. The trained hybrid model is later used to perform more accurate predictions of the thermal fields in the permafrost. 

The results of the trained hybrid model extrapolation into the upcoming century (up to 2100) show an increased predicted temperature when compared to the original PermaFOAM simulations. This increased temperature is systematic across all climate change scenarios. Moreover, the results highlight the possible higher-than-expected temperature increase of the permafrost and, therefore, a more aggressive than expected effect of the climate change scenarios on the permafrost behavior, mechanical properties, and melting time.

\hfill

\textit{Acknowledgments} \\
This work has been partially funded by the French Research Agency ANR (project HiPerBorea grant n ◦ ANR-19 CE46-0003-01 and project PERMACHANGE grant n ◦ ANR-25-CE46-5686), and beneﬁted from access to the supercomputers of CALMIP (project p12166) and GENCI (project A0060110794, CINES and TGCC).\\

\textit{Disclosure statement:} \\No potential conflict of interest is reported by the authors. 
%\break
%\printbibliography
\bibliographystyle{agsm}
\bibliography{Bibliography}

\end{document}